\documentclass[fleqn,usenatbib]{mnras}
\usepackage[utf8x]{inputenc}
\usepackage[T1]{fontenc}
\usepackage{ae,aecompl}
\usepackage[english]{babel}
\usepackage{lmodern}
\usepackage{enumitem}
\usepackage[squaren, Gray, cdot]{SIunits}
\usepackage{multirow}
\usepackage{color}
\usepackage{array}
\usepackage{setspace}
\usepackage{graphicx}
\usepackage{amsmath}
\usepackage{amssymb}
\usepackage{bm}
\usepackage[super]{nth}
\usepackage{cuted}

\definecolor{Blue}{rgb}{0,0.08,0.65}
\definecolor{Red}{rgb}{0.65,0.08,0.05}
\definecolor{Green}{rgb}{0.15,0.45,0.25}
\definecolor{Purple}{RGB}{153, 51, 153}

\newcommand \red[1] {{#1}}

\def\map{M_{\rm ap} }

\usepackage{mathtools, stmaryrd}
\usepackage{xparse} \DeclarePairedDelimiterX{\Iintv}[1]{\llbracket}{\rrbracket}{\iintvargs{#1}}
\NewDocumentCommand{\iintvargs}{>{\SplitArgument{1}{,}}m}
{\iintvargsaux#1} %
\NewDocumentCommand{\iintvargsaux}{mm} {#1\mkern1.5mu..\mkern1.5mu#2}

\title[Aperture mass PDF]{Probability distribution function of the aperture mass field with large deviation theory}

\author[A. Barthelemy et al.]{
Alexandre Barthelemy$^{1}$\thanks{E-mail: alexandre.barthelemy@iap.fr},
Sandrine Codis$^{1,2,3}$, Francis Bernardeau$^{1,2}$
\\ 
$^{1}$CNRS \& Sorbonne Universit\'e, UMR 7095, Institut d'Astrophysique de Paris, 75014, Paris, France\\
$^{2}$Institut de Physique Th\'eorique, Universit\'e Paris-Saclay,
CEA, CNRS, UMR 3681, 91191 Gif-sur-Yvette, France\\
$^{3}$AIM, CEA, CNRS, Universit\'e Paris-Saclay, Universit\'e Paris Diderot, Sorbonne Paris Cit\'e, 91191 Gif-sur-Yvette, France\\
}

\date{Accepted XXX. Received YYY; in original form ZZZ}

\pubyear{2020}

\begin{document}

\label{firstpage}
\pagerange{\pageref{firstpage}--\pageref{lastpage}}
\maketitle

\begin{abstract}
In the context of tomographic cosmic shear surveys, a theoretical model for the one-point statistics of the aperture mass (Map) is developed. This formalism is based on the application of the large deviation principle to the projected matter density field and more specifically to 
the angular aperture masses. The latter holds the advantage of being an observable that can be directly extracted from the observed shear field and to be, by construction, independent from the long wave modes. Furthermore we show that, with the help of a nulling procedure based on the so-called BNT transform, it is possible to build observables that depend only on a finite range of redshifts making them also independent from the small-scale modes. This procedure makes predictions for the shape of the one-point Probability Distribution Function of such an observable very accurate, comparable to what had been previously obtained for 3D observables. Comparisons with specific simulations reveal however inconsistent results showing that synthetic lensing maps were not accurate enough for such refined observables. It points to the need for more precise  
dedicated numerical developments whose performances could be benchmarked with such observables. We furthermore review the possible systematics that could affect such a formalism in future weak-lensing surveys like Euclid, notably the impact of shape noise as well as leading corrections coming from lens-lens couplings, geodesic deviation, reduced shear and magnification bias.
\end{abstract}


\section{Introduction}
The effect of weak gravitational lensing (WL) originates from the propagation of light rays through the inhomogeneous distribution of baryonic and dark matter which induces slight (de)magnification of the brightness of galaxies and distortion from their intrinsic shape \citep{1992ApJ...388..272K}. The statistics of WL fields provide a powerful tool for precision cosmology (see e.g. a review in \cite{kilbinger15}) and motivated the build up of new generation large galaxy surveys such as the Legacy Survey of Space and Time (LSST) \citep{LSST} or Euclid \citep{Euclid} which will provide data of unprecedented quality in the coming years. As such and in order to reach the percent precision and accuracy on the estimation of cosmological parameters, theorists need to build tools that can optimally extract information from those datasets and be able to provide accurate predictions in the non-linear regime of cosmic structure formation.

The most common approach is to focus on the information contained in the power spectra or equivalently their real-space counterparts the two-point correlation functions. Unfortunately these observables contain only complete statistical information for Gaussian random fields, a prescription valid with extremely good accuracy to describe primordial metric perturbations visible in the cosmic microwave background \citep{planck}. However, even starting from Gaussian initial conditions, the subsequent {\it non-linear} time-evolution of density fluctuations by means of the gravitational instability develops significant non-Gaussianities, in particular for small scales and late times. In this non-linear regime of structure formation, we observe both an increase in power in the power spectrum measurements (relative to linear evolution) and a generation of distinct non-Gaussianities in the late-time density field, which by projection also implies strong non-Gaussian features of the weak-lensing fields carrying non-negligible cosmological information. More quantitatively, \cite{DeepLearning} recently showed with a series of Deep Neural Networks how a significant amount of the cosmological information in the convergence field lies in the extreme rare events, that is the tails of the probability distribution function (PDF), and a Fisher analysis based on fast simulations in \cite{Patton17} demonstrated that the weak-lensing convergence PDF provides information complementary to the cosmic shear two-point correlation. \red{Note that this complementarity between two-point and non-Gaussian statistics is even more relevant in the presence of systematics such as shot and shape noise.} Still making use of Fisher analysis but this time using a "first principles" theoretical model of the convergence PDF, \cite{Boyle2020} also found that it provides tighter constraints for the equation of state of dark energy, the amplitude of fluctuations, the total matter fraction and the sum of neutrino masses, especially when performing a multi-scale analysis and in addition to the two-point correlation function. 
For the case of other non-Gaussian statistics, namely peak counts and Minkowski functionals, a recent analysis can be found in \cite{2020arXiv200612506Z} where it is shown that the Figure-of-Merit in the $\Omega_m - \sigma_8$ plane increases by a factor of 5 when adding those non-Gaussian statistics to the standard angular power spectrum. Note however that these quantitative analysis assume that non-Gaussian quantities can be predicted with great precision at the chosen filtering scales, which is not all a given especially when using un-tested -- because of the absence of a good theoretical model -- predictions from numerical simulations.

Because convergence maps are always only reconstructed up to a mass sheet degeneracy, \cite{Schneider1998} introduced the aperture mass field which boils down to filtering the convergence field with a compensated filter which also possesses a dual representation in the (almost) directly observable tangential-shear space. The study of the aperture mass can thus be performed directly from the measured data and treatments of masks in the field of view of a specific survey can for example be more controlled \citep{2020arXiv200608665P, 2016ApJ...819..158B}. Moreover the same non-Gaussian statistics (peaks, moments, PDF to name a few) can be used to probe the non-Gaussian features of the aperture mass field and are found, similarly to the convergence field, to provide complementary information to the power spectrum in different surveys \citep{2018MNRAS.474..712M, 2016MNRAS.463.3653K} and very recently in \cite{2020arXiv201007376M}. \red{It is to be noted that the recent introduction of the DES density-split statistics \citep{FriedrichDES17, GruenDES17} is another relevant method to treat the PDF of the tangential shear profiles at similar scales and redshifts than the present paper.}

In this paper we build a theoretical model for the aperture mass PDF. So far, only a few theoretical developments have been carried out in this direction: in the early 2000's \cite{BernardeauValageas} and following papers or \cite{2004MNRAS.350...77M} built the $\map$ PDF assuming different hierarchical models for the underlying density field. More recently \cite{paolo} used, as in this paper, large deviation theory \citep{seminalLDT} to compute the reduced-shear correction to the $\map$ PDF, but without accounting for the geometry of the past light-cone and projection effects nor comparing the predictions to numerical simulations, which are the main purposes of the present work.

The paper is organised as follows. Section~\ref{section::PDF} introduces all the formalism and procedure necessary to compute the aperture mass PDF with large deviation theory and the nulling procedure we use to reduce the sensitivity to very non-linear scales and baryonic effects. Section~\ref{section::comparison} describes the numerical simulations we use for comparison with our theoretical model and discusses diverse challenges we encountered. Section~\ref{noise} performs a succinct evaluation of the impact of shape noise, assuming a Euclid-like instrument, and cosmic variance on the aperture mass PDF thus giving an idea of the level of accuracy that theoretical models need to reach. Finally, section~\ref{conclusion} concludes. We give in the successive appendices many technical details on the procedure applied here, discuss the extension of the formalism to other filters than top-hats and also estimate the leading-order corrections coming from couplings between lenses, geodesic deviation, reduced shear and the magnification bias. \red{The last appendix of this paper presents an analytical estimate of the observability of non-Gaussian features in the Aperture mass PDF in realistic settings.}

\section{Aperture mass PDF}
\label{section::PDF}

\subsection{Aperture mass definition}
The convergence $\kappa$ can be interpreted as a line-of-sight projection of the matter density distribution between the observer and the source. More quantitatively it can be written as \citep{kappadef}
\begin{equation}
    \kappa(\bm{\vartheta}) = \int_0^{\chi_s} {\rm d}\chi \, \omega(\chi,\chi_s) \, \delta(\chi,\mathcal{D}\bm{\vartheta}),
    \label{def-convergence}    
\end{equation}
where $\chi$ is the comoving radial distance -- $\chi_s$ the radial distance of the source -- that depends on the cosmological model, and $\mathcal{D}$ is the comoving angular distance
\begin{equation}
    \mathcal{D}(\chi) \equiv\left\{
    \begin{aligned}{\frac{\sin (\sqrt{K} \chi)}{\sqrt{K}}}  {\text { for } K>0} \\ {\chi \qquad}  {\text { for } K=0} \\ {\frac{\sinh (\sqrt{-K} \chi)}{\sqrt{-K}}}  {\text { for } K<0}
    \end{aligned}\right. ,
\end{equation}
with $K$ the constant space curvature. The lensing kernel $\omega$ is defined as
\begin{equation}
\label{eq:weight}
    \omega(\chi,\chi_s) = \frac{3\,\Omega_m\,H_0^2}{2\,c^2} \, \frac{\mathcal{D}(\chi)\,\mathcal{D}(\chi_s-\chi)}{\mathcal{D}(\chi_s)}\,(1+z(\chi)).
\end{equation}

Note that equation~(\ref{def-convergence}) assumed no lens-lens couplings and no geodesic perturbations (Born approximation). Although they could affect the higher-order (joint) cumulants we are computing and the convergence PDF, it was showed analytically in \cite{Bernardeau1997} and numerically in \cite{Petri17} that the effect on the convergence skewness for the sources and scales of interest is negligible. A prescription to include those effects, mostly relevant in the context of CMB lensing, in the depicted formalism \textit{i.e} large deviation theory was given in \cite{Barthelemy20b}. 

The aperture mass $M_{\rm ap}$ is defined as a geometrical average of the local convergence with a window of vanishing average
\begin{equation}
    \map(\bm{\vartheta}) = \int {\rm d}^2\bm{\vartheta}' \, U_{\theta}(\vartheta') \, \kappa(\bm{\vartheta}' - \bm{\vartheta})
\end{equation}
with
\begin{equation}
    \int {\rm d}^2\bm{\vartheta}' \, U_{\theta}(\vartheta') = 0.
\end{equation}
Because convergence maps are always only reconstructed up to a mass sheet degeneracy, statistical quantities that can be measured in terms of convergence maps and which are not affected by this degeneracy can only be smoothed quantities with compensated filters as is the case for aperture mass maps. Moreover, aperture mass can be interestingly expressed as a function of the tangential component $\gamma_t$ of the shear \citep{kkaiser1994,schneider1996}
\begin{equation}
    M_{\mathrm{ap}}(\bm{\vartheta})=\int \mathrm{d}^{2} \bm{\vartheta}^{\prime} Q_\theta\left(\vartheta^{\prime}\right) \gamma_{t}\left(\bm{\vartheta}-\bm{\vartheta}^{\prime}\right),
    \label{Mapshear}
\end{equation}
where
\begin{equation}
    Q_\theta(\vartheta)=-U_\theta(\vartheta)+\frac{2}{\vartheta^{2}} \int_0^{\vartheta} \mathrm{d} \vartheta^{\prime} \vartheta^{\prime} U_\theta\left(\vartheta^{\prime}\right),
\end{equation}
thus rendering the aperture mass a direct observable up to a reduced shear correction but which can be accounted for as discussed in appendix~\ref{2ndOrder}.

\begin{figure}
    \centering
    \includegraphics[width = \columnwidth]{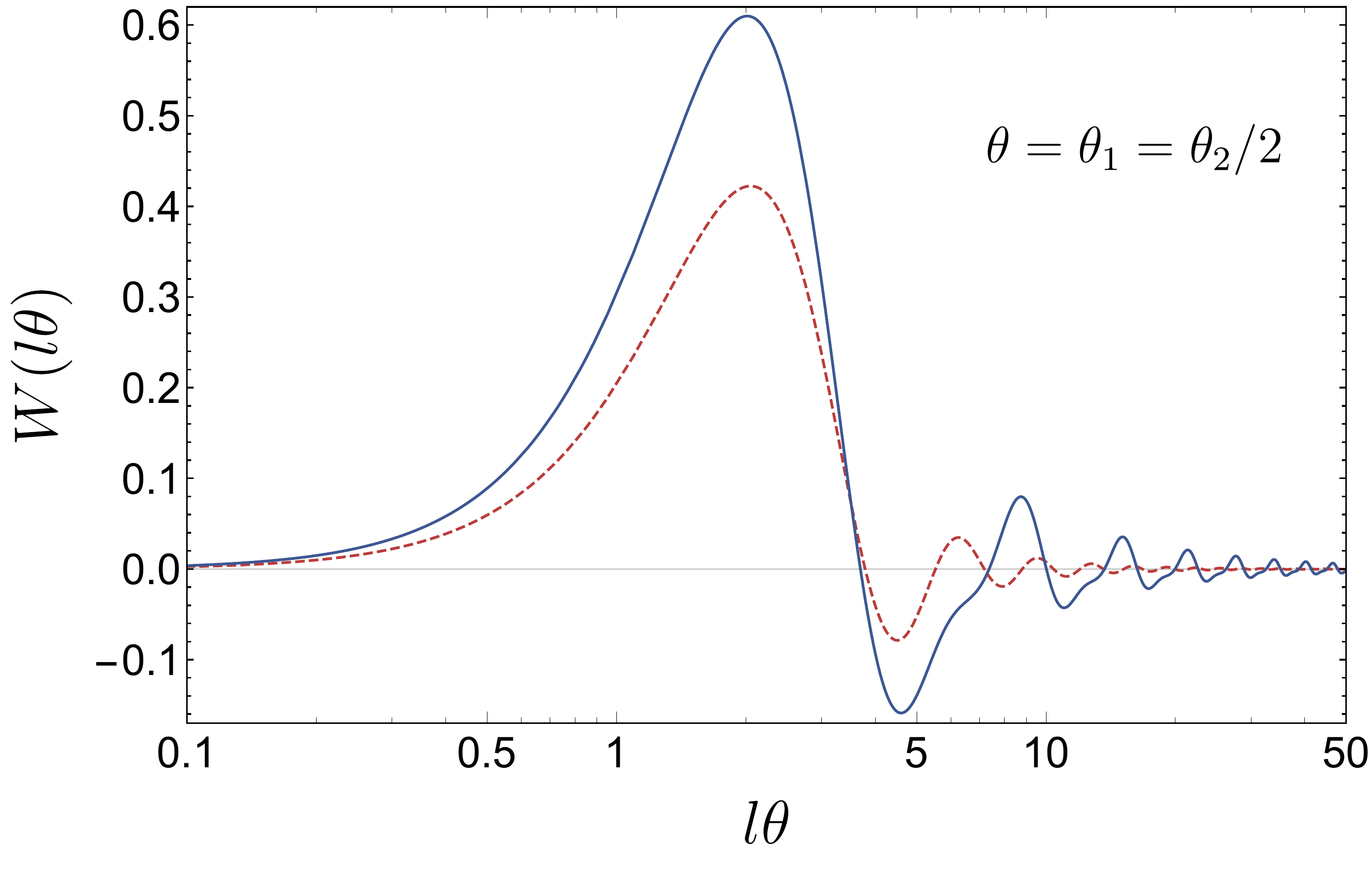}
    \caption{Comparison of the filter given in equation~(\ref{filter_comp}) (blue solid line multiplied by -1) and the one used in \protect\cite{Schneider1998} (red dashed line).}
    \label{Filter}
\end{figure}

In principle, the large deviation formalism could be used with any filter function as was shown in \cite{seminalLDT}, \cite{paolo} and in appendix~\ref{sensitivity}. However, we will here adopt a simple prescription
\begin{equation}
    \map(\bm{\vartheta}) = \kappa_{<\theta_2}(\bm{\vartheta}) - \kappa_{< \theta_1}(\bm{\vartheta}),
    \label{filter_comp}
\end{equation}
where $\kappa_{<\theta_{1,2}}$ denotes the convergence field filtered by a top-hat window of angular radius $\theta_1$ and $\theta_2 = 2 \, \theta_1$. This choice is both motivated by the relative simplicity of obtaining statistics of concentric disks/spheres within the large deviation formalism -- other choices of compensated filters are often used in the literature (see for starter \cite{Schneider1998}) -- and also because top-hat filtering allows for a more rigorous assessment of the scales correctly described by the theory presented in this paper. We show the shape of our filter in Fig.~\ref{Filter} compared to the one used in \cite{Schneider1998}.

\subsection{Projection formula}
\begin{figure}
    \centering
    \includegraphics[width = \columnwidth]{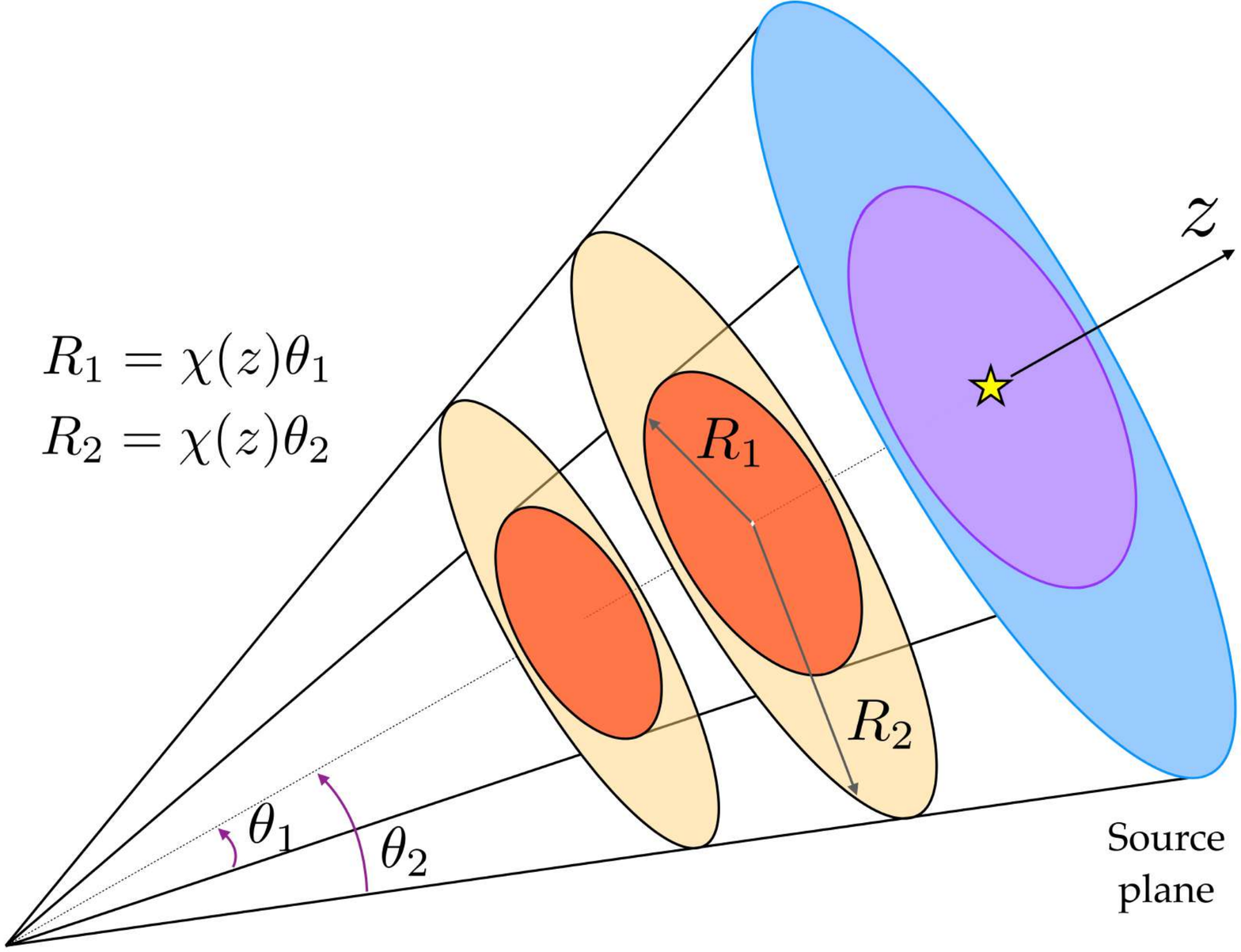}
    \caption{Schematic view of our procedure to predict the aperture mass one-point statistics. Here projected quantities are seen as a superposition of thin and statistically independent slices of the underlying 3D density field along the line of sight (equation~(\ref{projection})). Since the dynamics of disks inside a slice is on average well-described by cylindrical collapse, we use it to study the joint statistics of the 2D density field at two different scales (yellow and orange) at every redshift up to the source plane (section~\ref{LDT}). The scales studied at each redshift allow us to take into account the geometry of the light-cone.}
    \label{schema}
\end{figure}

Under small-angle/Limber approximation, it has been showed in \cite{BernardeauValageas,Barthelemy20a} that the convergence field filtered in an angular top-hat window function can be seen as a juxtaposition of statistically independent 2D slices of the underlying density field. Filtering the field using equation~(\ref{filter_comp}) does not change the demonstration and thus the cumulants of the aperture mass lensing field are given by
\begin{equation}
    \langle \map^p \rangle_c = \int_0^{\chi_s} {\rm d}\chi \, \omega^p(\chi,\chi_s) \, \langle (\delta_{<\mathcal{D}(\chi)\theta_2}-\delta_{<\mathcal{D}(\chi)\theta_1})^p \rangle_c,
    \label{projection}
\end{equation}
where $\delta_{<\mathcal{D}(\chi)\theta_2}-\delta_{<\mathcal{D}(\chi)\theta_1}$ is a random variable defining the density slope between two concentric disks of radii $\mathcal{D}(\chi)\theta_2$ and $\mathcal{D}(\chi)\theta_1$ at comoving radial distance $\chi$. Equation~(\ref{projection}) thus reduces the complexity of the problem down to computing the one-point statistics of the density slope in each two-dimensional slice (equivalently the slope between infinitely long cylinders at the same redshift) along the line-of-sight. To that aim, we will first recall the mathematical definition for cumulants, generating functions, PDF and the relationships between them, before turning to the one-point statistics of the 2D density slope obtained via large-deviation theory. Using this result, we will then build the non-linear cumulant generating function of the aperture mass and its PDF. A schematic representation of our procedure can be found in Fig.~\ref{schema}.

\subsection{Generating functions and PDF}
Throughout this work, we make use of different statistical quantities that we briefly introduce here for clarity.
From the PDF ${\mathcal P}_X$ of some continuous random variable $X$ one can define the moment generating function as the Laplace transform of the PDF
\begin{equation}
    M_X(\lambda) =\mathrm{E}\left(e^{\lambda X}\right) =  \int_{-\infty}^{+ \infty} e^{\lambda x} {\mathcal P}_X(x) {\rm d}x,
    \label{laplace}
\end{equation}
or equivalently as the expectation value\footnote{Note that we make use throughout this work of the ergodicity hypothesis where one assumes that ensemble averages are equivalent to spatial averages ($E(.)\rightarrow \left\langle.\right\rangle$) over one realisation of a random field at one fixed time. This requires that spatial correlations decay sufficiently rapidly with separation such that one has access to many statistically independent volumes in one realisation.} of the random variable $e^{\lambda X}$. The moment generating function, as its name implies, can be used to find the moments of the distribution as can be seen from the series expansion of the expectation of $e^{\lambda X}$,
\begin{equation}
\begin{aligned}
    M_{X}(\lambda)\!&=\!\mathrm{E}\left(e^{\lambda X}\right)\!=\!1 \!+\!\lambda \mathrm{E}(X)\!+\!\frac{\lambda^{2} \mathrm{E}\left(X^{2}\right)}{2 !}\! +\!\frac{\lambda^{3} \mathrm{E}\left(X^{3}\right)}{3 !}\!+\!\cdots 
    \\
    &= \sum_{n = 0}^{+ \infty} \frac{\lambda^{n} \mathrm{E}\left(X^{n}\right)}{n !},
\end{aligned}  
\end{equation}
so that the $n$-th derivative of the moment generating function in $\lambda=0$ is equal to the $n$th order moment, $\mathrm{E}\left(X^{n}\right)$.
The logarithm of the moment generating function is the cumulant generating function (CGF)
\begin{equation}
    \phi_X(\lambda) = \log(M_X(\lambda)) = \sum_{n=1}^{+ \infty} k_{n} \frac{\lambda^{n}}{n !}
    \label{log}
\end{equation}
where $k_n$ are the cumulants (i.e the connected moments) of the distribution.

These definitions can of course be extended to the case of joint probabilities. For the case of two continuous random variables $X$ and $Y$ relevant to this paper, the joint cumulant generating function reads
\begin{eqnarray}
    \phi_{X,Y}(\lambda_1,\lambda_2)\!\!\!\!\! \!&=&\!\!\!\!\!\! \log(M_{X,Y}(\lambda_1,\lambda_2)) = \log\left[E\left(e^{\lambda_1X+\lambda_2Y}\right)\right]\nonumber \\ &=&\!\!\! \sum_{p,q=0}^{+ \infty} \langle X^p Y^q \rangle_c \frac{\lambda_1^{p}\lambda_2^{q}}{p!q!} - 1,
    \label{jointCGF}
\end{eqnarray}
which in particular allows us to straightforwardly define the CGF of any linear combination of random variables from their joint CGF \red{which is useful in our context as shown in equation~\ref{projection}}. More precisely, noticing that $\phi_{X+Y}(\lambda)=\phi_{X,Y}(\lambda,\lambda)$ allows us to generalise the famous relation ${\rm Var}(X+Y) = {\rm Var}(X) + {\rm Var}(Y) + 2{\rm Cov}(X,Y)$ to any cumulant
\begin{equation}
    k_{n, X + \alpha Y} = \sum_{j=0}^n \dbinom{n}{j} \alpha^{n-j} \langle X^j Y^{n-j} \rangle_c.
    \label{knXY}
\end{equation}
It turns out that the quantities
\begin{equation}
    S_n = \frac{k_n}{k_2^{n-1}},
    \label{Sp}
\end{equation}
called reduced cumulants and where $k_2$ is the variance, are of importance in our context as the ratios $S_n$ of the cosmic matter density field were indeed shown to be independent from the variance (and therefore redshift) down to mildly non-linear scales \citep{Peebles,1995MNRAS.274.1049B}. We thus also define the scaled cumulant generating function (SCGF hereafter) as
\begin{equation}
    \varphi_X(\lambda) = \lim_{k_2 \rightarrow 0} \sum_{n=0}^{+\infty} S_n \, \frac{\lambda^n}{n!}=\lim_{k_2 \rightarrow 0} k_2 \,\phi_X\left(\frac{\lambda}{k_2}\right),
    \label{defscgf}
\end{equation}
that we will in our context extrapolate to non-zero values of the variance. Eventually, one can then reconstruct the PDF for the random variable $X$ as an inverse Laplace transform (inverting equation~(\ref{laplace})) given by
\begin{equation}
    {\mathcal P}_X(x) = \int_{-i\infty}^{+i\infty} \frac{{\rm d}\lambda}{2\pi i} \, \text{exp}\left(-\lambda x + \phi_X(\lambda)\right).
    \label{eq::laplace}
\end{equation}

\subsection{Statistics of the 2D density slope obtained via large deviation theory}
\label{LDT}

Let us now recall some of the results of LDT for matter densities in disks (equivalently long cylinders). For more details, we refer the reader to \cite{seminalLDT} and \cite{cylindres}. We will first give general notions on random variables admitting a large deviation principle then move on to the specific case of the matter density field.

A set of random variables $\{\rho_i^\epsilon\}$, $1\leq i \leq N$, with joint PDF ${\mathcal P}_{\epsilon}(\{\rho_i^\epsilon\})$ is said to satisfy a large deviation principle if the limit
\begin{equation}
    \Psi_{\{\rho_i^\epsilon\}}(\{\rho_i^\epsilon\}) = - \lim_{\epsilon \rightarrow 0} \epsilon \log\left[{\mathcal P}_{\epsilon}(\{\rho_i^\epsilon\})\right]
    \label{LDP}
\end{equation}
exists, where $\epsilon$ is the \textit{driving parameter}. This driving parameter indexes the set of random variables with respect to some evolution, for example a time evolution. 
For example, a common example of a random variable satisfying a large deviation principle is the sum of successive coin tosses where the driving parameter is one over the number of tosses entering the sum. In the case of the matter density field at a single scale this driving parameter is its variance which acts as a clock from initial to late times. For the joint statistics of concentric disks of matter, the common driving parameter could be the variance at any radius/scale since all variances behave the same in the 0 limit, being proportional to the growth rate of structure in the linear regime such that the SCGF -- or equivalently cumulants at tree order -- is not affected by this choice. We now omit the $\epsilon$ sub/superscripts in our notation for simplicity.

The existence of a large deviation principle for the set of random variables $\{\rho_i\}$ implies that their SCGF $\varphi_{\{\rho_i\}}$ is given through Varadhan's theorem as the Legendre-Fenchel transform of the rate function $\Psi_{\{\rho_i\}}$
\begin{equation}
    \varphi_{\{\rho_i\}}(\{\lambda_i\}) = \sup_{\{\rho_i\}} \,\left[\sum_i \lambda_i\rho_i - \Psi_{\{\rho_i\}}(\{\rho_i\})\right],
    \label{varadhan}
\end{equation}
where the Legendre-Fenchel transform reduces to a simple Legendre transform when $\Psi_{\{\rho_i\}}$ is convex. In that case, 
\begin{equation}
    \varphi_{\{\rho_i\}}(\{\lambda_i\}) =  \sum_i \lambda_i\rho_i - \Psi_{\{\rho_i\}}(\{\rho_i\}),
    \label{Legendre}
\end{equation}
where $\{\rho_i\}$ are a function of $\{\lambda_i\}$ through the stationary conditions 
\begin{equation}
   \lambda_k = \frac{\partial \Psi_{\{\rho_i\}}(\{\rho_i\})}{\partial \rho_k}\,,\quad  \forall k \in \{1,\cdot,N\}.
    \label{stationnary}
\end{equation}
Another consequence of the large-deviation principle is the so-called contraction principle.
This principle states that for a set of random variables $\{\tau_i\}$ satisfying a large deviation principle and related to $\{\rho_i\}$ through the continuous map $f$, then the rate function of $\{\rho_i\}$ can be computed as
\begin{equation}
    \Psi_{\{\rho_i\}}(\{\rho_i\}) = \inf_{\{\tau_i\}:f(\{\tau_i\}) = \{\rho_i\}} \Psi_{\{\tau_i\}}(\{\tau_i\}).
    \label{contraction}
\end{equation}
This formula is called the contraction principle because $f$ can be many-to-one in which case we are {\it contracting} information about the rate function of one random variable down to the other. In physical terms, this states that an improbable fluctuation of $\{\rho_i\}$ is brought about by the most probable of all improbable fluctuations of $\{\tau_i\}$.

For the case of the matter density field and starting from Gaussian initial conditions\footnote{Primordial non-Gaussianities could also straightforwardly be accounted for in this formalism as shown by \cite{NonGaussianities}.}, the rate function of the linear field is simply given by a quadratic term. Using the contraction principle, the rate function of the late-time density field at different scales can then be computed from the initial conditions if the most likely mapping between the two is known, that is if one is able to identify the leading field configuration that will contribute to the infimum of equation~(\ref{contraction}). In cylindrically symmetric configurations, which is the case for a disk of radius $R_k$ in a slice at redshift $z$, one could conjecture \citep{Valageas} that the most likely mapping between initial and final conditions is cylindrical collapse (similarly to spherical collapse being the most likely dynamics for 3D density fluctuations). Then the rate function of the late-time density field in concentric disks of radii $R_i$ is given by
\begin{equation}
    \Psi_{\rm cyl}(\{\rho_{i}\})=\frac{\sigma^2_{R_1}}{2} \sum_{k,j}\Xi_{kj}(\{\tau_{i}\})\bar{\tau}_{k}\bar{\tau}_{j},
    \label{psicyl}
\end{equation}
where $\sigma^2_{R_1}$ -- our driving parameter -- is the variance within the smallest disk, $\Xi_{kj}(\{\tau_{i}\})$ is the inverse of the covariance matrix between the linear density field inside the initial disks (before collapse) of radii $R_{k} \, \rho_{k}^{1/2}$ (given thanks to mass conservation in each collapsing disk), and $\bar{\tau}_k$ are the linear density contrasts obtained through the most probable mapping between the linear and late-time density fields. This mapping is given by the 2D spherical (cylindrical) collapse for which an accurate parametrisation is given by\footnote{This parametrisation was first proposed by \cite{Bernardeau1995} and can be shown to provide a very accurate approximation to the true spherical collapse dynamics so that the effect on the PDF for the 3D matter density field is much smaller than the difference between the theory as it is and the measurement in simulations.}
\begin{equation}
    \zeta(\bar{\tau}_k) = \rho_k =  \left(1 - \frac{\bar{\tau}_k}{\nu} \right)^{-\nu}.
    \label{collapse}
\end{equation}
In the spirit of previous works involving the density filtered in spherical cells, the value of $\nu$ in this parametrisation of $\zeta$ is chosen to be $\nu = 1.4$ so as to reproduce the value of the tree-order skewness in cylinders as computed from perturbation theory \citep{cylindres}.

Finally, as a straightforward consequence of the contraction principle, the rate function given by equation~(\ref{psicyl}) is also the rate function of any monotonic transformation of $\rho$, such that for the density contrast $\delta = \rho - 1$, we have $\Psi_{\delta}(\delta) = \Psi_{\rho}(\rho(\delta))$. Thus plugging equation~(\ref{psicyl}) in equation~(\ref{Legendre}) gives us the joint SCGF of concentric disks of the density field at redshift $z$.
Since we are, as shown in equation~(\ref{projection}), interested in the density slope between two concentric disks, all that remains is to recall that its SCGF is easily expressed from its joint statistics through 
\begin{equation}
    \varphi_{\delta_2-\delta_1} (\lambda) = \varphi_{\delta_1,\delta_2} (-\lambda,\lambda).
    \label{slopeSCGF}
\end{equation}
Then, the CGF can be derived and eventually the statistics of the aperture mass field can be obtained through the projection formula given by equation~(\ref{projection}).

\subsection{Non-linear cumulant generating function}
\label{section::nonlinearCGF}

Large deviation theory in the context of cosmic structure formation is strictly speaking only valid in the regime where the variance goes to zero. However a key point of its application to make useful predictions is to extrapolate these asymptotic results (in particular the CGF) to finite values of the variance. As noted in appendix~\ref{nonlinearvariance}, one thus needs a prescription to compute the (co)variances for example appearing in equation~(\ref{psicyl}). We use the non-linear power spectrum coming from Halofit \citep{Halofit} for computation of these covariances. As another sanity check to ensure that any possible discrepancies with the numerical simulation are not too much influenced by this choice, we also re-scale the projected CGF by the measured variance $\sigma^2_{\map, {\rm sim}}$ instead of the one computed with Halofit $\sigma^2_{\map, {\rm hfit}}$,
\begin{equation}
    \phi_{\map}(\lambda) = \frac{\sigma_{\map, {\rm hfit}}^2}{\sigma_{\map,{\rm sim}}^2}\phi_{\map}\left(\lambda \frac{ \sigma_{\map,{\rm sim}}^2}{\sigma_{\map, {\rm hfit}}^2}\right),
    \label{rescaling}
\end{equation}
so that our resulting $\map$ CGF contains the exact variance and all higher-order cumulants given by the non-linear collapse dynamics. \red{Note however that the agreement between the covariances predicted with Halofit and the ones measured in the simulation is to the percent, except when using the BNT transform that we introduce in section \ref{section::nulling}. A more quantitative assessment of the agreement between the measured and predicted quantities with Halofit can be found in section \ref{nulledCL} and more specifically in figures~\ref{clzs1} and \ref{clnull} where we discuss in more details some of the found discrepancies.} 

The need for an exterior input of the non-linear variance along the line of sight does weaken a bit the "from first principles" quality that large-deviation frameworks in Cosmology usually exhibit but nonetheless note that such contributions are contained in the non-linear power spectrum which has focused a lot of attention in the recent years and is very reliably modelled now \citep{Halofit,EuclidEmulator}. Eventually, this comes with the modelling of a true non-Gaussian observable (up to a reduced shear correction but as opposed to the convergence which needs reconstruction of mass maps) which is sensitive to multiple scales at once, a property usually very useful to break degeneracy between cosmological parameters \citep{Boyle2020}.

We now have all the tools to successfully compute the non-linear cumulant generating function of the aperture mass: i) Given a non-linear prescription for the power spectrum we can compute the covariance matrix at redshift $z$ between any two disks of radius $R_1$ and $R_2$
\begin{equation}
    \sigma^2\!(R_1,R_2;\!z) \!=\!\!\!\!\int \!\!\frac{{\rm d}^2\bm{k}_{{\perp}}}{(2\pi)^2}  \!P(k_{{\perp}};z)  W_{TH}(R_1 k_{\perp}\!)  W_{TH}(\!R_2 k_{\perp}\!),
    \label{eq::covariance}
\end{equation}
where $W_{TH}(l) = 2J_1(l)/l$ and $J_1$ is the first Bessel function of the first kind; ii) this enables to compute the rate function (\ref{psicyl}) for any values of the densities inside each disk in a given slice along the line of sight; iii) numerically inverting the stationary condition (\ref{stationnary}) and  using equations~(\ref{Legendre}) and (\ref{slopeSCGF}), we can now compute for any $\lambda$ the CGF of the 2D density slope within each slice; iv) Using the projection formula~(\ref{projection}) and equation~(\ref{rescaling}) we finally compute the non-linear CGF of the aperture mass.

\subsection{Analytical cumulant generating function}
\label{analyticalCGF}

Now that we can compute the non-linear CGF of the aperture mass we would like to compute its PDF. However, the Laplace transform in equation~(\ref{eq::laplace}) requires to have an analytical expression of the integrand, in particular the CGF, so that it can be continued in the complex plane. Unfortunately this is not the case with the described formalism, the crux of the matter residing in inverting the stationary condition (\ref{stationnary}) for complex $\lambda$ values given that the covariance~(\ref{eq::covariance}) is only defined numerically and that no explicit solution is known for a generic power spectrum except the simplistic case of power-law power spectra. 

\red{For a different approach to tackle this issue than used in this paper we refer for example to \cite{Barthelemy20a}.} 
The solution we propose here was first used in \cite{BernardeauValageas} \red{and then later in \cite{FriedrichDES17} for example.} It consists in fitting an effective mapping, $\zeta(\tau_{\rm eff})$, between an effective un-smoothed Gaussian initial field and the aperture mass field whose PDF we want to compute. We hence re-write the aperture mass SCGF, still given by the Legendre transform of the effective rate function
\begin{equation}
    \varphi_{\map}(\lambda) = \lambda \zeta(\tau_{\rm eff}) - \frac{1}{2}\tau_{\rm eff}^2 
    \label{effectiveRF}
\end{equation}
with the stationary condition written as
\begin{equation}
    \lambda = \frac{\rm d}{\rm d \zeta} \frac{\tau_{\rm eff}^2}{2} = \tau_{\rm eff} \left(\frac{{\rm d}\zeta(\tau_{\rm eff})}{{\rm d}\tau_{\rm eff}}\right)^{-1}.
    \label{EffectiveStationary}
\end{equation}
Note that now the effective mapping
\begin{equation}
    \zeta(\tau_{\rm eff}) = \sum_{k = 0}^n \frac{\mu_{k}}{k!} \tau_{\rm eff}^k,
\end{equation}
where $\mu_{0} = 0$, $\mu_{1} =1$ and the other coefficients will be fitted, makes for an easy analytic continuation of the mapping to the complex plane which in turns allows us to invert the stationary condition for complex values of $\lambda$.

All that remains is to fit the values of the $\mu$ coefficients. First let us notice that by definition of the Legendre transform we have
\begin{equation}
    \frac{{\rm d}\varphi_{\map}(\lambda)}{{\rm d}\lambda} = \zeta(\tau_{\rm eff}),
\end{equation}
and thus from equation~(\ref{effectiveRF})
\begin{equation}
    \frac{1}{2}\tau_{\rm eff}^2 = \lambda\frac{{\rm d}\varphi_{\map}(\lambda)}{{\rm d}\lambda}-\varphi_{\map}(\lambda).
\end{equation}
Then having already computed the SCGF for real values of $\lambda$ one can easily produce a table of both $\zeta(\tau_{\rm eff})$ and $\tau_{\rm eff}$ and finally fit the $\mu$ coefficients. In practice, we typically choose a polynomial mapping of odd degree higher than 5 which reproduces very well the real generating function. Note that the same procedure can be applied to directly fit the extrapolated aperture mass CGF, the $\mu$ coefficients now taking the values
\begin{equation}
    \mu_{k}^{\rm CGF} = \sigma_{\map}^k \, \mu_{k},
\end{equation}
and the inverse Laplace transform of the CGF in equation~(\ref{eq::laplace}) can now be performed numerically -- for example using Simpson's method along the imaginary axis -- without any additional complication. We finally give in appendix~\ref{EffectiveNu} the link that can be made between the $\mu$ coefficients and the cumulants of the field. Given that the fitted $\mu$ values and the computed cumulants agree very well, computing cumulants to obtain those coefficients instead of fitting them might become a viable option, especially in cases where the large deviation formalism is used for filters other than top-hats and where imposing the stationary condition even to obtain the real space CGF is a numerical challenge in itself \citep{paolo}.

\begin{figure}
    \centering
    \includegraphics[width=\columnwidth]{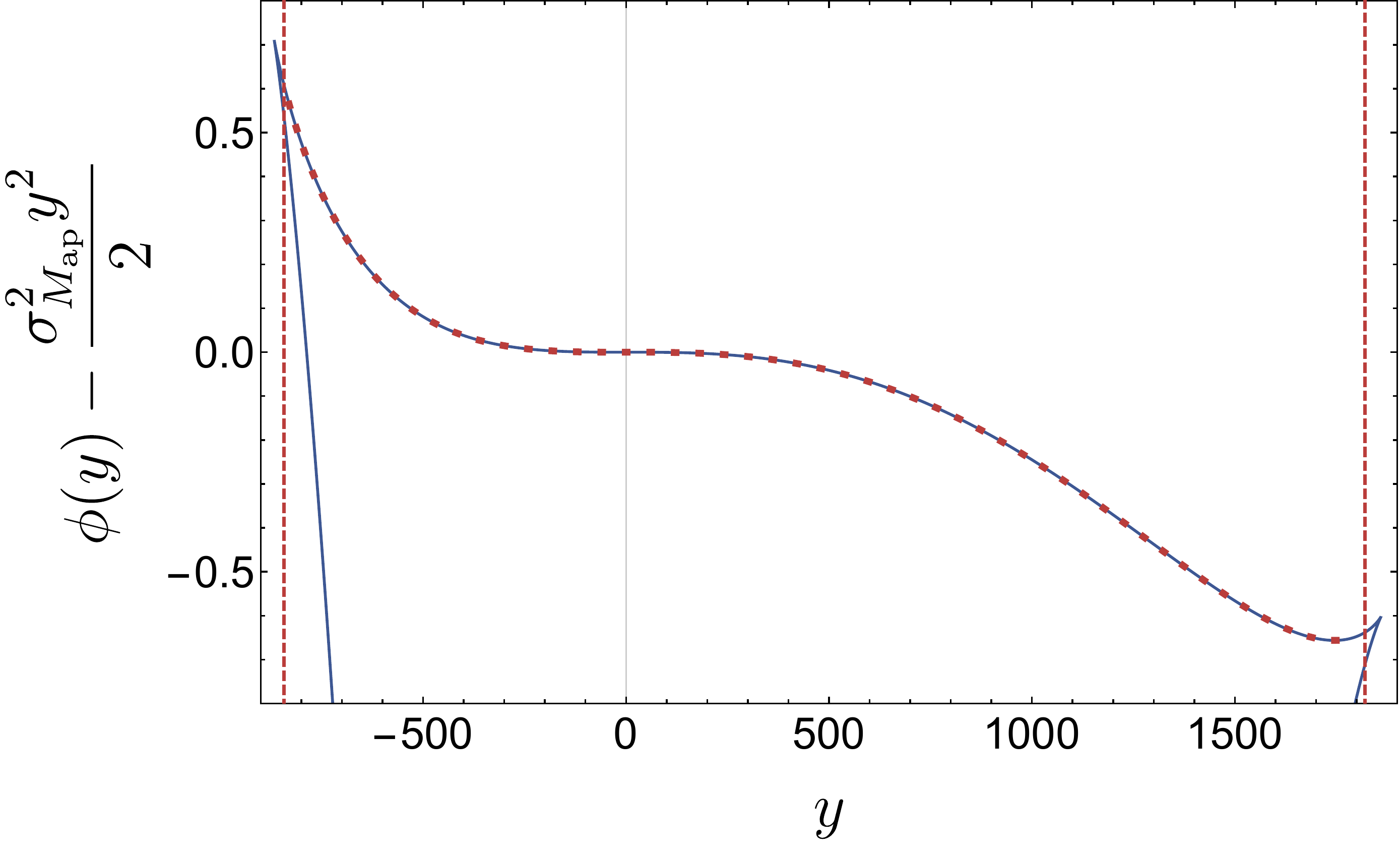}
    \caption{Cumulant generating function of the aperture mass at redshift $z_s = 1$,  $\theta_1 = 15$ and $\theta_2 = 30$ arcmin. The dashed red line is the CGF as computed with Large deviation theory and the blue line is the one computed with a fitted effective collapse of degree 7. The dashed red vertical lines indicate the position of the critical points of the CGF as computed with LDT. The successive derivatives in zero (cumulants) are perfectly reproduced, as well as the general shape and location of critical points. \red{The $\sigma^2_{\map} y^2/2$ term was substracted here to remove the quadratic contribution to the CGF and better display the part, corresponding to high-order cumulants, modelled by our formalism. We subtract the exact same quantity to the 2 curves so that the agreement displayed is not affected by that choice.}}
    \label{effective_CGF}
\end{figure}

Overall Fig.~\ref{effective_CGF} illustrates how well this approach reproduces the $\map$ cumulant generating functions computed with large deviation theory. In particular this procedure being from a mathematical point of view strictly identical to the usual large deviation approach applied to the matter density field one-point statistics, the critical behaviour that the (S)CGF exhibits along the real axis and that is the result of a change of convexity of the rate function or equivalently multiple solutions to the stationary condition \red{(which are apparent on the blue line of Fig.~\ref{effective_CGF})}, will be also present in the case of reconstructed generating functions via an effective mapping. Finally note that though the procedure we described in this section is the one that we implement to construct the aperture mass PDF, we also give some more technical comments on the approach itself for projected quantities in general in appendix~\ref{technical}.

\subsection{Aperture mass PDF}
\label{PDFplots}

\begin{figure}
    \centering
    \includegraphics[width = \columnwidth]{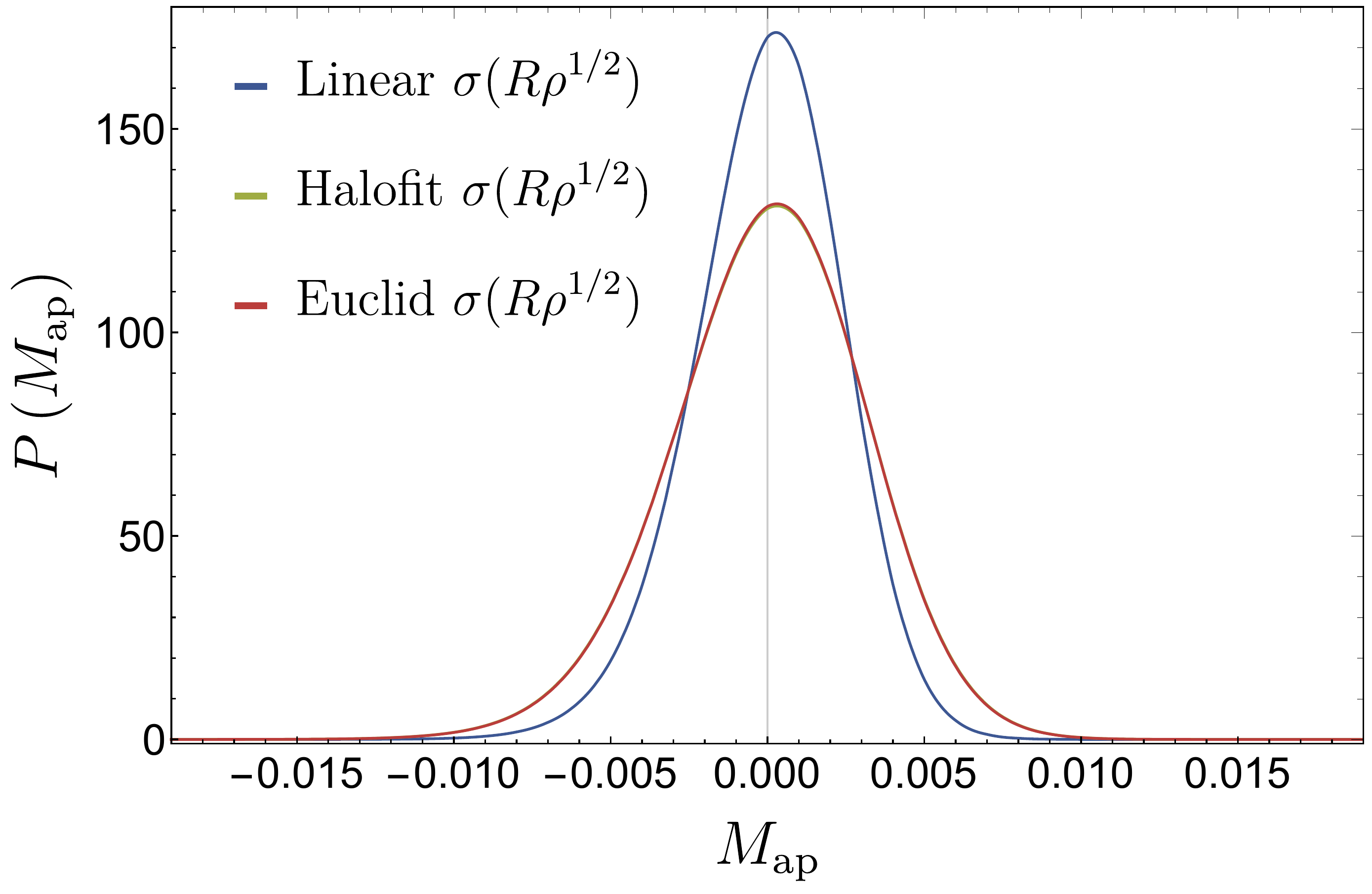}
    \includegraphics[width = \columnwidth]{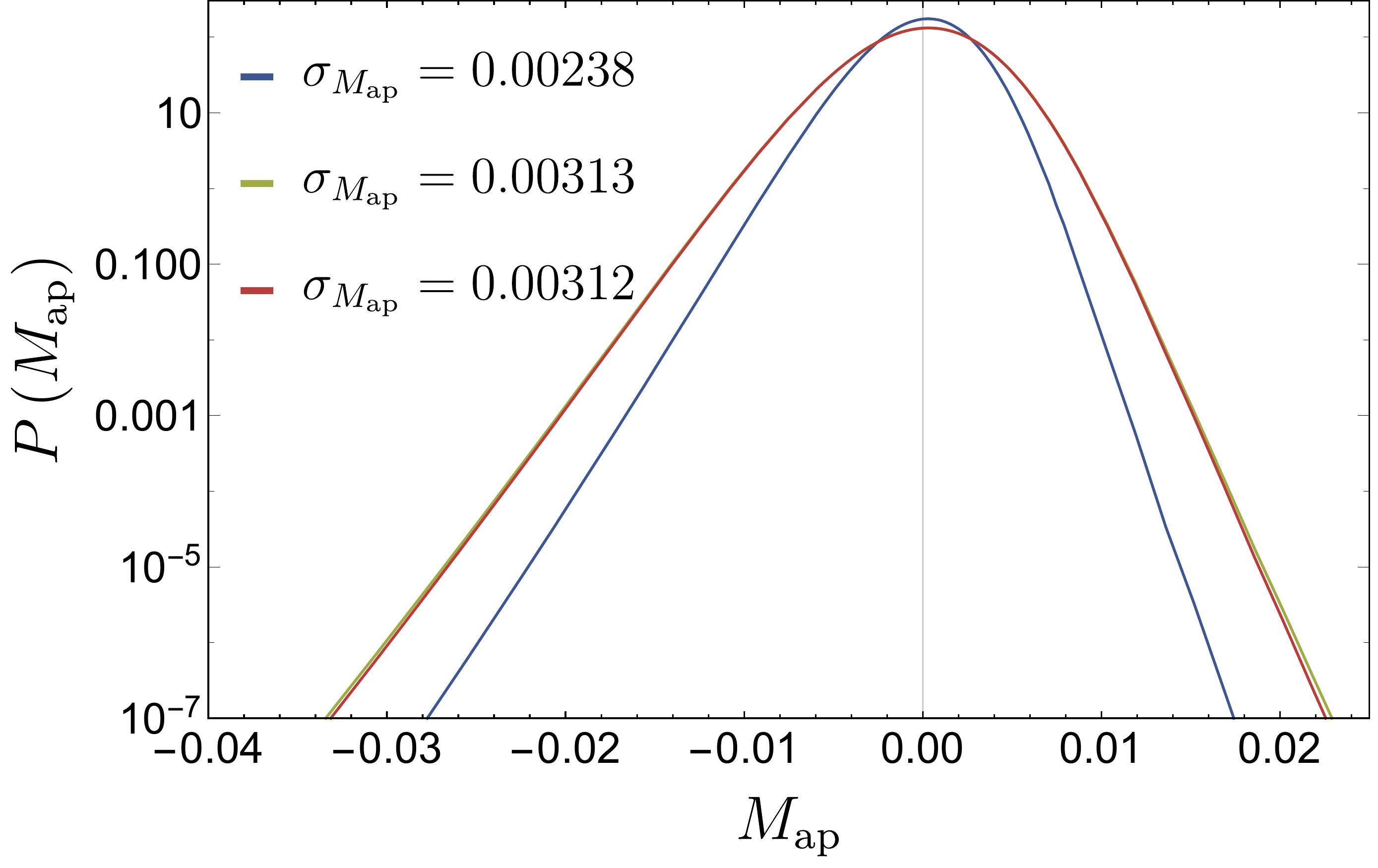}
    \caption{Aperture mass PDF at $z_s = 1$, $\theta_1 = 15$ and $\theta_2 = 30$ arcmin. PDFs are obtained fitting an effective collapse of degree 7 and we compare different prescriptions for the non-linear covariance of disks in redshift slices along the line of sight. The blue line is the traditional 1-cell approach where the CGF computed with the linear covariance is re-scaled by the non-linear driving parameter. The green and red solid lines are the PDF obtained from the full non-linear CGF with different prescriptions for the non-linear covariances. The green and red curves are almost indistinguishable which illustrates that the $\map$ PDF is not sensitive to the detailed modelling of the non-linear matter power spectrum.}
    \label{linVShalo}
\end{figure}

As an illustration, Fig.~\ref{linVShalo} displays the resulting prediction for an aperture mass PDF for a single source redshift at $z_s = 1$. The opening angle is chosen to be $\theta_1 = 15$ arcmin and the non-linear covariance \red{of equation~(\ref{psicyl})} is treated in three different ways, namely i) the linear prediction with a re-scaling of the obtained SCGF by the non-linear driving parameter -- the variance of the field at the smallest scales -- inside each redshift slice along the line of sight, ii) using the full Halofit power spectrum as input [our baseline approach] and iii) using the Euclid emulator of the non-linear power spectrum \citep{EuclidEmulator} for comparison of different non-linear prescriptions. The blue solid line shows that indeed re-scaling the SCGF by the driving parameter as in the 1-cell case does not lead to the correct non-linear variance of the aperture mass \red{(we assume in this subsection that it is equal to the Halofit/Euclid emulator prediction)} which would lead to a major source of disagreement between this model and the \red{measured/simulated} PDF.
This further illustrates the discussion in section \ref{section::nonlinearCGF} where the importance of using the full non-linear power spectrum was underlined. However, the $\map$ PDF does not appear to be very sensitive to the precision in the modelling of the non-linear power spectrum as the very good agreement between the Halofit and Euclid Emulator prescriptions implies. \red{Indeed, and as illustrated in figure 8 of \cite{EuclidEmulator}, the scale-dependence of the two power spectra is not exactly the same though this does not seem to affect significantly the values of the high-order cumulants in the PDF of the aperture mass.}

As expected from our formalism, we observe two exponential cut-offs on each side of the PDF, the positive tail being driven by large values of the convergence field filtered at the scale $\theta_2$ and the negative tails by large values of the convergence field filtered at the scale $\theta_1$. The convergence field becoming more and more skewed with decreasing smoothing scale, we expect that the negative tail of the aperture mass is the most prominent one, which also implies a shift of the most likely value towards positive values. However note that those considerations are only rough approximations that help us to understand the general shape of the aperture mass PDF. In particular, the tails of the $\map$ PDF are not at all identical -- \red{different power laws} -- to the respective tails of $P(\kappa_{< \theta_1})$ and $P(\kappa_{< \theta_2})$ \red{which are the tails of the convolution of the 2 convergence PDFs if the two scales were independent.} Overall the displayed PDF is far from Gaussian thus highlighting how non-negligible non-Gaussian features of the aperture mass distribution are in this regime.

\subsection{${\map}$ PDF with a nulling strategy}
\label{section::nulling}

One of the important issues faced by theoretical approaches that aim at describing quantities projected along the line of sight, is the mixing of both very non-linear scales not accurately probed by standard first principles perturbative approaches such as ours, and reasonably larger (quasi-linear) scales more accessible to the theory. As such usual weak-lensing statistical probes are often modelled by more phenomenological approaches such as halo models that can also take into account baryonic physics which becomes important at small scales \citep{HMcode}, and even more so making use of numerical simulations \citep{Baryonification}. However those simulations are not always tested in fine details, especially for higher-order non-Gaussian statistics. 

Alternatively, a theoretical strategy to disentangle scales in lensing quantities known as the Bernardeau-Nishimichi-Taruya (BNT) transform or nulling strategy was proposed by \cite{Nulling} and  allows for very accurate theoretical predictions in the context of power spectrum analysis or more recently the convergence PDF \citep{Barthelemy20a}. This nulling strategy was used very recently in \cite{x-cut} to remove the sensitivity to the poorly modelled small scales for the two-point cosmic shear signal, and therefore improve cosmological constraints using the Dark Energy Survey shear data. This will become even more relevant for future lensing experiments with better knowledge of redshifts.

This BNT transform can only be used in the context of a tomographic analysis of at least 3 source redshifts (or redshift bins, although not treated here) and is a linear transformation $M$ applied to the set of lensing kernels $\omega_i \equiv \omega(\chi,\chi_{s,i})$ giving rise to a new set of re-weighted kernels
\begin{equation}
    \Tilde{\omega}^j = M^{ij}\omega_i.
\end{equation}
For a set of 3 source planes labeled from $j= i-2$ to $j = i$ arranged by ascending source redshift, it was showed in \cite{Nulling} that $M$ must satisfy the system
\begin{equation}
    \begin{dcases}
    \sum_{j=i-2}^{i} M^{j i}=0, \\
    \sum_{j=i-2}^{i} \frac{M^{j i}}{\chi_{s,j}}=0,
    \end{dcases}
\end{equation}
which is under-constrained so that we also impose by convention $M^{ii} = 1$. The elements of $M$ can thus be computed considering sequential triplets of tomographic bins, going from the lowest to the highest redshift, such that
\begin{align}
    M^{i-2,i} = \frac{\chi_{i-2}(\chi_{i-1}-\chi_{i})}{\chi_{i}(\chi_{i-2}-\chi_{i-1})},\\
    M^{i-1,i} = \frac{\chi_{i-1}(\chi_{i}-\chi_{i-2})}{\chi_{i}(\chi_{i-2}-\chi_{i-1})}.
\end{align}

\begin{figure}
    \centering
    \includegraphics[width = \columnwidth]{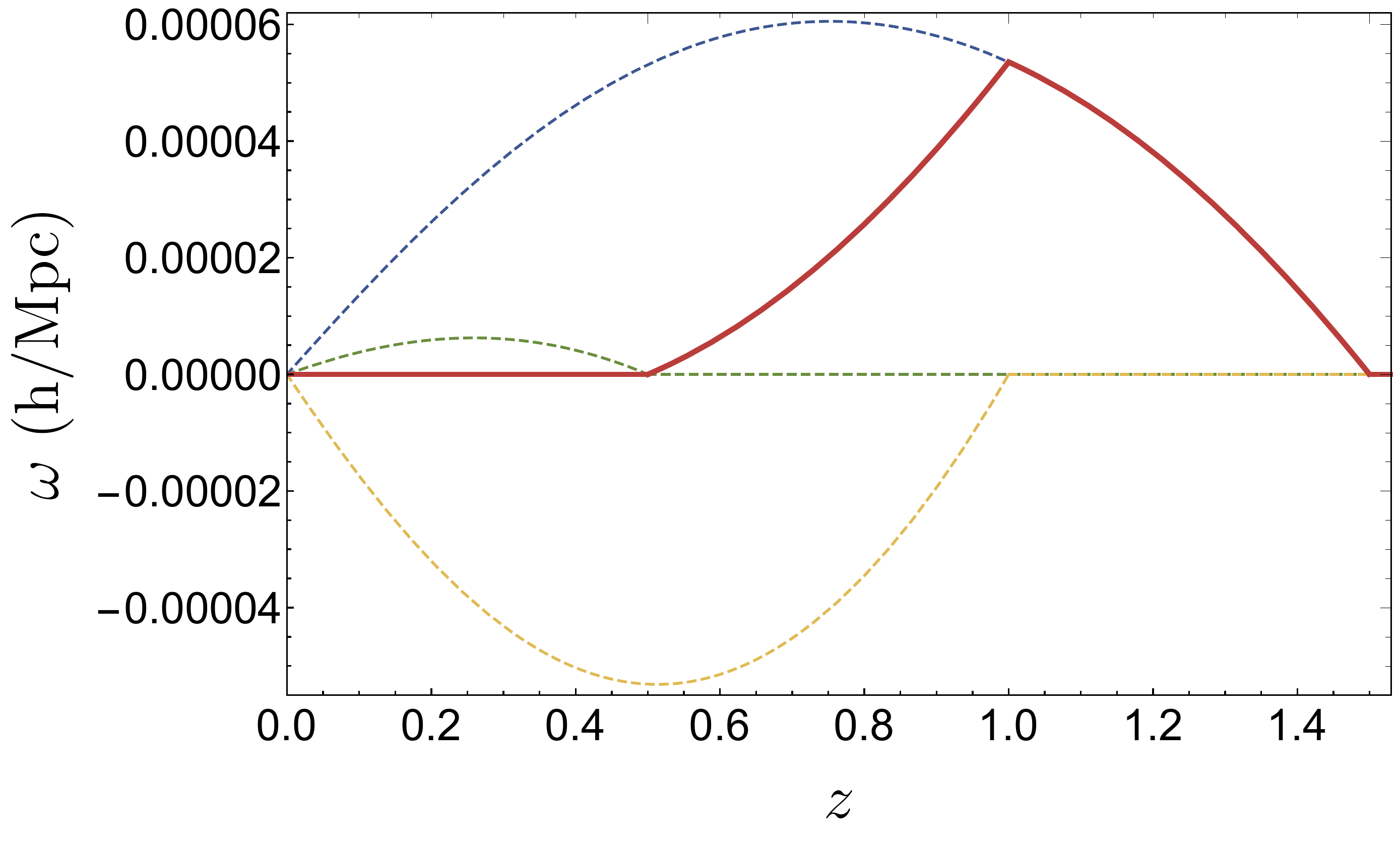}
    \caption{Illustration of the effect of the BNT transform on lensing kernels. The green, yellow and blue dashed lines are the kernels up to $z_s = 0.5, 1, 1.5$ respectively re-weighted by their appropriate BNT coefficients, $M^{ij} = [0.324, -1.324, 1]$ where $j$ is fixed and equal to 3 if the blue kernel is the third of a tomographic analysis. The thick red line is the sum of the 3 re-weighted kernel. The effect of nulling is to set to zero the contribution of all lenses below the closer plane.}
    \label{null_kernel}
\end{figure}

We display in Fig.~\ref{null_kernel} an example for a set of 3 source planes located at $z_s = 0.5, 1, 1.5$. The green, yellow and blue dashed lines are the kernels up to $z_s = 0.5, 1, 1.5$ respectively re-weighted by their appropriate BNT coefficients while the thick red line is the sum of the 3 re-weighted kernels. Note that the blue dashed line is also the original kernel since its BNT coefficient is set to 1. One can thus clearly see that the effect of nulling is to set to zero the contribution of all lenses below the closer plane and thus to cancel out the contribution of small scales which are very non-linear and where the effect of baryonic physics becomes non-negligible.

\begin{figure}
    \centering
    \includegraphics[width = \columnwidth]{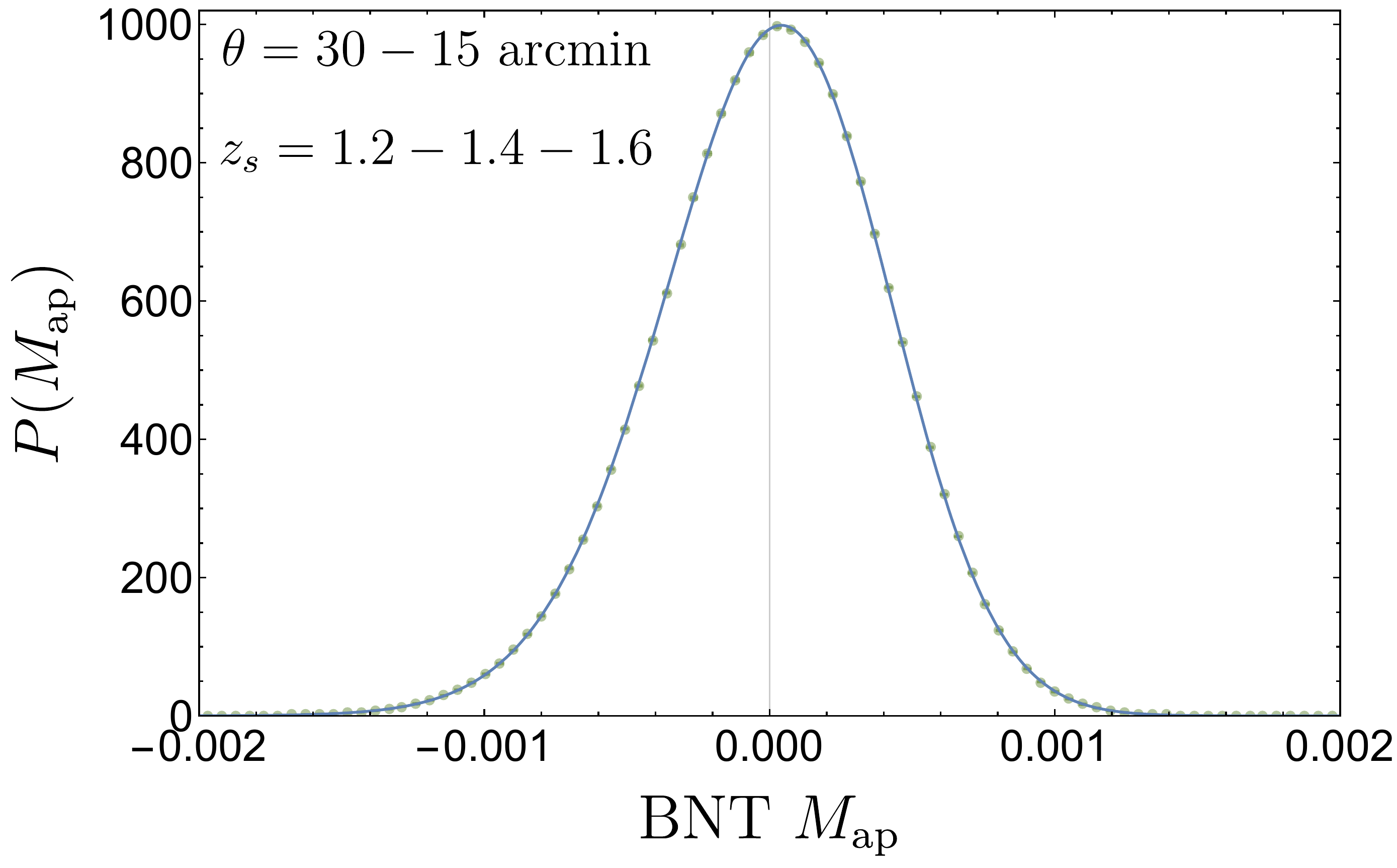}
    \includegraphics[width = \columnwidth]{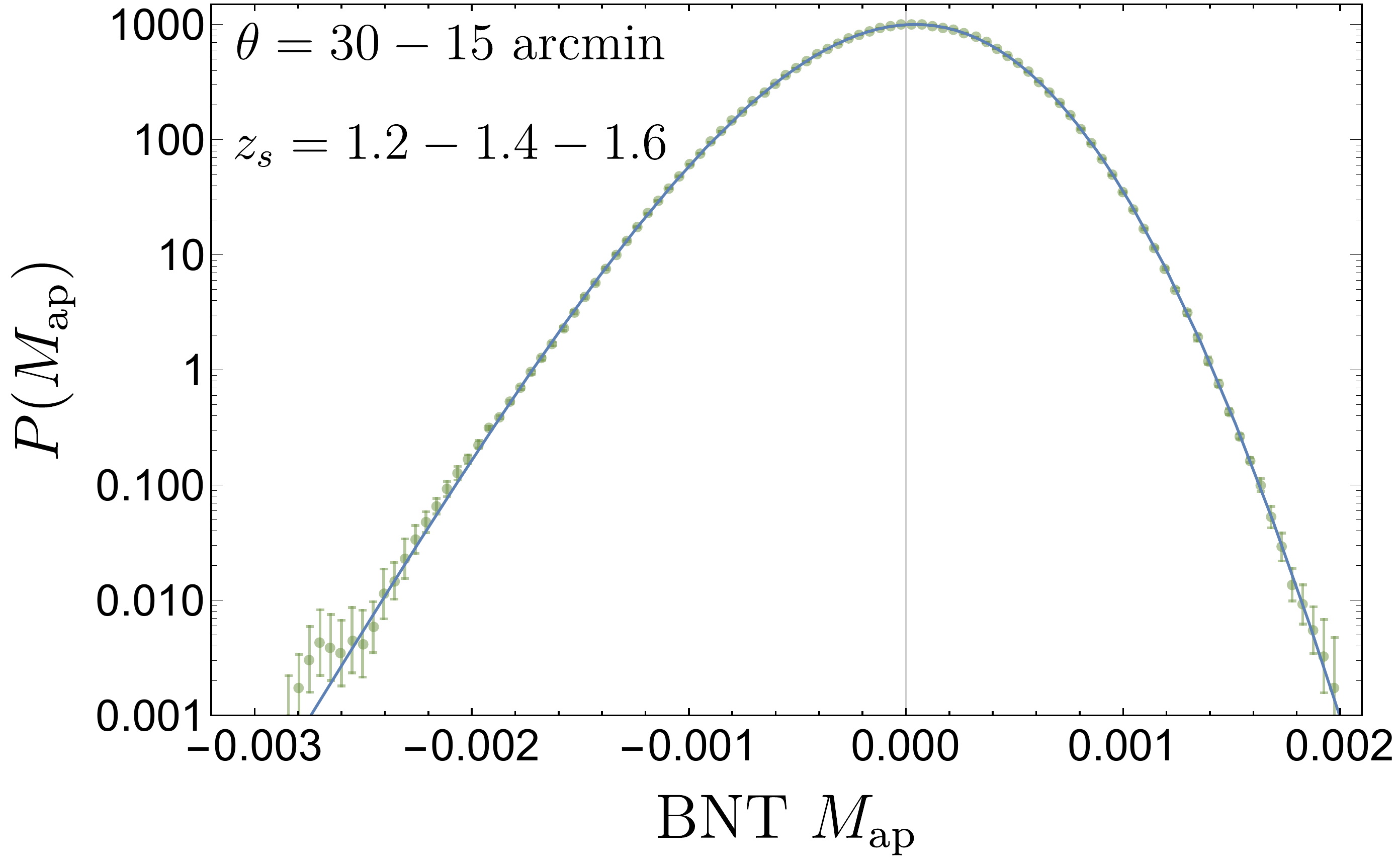}
    \includegraphics[width = \columnwidth]{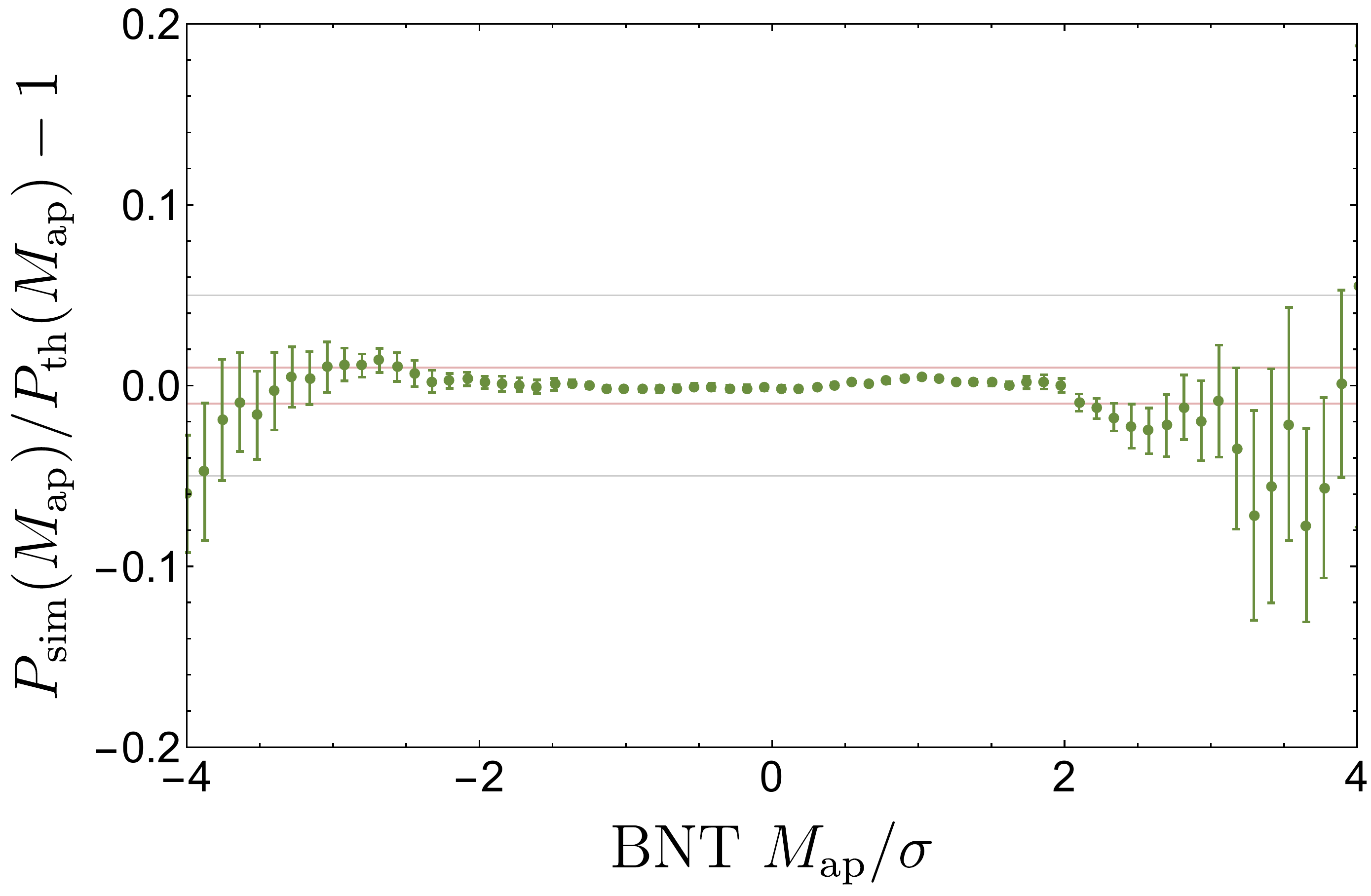}
    \caption{Theoretical BNT $\map$ PDF compared to one measured in the numerical simulation described in section \ref{numerical_sims}. The expected exponential cut-offs in the tails are well observed when one restricts the lensing kernel to physical scales accessible to perturbation theory. The hereby formalism also agrees very well with the measured PDF (from one realisation of the map at the lowest resolution). Note that the error bars represent the standard error-on-the-mean computed from 8 sub-samples of the full-sky. The red horizontal lines indicate the region of $\pm$ 1\% residual and the grey ones $\pm$ 5\%.}
    \label{NulledPDF}
\end{figure}

For our purpose, the BNT transform -- which boils down to a simple linear combination of the maps -- is also straightforward to implement in our theoretical approach to the aperture mass PDF since we only need to replace the original kernel with its nulled counterpart. We finally show in Fig.~\ref{NulledPDF} how this construction allows for a very effective description of the $\map$ one-point statistics by comparing our formalism \red{-- case ii) of section~\ref{PDFplots} and Fig.~\ref{linVShalo} --} to measurements made in the numerical simulation described in the following section. One can appreciate that the exponential cut-off in the tails of the PDF, a prediction of our formalism, is well-observed once one reduces the lensing kernel down to scales accessible to first-principles theoretical modelling (\textit{i.e} perturbation theory). Apart from the general shape, one can also see that the theoretical PDF agrees really well with the measured one, way below the percent in the bulk and within at least 5\% in the $\pm$ 4$\sigma$ region around the peak. 

\section{Theoretical predictions and numerical simulations}
\label{section::comparison}

To make a precise assessment of the validity domain of such an approach -- for instance in terms of angular scales -- comparisons with numerical simulations are mandatory.
It is to be stressed that past applications of the large deviation principle agreed remarkably well with numerical results derived from simulations \citep{2014PhRvD..90j3519B,cylindres} and we have no reason to believe it would not be the case in this specific case. Such comparisons are actually interesting in both ways as they can be used to assess the validity regime of such theories but also to validate the accuracy of the simulations
which are usually not tested against non-Gaussian statistical properties.
In the case of the aperture mass, we show in this section that there are issues in the measurement of its one-point PDF in state-of-the-art full sky weak-lensing numerical simulations, which prevents a precise assessment of the validity regime of our implementation.

\subsection{Numerical data}
\label{numerical_sims}
\begin{table}
    \centering
    \bgroup
    \def\arraystretch{1.5}
    \begin{tabular}{|c|c|c|c|c|c|c|}
   \hline
    $\Omega_m$ & $\Omega_{\Lambda}$ & $\Omega_{\rm cdm}$ & $\Omega_b$ & h & $\sigma_8$ & $n_s$  \\
    \hline
    0.279 & 0.721 & 0.233 & 0.046 & 0.7 & 0.82 & 0.97 \\
    \hline
    \end{tabular}
    \egroup
    \caption{Cosmological parameters used throughout this paper.}
    \label{table1}
\end{table}
We consider a set of state-of-the-art full-sky gravitational lensing simulations generated by \cite{Simulation}. Note that the simulations being full-sky is important in the comparisons to our theoretical formalism since we both need a sufficient statistics and also long wave-modes to be present in the simulated data. There, 14 boxes with side lengths of $L = 450 $Mpc$/h$, $2L \cdots 14L$ were prepared along with 6 independent copies.
The number of particles for each box was $2048^3$, making the mass and spatial resolutions better for smaller boxes. These boxes were placed around a fixed vertex representing the observer’s position while each box was duplicated eight times and placed around the observer using periodic boundary conditions. Spherical lens shells with width of 150 Mpc$/h$ (3 per box) were then considered to trace the resulting light-ray paths from the observer to the last scattering surface. 
Each box was evolved in a periodic cosmological N-body simulation following the gravitational evolution of dark matter particles without baryonic processes using {\sc gadget2}. The initial conditions were generated from second-order Lagrangian perturbation theory with the initial linear power spectrum calculated using the Code for Anisotropies in the Microwave Background ({\sc camb}, \cite{camb}). 
It was checked that the matter power spectra agreed with theoretical predictions of the revised Halofit \citep{Halofit} and ray-tracing was performed using the public code {\sc graytrix} which follows the standard multiple-lens plane algorithm in spherical coordinates using the {\sc healpix} algorithm. The data set eventually includes full-sky convergence maps from redshifts $z= 0.05$ to 5.3 at intervals of 150~Mpc$/h$ comoving radial distance and are freely available for download\footnote{\url{http://cosmo.phys.hirosaki-u.ac.jp/takahasi/allsky_raytracing/}}. The adopted cosmological parameters are consistent with the WMAP-9 year result and shown in Table~\ref{table1}. The pixelization of the full-sky maps follows the {\sc healpix} ring scheme with available resolutions of {\sc nside} = 4096, 8192 and 16384.

\subsection{Nulled convergence power spectrum}
\label{nulledCL}

\begin{figure}
    \centering
    \includegraphics[width = \columnwidth]{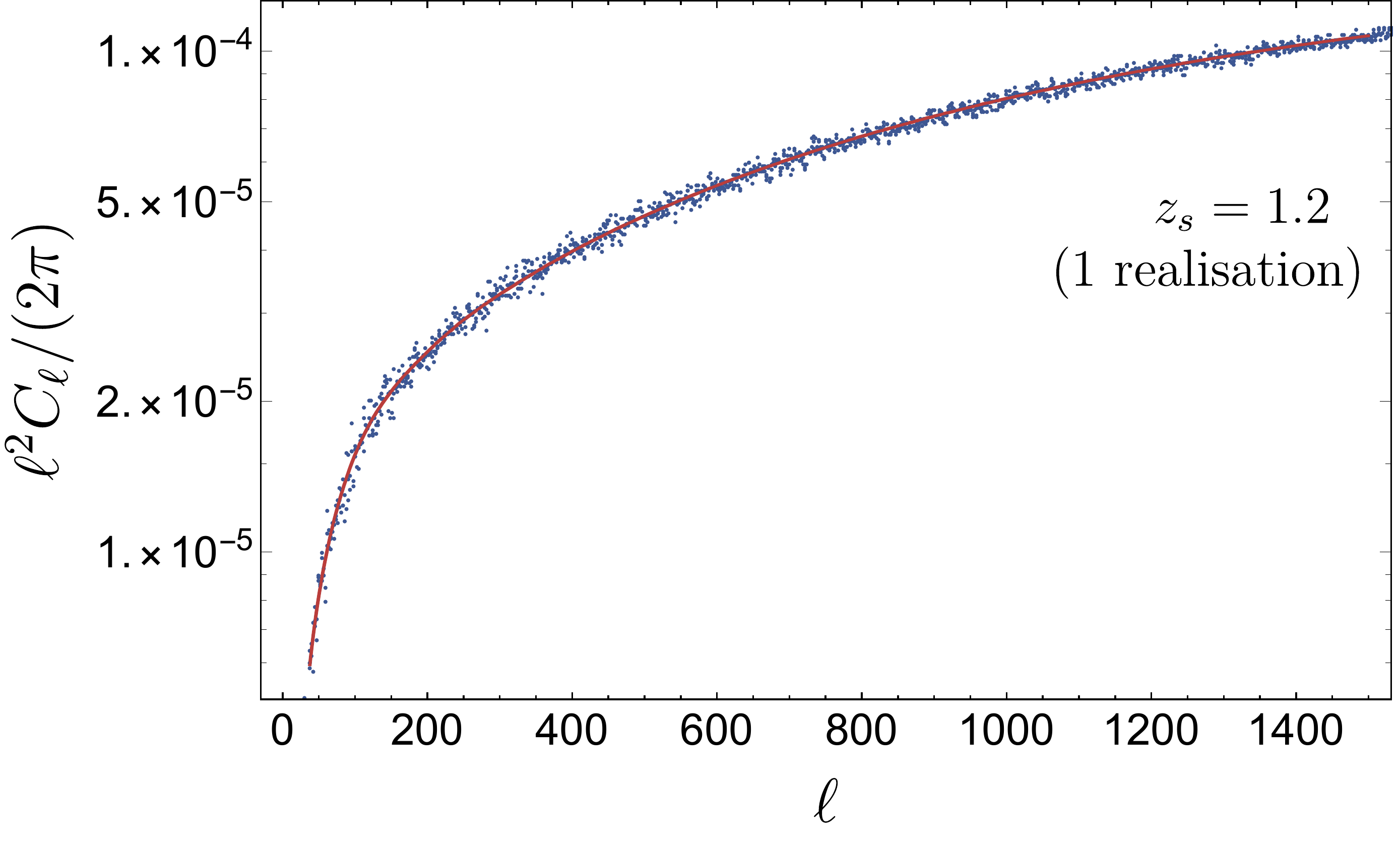}
    \includegraphics[width = \columnwidth]{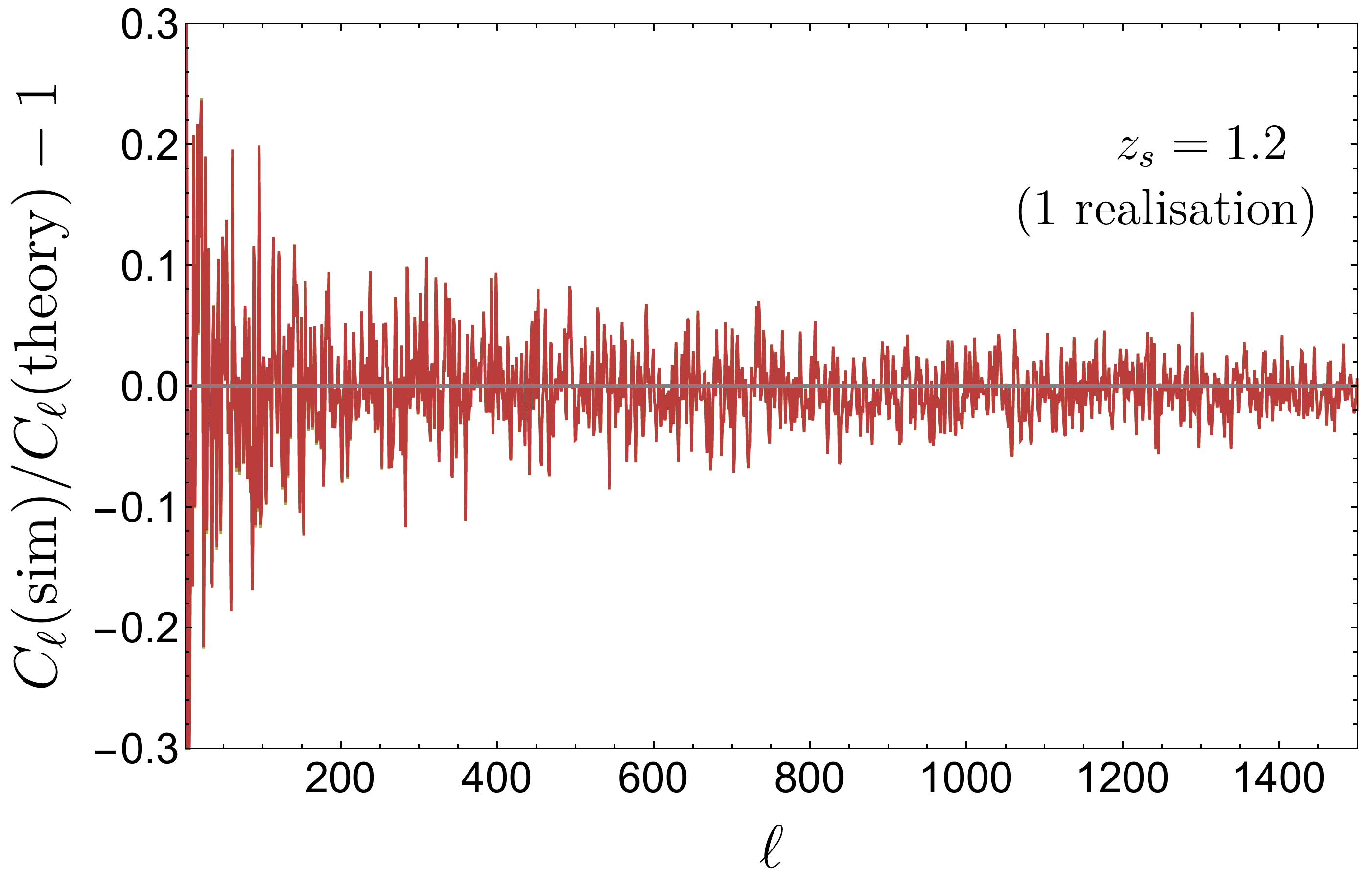}
    \caption{Power spectrum of the convergence field at source redshift $z_s = 1.2$. The red solid line is the theory as computed with equations~(\ref{eq::Cl}) and (\ref{eq::lensshell}). Equation~(\ref{eq::Cldiscrete}) is also implemented in green but not visible since the agreement with the red line is very good. The blue points are the Cls as measured in 1 full-sky realisation. The agreement between the theory and the measurements is very good.}
    \label{clzs1}
\end{figure}

Let us first study the agreement between the measured and computed nulled convergence power spectrum. To that end we follow the prescription of \cite{Simulation} and define the power spectrum as
\begin{equation}
    (1+ (\ell/\ell_{\rm res})^2)C_{\ell}^{\kappa}= \int_{0}^{\chi_s} {\rm d} \chi \frac{\omega(\chi,\chi_s)^2}{\chi^2} P_{z}\left(k=\frac{\ell}{\chi}\right).
    \label{eq::Cl}
\end{equation}
Here $\ell_{\rm res} = 1.6 \times$NSIDE and the factor in front of $C_{\ell}^{\kappa}$ accounts for the finite angular resolution of the maps. The effect of lens-shell thickness is also taken into account in \cite{Simulation} by replacing the matter power spectrum by
\begin{equation}
    P_z(k) \rightarrow \frac{(1+c_1 k ^{-\alpha_1})^{\alpha_1}}{(1+c_2 k ^{-\alpha_2})^{\alpha_3}} P_z(k),
    \label{eq::lensshell}
\end{equation}
where the additional parameters are simulation specific and equal to $c_1=9.5171 \, 10^{-4}, \ c_2 =5.1543 \, 10^{-3}, \ \alpha_1 =1.3063, \ \alpha_2 =1.1475 \text{ and} \ \alpha_3 =0.62793 $. Note that we find that those two effects -- lens-shell thickness and finite angular resolution -- have very little effect at the scales we are interested in. 
Moreover we theoretically computed for the variance at many different scales that the so-called source-plane bias (see section 3 of \cite{Takahashi11} for details), which states that a re-weighting of the convergence pixels by the inverse magnification should be performed before any cumulant measurements, only has a sub-percent influence on the values. 
For reference we plot in Fig.~\ref{clzs1} the convergence power spectrum for a source redshift at $z_s = 1.2$. There, as expected from \cite{Simulation}, the agreement between the theoretical model -- with a Halofit matter power spectrum as input -- and the measurement in the simulation is excellent. 

\begin{figure}
    \centering
    \includegraphics[width = \columnwidth]{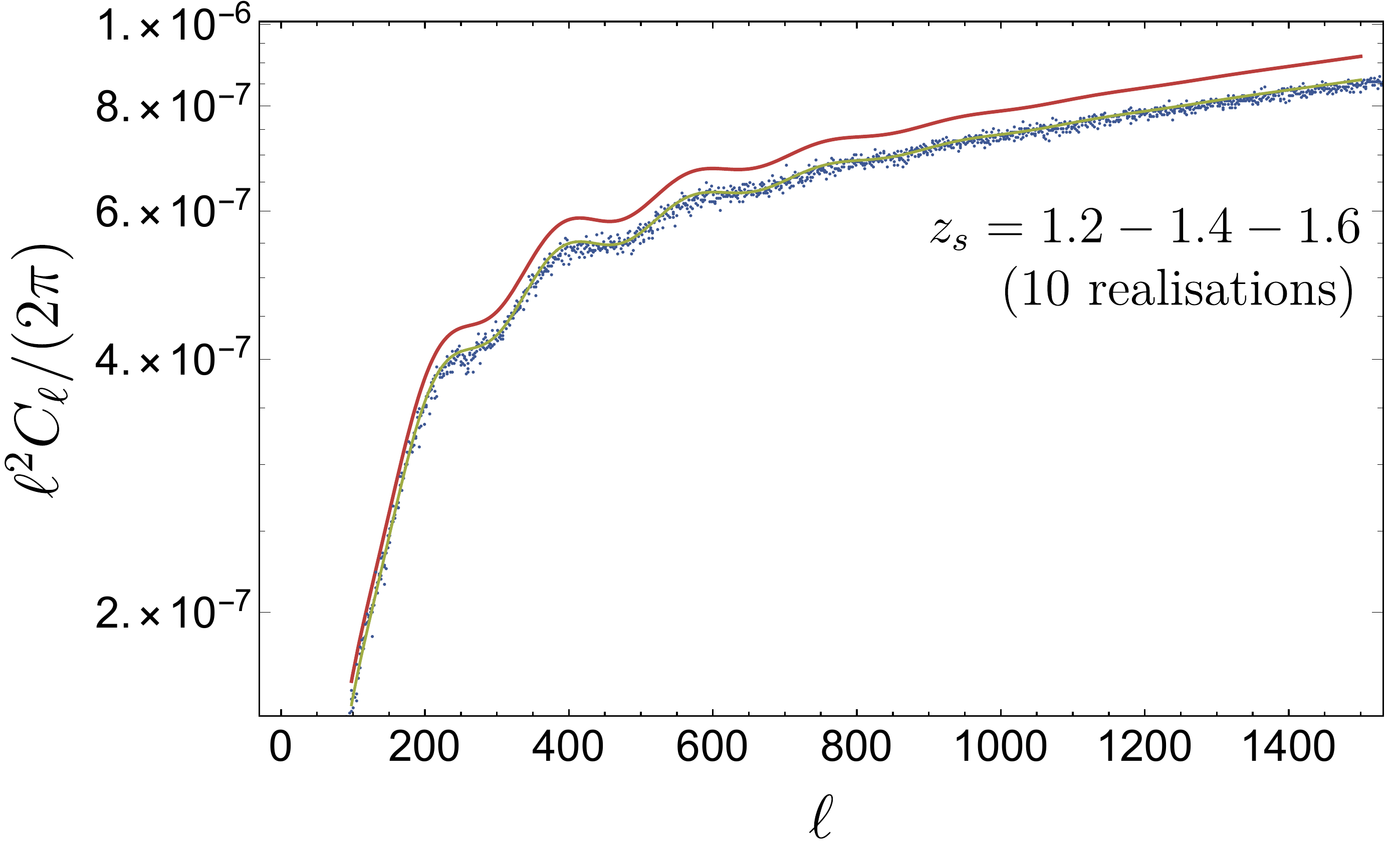}
    \includegraphics[width = \columnwidth]{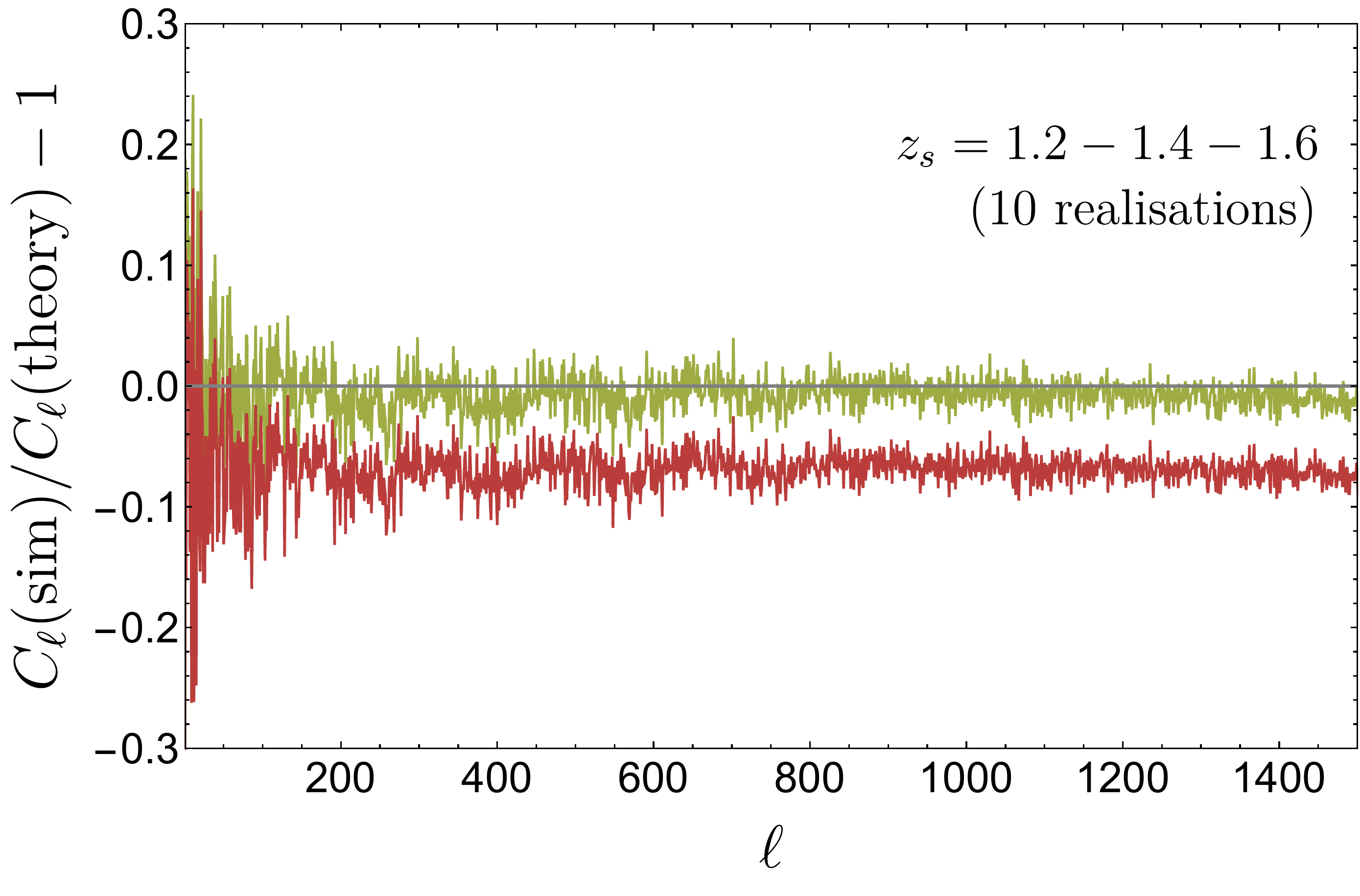}
    \caption{Power spectrum of the nulled convergence field with source redshifts located at $z_s = 1.2, 1.4 \ \& \, 1.6$. The red (resp. green) solid line is the theory as computed with equations~(\ref{eq::Cl}) (resp. (\ref{eq::Cldiscrete})) and (\ref{eq::lensshell}). The blue points are the average of the Cls as measured in 10 full-sky independent realisations. A constant bias of roughly 7\% is observed when not accounting for the discreteness of lens planes.}
    \label{clnull}
\end{figure}

We then show in Fig.~\ref{clnull} the computed and measured nulled power spectrum of the convergence for source redshifts located at $z_s =$ 1.2, 1.4 and 1.6. There, a constant bias is observed at all scales and whose amplitude is directly linked to the width of the nulled convergence lensing kernel, the narrower the greater the bias. This bias can be fully explained by taking into account the discreteness of contributing lens planes in the computation of the nulled $C_{\ell}^\kappa$ 
\begin{equation}
    (1+ (\ell/\ell_{\rm res})^2)C_{\ell}^{\kappa}= \sum_i  \Delta_i \frac{\omega(\chi_i,\chi_s)^2}{\chi_i^2} P_{z}\left(k=\frac{\ell}{\chi_i}\right),
    \label{eq::Cldiscrete}
\end{equation}
where $\Delta_i$ are the width of the lens planes, $\Delta_i = 150$ Mpc$/h$, and $\chi_i$ are the comoving distances from the observer to the centres of those planes, $\chi_i = 150(i-0.5)$ Mpc$/h$. This correction becomes more relevant here than in the no-nulling case since fewer planes contribute. \red{In our case with source redshifts located at $z_s =$ 1.2, 1.4 and 1.6, the effective thickness of the nulled kernel is 600 Mpc$/h$ with thus only 4 lens planes contributing to the total effect which is not enough to mimic with sufficient accuracy the continuous line-of-sight integral.}

As a sanity check we also computed the leading correction to the Limber approximation given in \cite{ExtendedLimber} and did not find any difference. This is not so surprising since, though nulling kernels are somewhat narrower, the $\ell$ values for which the limber approximation is supposedly valid are $\ell \gg \Bar{\chi}/\Delta\chi \simeq 5$. This still adds the benefit of effectively checking that the Limber approximation is still valid for a narrow nulled lensing kernel.

Fortunately, the bias that thus appears in the naive estimation of the variance of the field is roughly constant across all scales for a given nulled lensing kernel and thus cancels out in the estimation of reduced high-order statistical quantities. We checked it for example on the reduced skewness of the aperture mass in Fig.~\ref{measuredS3} where replacing the integrations along the line of sight by summations over the lens planes only affected the value by less than half a percent. This is also clear when comparing the line of sight integrations to measurement in the simulation as shown in Fig.~\ref{measuredkXY}.

Finally, the reason for this constant bias can be understood in the following way: The integrand of equation (\ref{eq::Cl}) is the product of i) a lensing kernel term that solely depends on the redshift of contributing lenses and not on the scale $\ell$, and ii) a matter power spectrum term, that since nulling makes us probe a narrow range of both redshifts and physical scales can be approximated by $k^{n_s}$ times some redshift dependence. Now $n_s$ would obviously change for different values of $\ell$ but since the nulled lensing kernel varies very rapidly the resulting $1/\chi^{n_s}$ term in the integrand will merely act on its amplitude rather than its shape up to quite significant values of $n_s$. This makes the integrand of equation (\ref{eq::Cl}) roughly dependent on $\ell$ only through a multiplicative term and thus the error one commits on the integral replacing (\ref{eq::Cl}) by (\ref{eq::Cldiscrete}) is proportional to this multiplicative term which results in a constant bias for the $C_{\ell}^{\kappa}$ across all scales. This also means that one can actually predict the amplitude of the resulting bias $b$ simply by computing
\begin{equation}
	b = \frac{\sum_i \Delta_i \, \omega(\chi_i,\chi_s)^2/\chi_i^2}{\int {\rm d}\chi \omega(\chi,\chi_s)^2/\chi^2}.
\end{equation}
We thus recover the $\sim 6\%$ difference that we found taking into account the discreteness of the lens planes. Varying the value of $n_s$ from 0 to -10, which are both very extreme and un-realistic values since we would expect $n_s \sim -1.5$ for our scales of interest, we find that the value of $b$ only changes by 1/1000 thus confirming our formula for the bias.


\subsection{Filtering methods}

To convolve those maps with a difference of top-hat windows of the desired angular radii and thus access simulated $\map$ statistics, we used 2 different methods that proved to give equivalent results at more than the percent precision for skewness and kurtosis and indistinguishable by eye PDFs, thus giving us confidence in the filtering schemes. Our \textit{real-space} method consists in using the \textit{query\_disc} function of {\sc healpy} to find all pixels whose centres are located within a disk centred at one specific pixel $p$ which allows to reassign the value of $p$ as being the mean of all the pixels inside the disk. Our \textit{dual-space} method consists in convolving the convergence map with the appropriate filter by decomposing both the map and filter in spherical harmonics using the \textit{map2alm} and the \textit{beam2bl} functions of {\sc healpy}, convolving the map and filter in this space and then going back to pixel-space via the \textit{alm2map} function. 
The 2 methods were found to agree at the level of each map and for all scales which was not a given since i) the \textit{real-space} procedure does not yield an exact top-hat and ii) the $a_{lm}$ transform has a non empty kernel, meaning that a randomly generated map in pixel-space, sent to $a_{lm}$-space and back might be significantly different from the original one.

\subsection{Measured BNT $\map$ skewness}
\label{section::Mapskewness}

In principle, we would expect the large deviations + nulling formalism presented in this paper to perform extremely well as the BNT transform localises the lensing kernel to a finite range of redshifts and therefore physical scales thus making the $\map$ one-point statistics closer to the density slope in long cylinders for which large deviation formalism was already proven very efficient for example in \cite{cylindres}. 
Moreover this strategy applied to convergence maps was also found to be very effective \citep{Barthelemy20a}. Also note that a very satisfying agreement between the theory and this simulation suite was already found in Fig.~\ref{NulledPDF} and that we would merely like to extend this result down to smaller scales \red{and different ratios between the filtering scales}.

\begin{figure}
    \centering
    \includegraphics[width = \columnwidth]{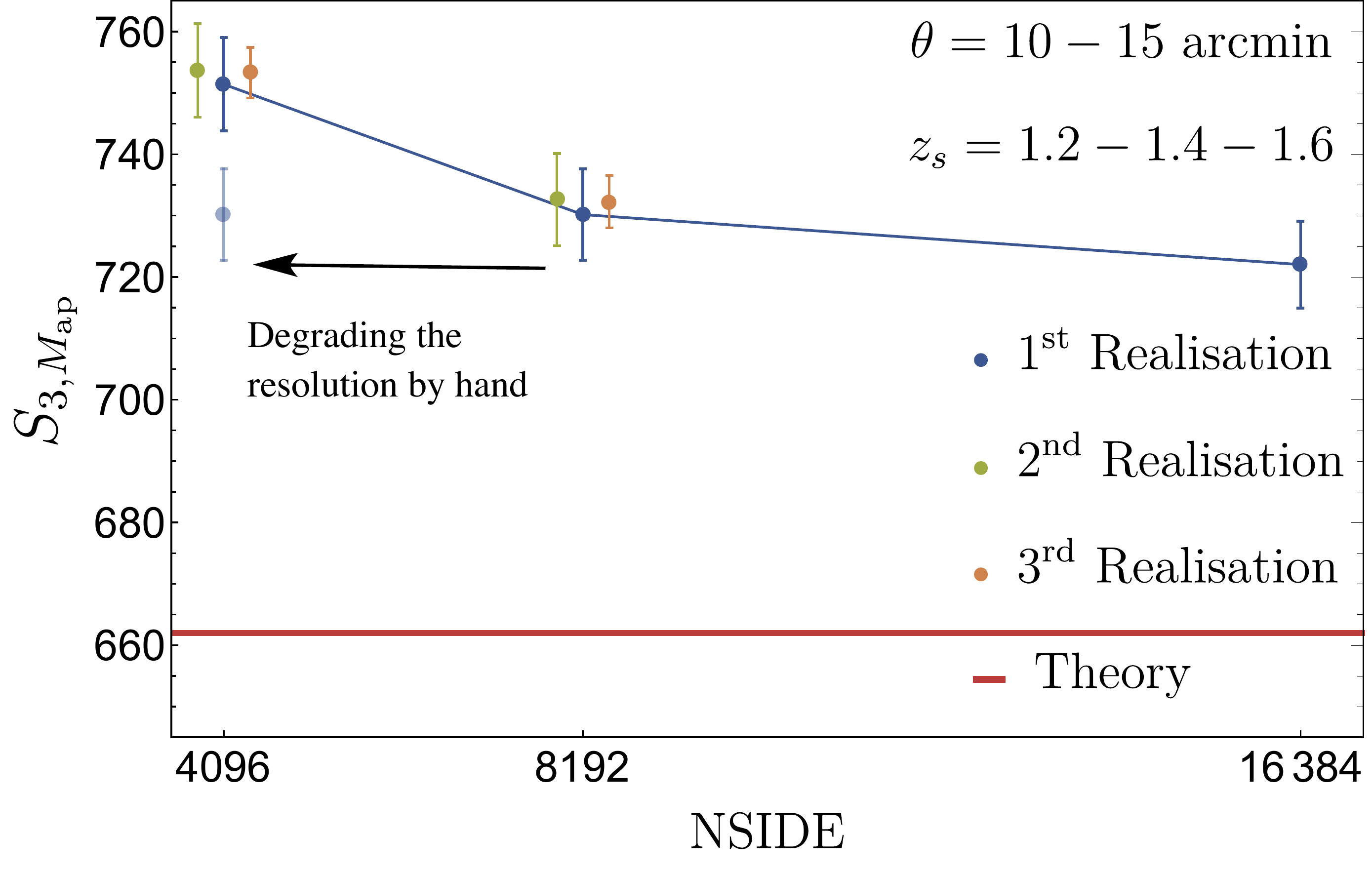}
    \caption{Different BNT $\map$ skewness values as measured in the simulation as a function of the map resolution NSIDE for three different realisation with respectively blue, green and orange error bars. For comparison, the corresponding prediction from tree-order perturbation theory is displayed using a red solid line.}
    \label{measuredS3}
\end{figure}

The difficulty that has been encountered when trying to assess the validity regime of our theoretical approach lies in the difficulty to precisely measure and assign error bars to measured $\map$ quantities. In the context of the PDF, this can be further exemplified on the measurement of the skewness, a single number but that determines the first and foremost (thanks to cumulant hierarchy) non trivial contribution to non-gaussianity in the PDF. To this purpose, we study the BNT $\map$ skewness with opening angles of $\theta_2 = 15$ and $\theta_1 = 10$ arcmin and source planes respectively located at redshift $z_s = 1.2 - 1.4 - 1.6$. For 3 different realisations of the full sky maps -- except for the greatest resolution which only offers 1 realisation -- we apply our filtering scheme, measure the BNT $\map$ skewness \red{in the map} and measure error bars as standard error on the mean computed among 8 subsamples of the full sky. \red{We find that:
i) the 2 different filtering schemes give similar results with most of the time much more that the percent precision.}
As shown on Fig.~\ref{measuredS3},  
ii) different realisations at the same resolution agree very well being statistically compatible with one another as probed by the error bars and with mean values very close to each other;
iii) there is a visible shift of the value with increasing resolution to the point that the lowest and highest resolution -- though coming from the same realisation -- do not seem compatible, which also does not seem to come from resolution itself since degrading the map by hand to a lower resolution, filtering it and then re-measuring the skewness leads to very similar results;
iv) none of the measured values agree with the theoretical prediction ($\lesssim 10 \%$ relative difference).
\red{This is reminiscent of for example section 4.2 of \cite{Uhlemann19} where the impact of resolution of the N-body simulation observed for the reduced skewness of the 3D matter density field was similar though the 2 cases are not exactly comparable.}
Also note that to some extent, those observations are still valid in the case presented previously in Fig.~\ref{NulledPDF} but smaller. There, the PDF was measured from one single realisation of the map at the lowest resolution and proved to perform very well. 

Additionally, note that the discrepancy between the measured $\map$ skewness for different resolutions -- but still the same realisation -- of the nulled convergence map is mainly sourced by the fact that the measured $\map$ variance varies between resolutions.

\begin{figure}
    \centering
    \includegraphics[width = \columnwidth]{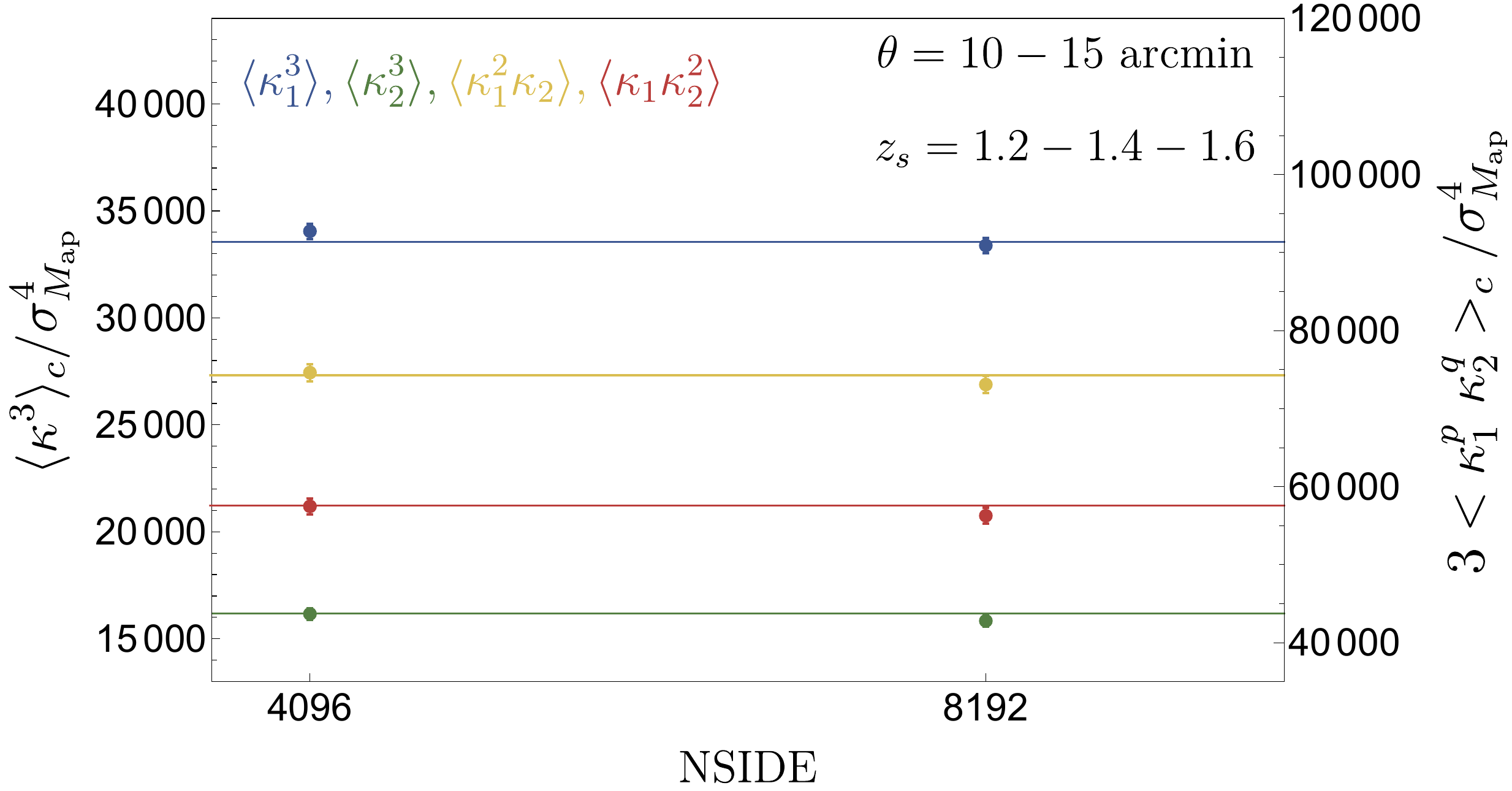}
    \caption{Reduced cross-cumulants of the nulled convergence field appearing in equation~(\ref{decomposeS3}). Points with error bars are measured in the simulation's first realisation and solid lines represent the corresponding theoretical predictions. Even though the values tend to be quite sensitive to the resolution, a sub-percent agreement with the theoretical predictions is found.}
    \label{measuredkXY}
\end{figure}

A more careful examination of the BNT $\map$ skewness value can also be performed decomposing it as cross-cumulants of the BNT convergence field filtered at 2 different scales. Following equation~(\ref{knXY}) it is written as
\begin{eqnarray}
    S_{3,\map} &=& \frac{\left\langle\map^3\right\rangle_c}{\sigma^4_{\map}}\\ &=&\frac{\left\langle\kappa_2^3\right\rangle_c \!-\! \left\langle\kappa_1^3\right\rangle_c \!+\! 3 \left\langle\kappa_1^2 \kappa_2\right\rangle_c \!-\! 3 \left\langle\kappa_1 \kappa_2^2\right\rangle_c }{\sigma^4_{\map}}.
    \label{decomposeS3}
\end{eqnarray}
We then measured each cross-cumulants in the first realisation at the lowest and medium resolutions and compared them to their respective theoretical predictions as shown in Fig.~\ref{measuredkXY}. The agreement for those cross-cumulants is found to be excellent (to the percent precision and very close to the 1-$\sigma$ error bar), allowing us to extend the results obtained on the 1-point convergence to a multi-scale analysis. 
Let us notice that in eq.~(\ref{decomposeS3}) numbers of similar orders of magnitude subtract each other which in turn tends to decrease the precision of the prediction of the BNT $\map$ skewness since absolute difference tend to stay the same while the subtraction of 2 numbers of similar magnitude results in a smaller number which finally increases the relative difference between the theory and the numerical simulation. 
We are thus looking at some very subtle effects in the convergence field itself which tend to have a significant impact on the quantities we are interested in. This could be the sign of higher order perturbative corrections (loop corrections) but given that observables similar to the BNT $\map$ such as densities in cylinders were already studied and found to be accurate at tree-order \citep{cylindres}, 
a more likely explanation is to be found on the simulation side. Indeed subtle numerical artefacts could show up in the BNT $\map$ such as resolution or discretisation effects, convergence of the N-body algorithm etc.
In fact, since the relative incoherences observed on the skewness when increasing the maps resolution are not at all seen on the convergence field itself, it could indeed be possible that small errors in the simulation are amplified when looking at $\map$ statistics.

Finally note that we also tried to perform -- sometimes successfully -- comparison of our theoretical formalism to other numerical suites. However other issues such as the evaluation of the degree of independence between realisations of randomised lines of sight in replicates of a small N-body box, and other simulation-specific issues prevented us from doing any more refined comparison to available ray-tracing simulations. There is therefore a dire need for future code comparisons, validations and improvements in the line of \cite{2020MNRAS.493..305H} that specifically target higher-order statistics. Such future developments could then allow to test in more details the validity regime of our theoretical model for the BNT $\map$ statistics. In any case, exploitation of the $\map$ PDF as an observable should not, in light of those findings, be done assuming that numerical simulations are sufficiently accurate, and physical formalism just like ours should be considered to at least test the different numerical schemes in cosmological analysis of non-Gaussian statistics.

\subsection{Resolution effects} 
Before concluding, let us mention two additional tests that were done on the theory side to try to mend the discrepancy  between the theory and the simulation. We separate those since they are not tied at all to the comparison to simulations in general.  First, we tried to introduce a $l$-cut due to the resolution in the computation of the skewness itself which is done by introducing the corresponding $k$-cut at each slice along the line of sight thus modifying the variance and its derivatives in equation~(\ref{mapskewness}). 
Second, we considered the effect of having a non-exact top-hat filter in appendix~\ref{sensitivity} choosing our $L$ parameter -- typically $L = 20$ for the lowest resolution -- to smooth our filter over 2 pixels. Those two prescriptions were shown to have non significant impact on the skewness and therefore are unlikely explanations for the observed differences.

\section{Shape noise}
\label{noise}

Since the weak-lensing aperture mass is obtained from cosmic shear measurements by means of the measured shape of galaxies which themselves are intrinsically elliptical, the observed shear is the sum of contributions from weak-lensing and the intrinsic galactic ellipticities. Shape noise is caused by the variance of the  intrinsic ellipticity, which is the dominant source of noise in shear measurements and impacts the convergence field with a random noise that can be modelled by a Gaussian with zero mean and variance $\sigma_{SN}^2$. To estimate the variance of the shape noise distribution in a Euclid-like configuration, we assume
\begin{equation}
\sigma_{SN}^2=\sigma_\epsilon^2 N_{\rm bin}/(n_{g_s}\Omega_{\theta}),
\label{sigmaSN}
\end{equation}
where  $\sigma_\epsilon^2 = 0.3$ is the Gaussian noise on the determination of the lensing-induced ellipticity of each galaxy, $\Omega_{\theta}$ is a solid angle in units of~arcmin$^2$ -- the area of a pixel -- $n_{g_s} = 30$ arcmin$^{-2}$ is the normalised mean number of observed galaxies, and $N_{\rm bin} = 10$ is the number of equally populated redshift bins in the survey \citep{2020A&A...636A..95D}. 
The resulting Gaussian noise that this procedure thus induces on the constructed $\map$ field is then given by
\begin{equation}
    \sigma^2_{SN, \map} = \frac{\alpha \sigma_{\epsilon}^2 N_{\rm bin}}{\pi n_{g_s}}\left(\frac{1}{\theta_1^2}-\frac{1}{\theta_2^2}\right),
    \label{sig2SN}
\end{equation}
where $\alpha = 1$ for regular $\map$ and $\alpha = \sum_{i=j-2}^{j} \left(M^{ij}\right)^2$ for BNT $\map$. We would then obtain the resulting noisy theoretical $\map$ PDF by convolving the previously computed PDFs with a zero-mean Gaussian of the appropriate variance
\begin{multline}
    P_{SN}(\map) = \frac{1}{\sigma_{SN, \map} \sqrt{2\pi}}\int_{-\infty}^{+\infty} {\rm d}\hat{\map} P(\hat\map) \\ \exp\left(-\frac{(\map - \hat{\map})^2}{2 \sigma^2_{SN, \map}}\right).
    \label{Pnoise}
\end{multline}

\begin{figure}
    \centering
    \includegraphics[width = \columnwidth]{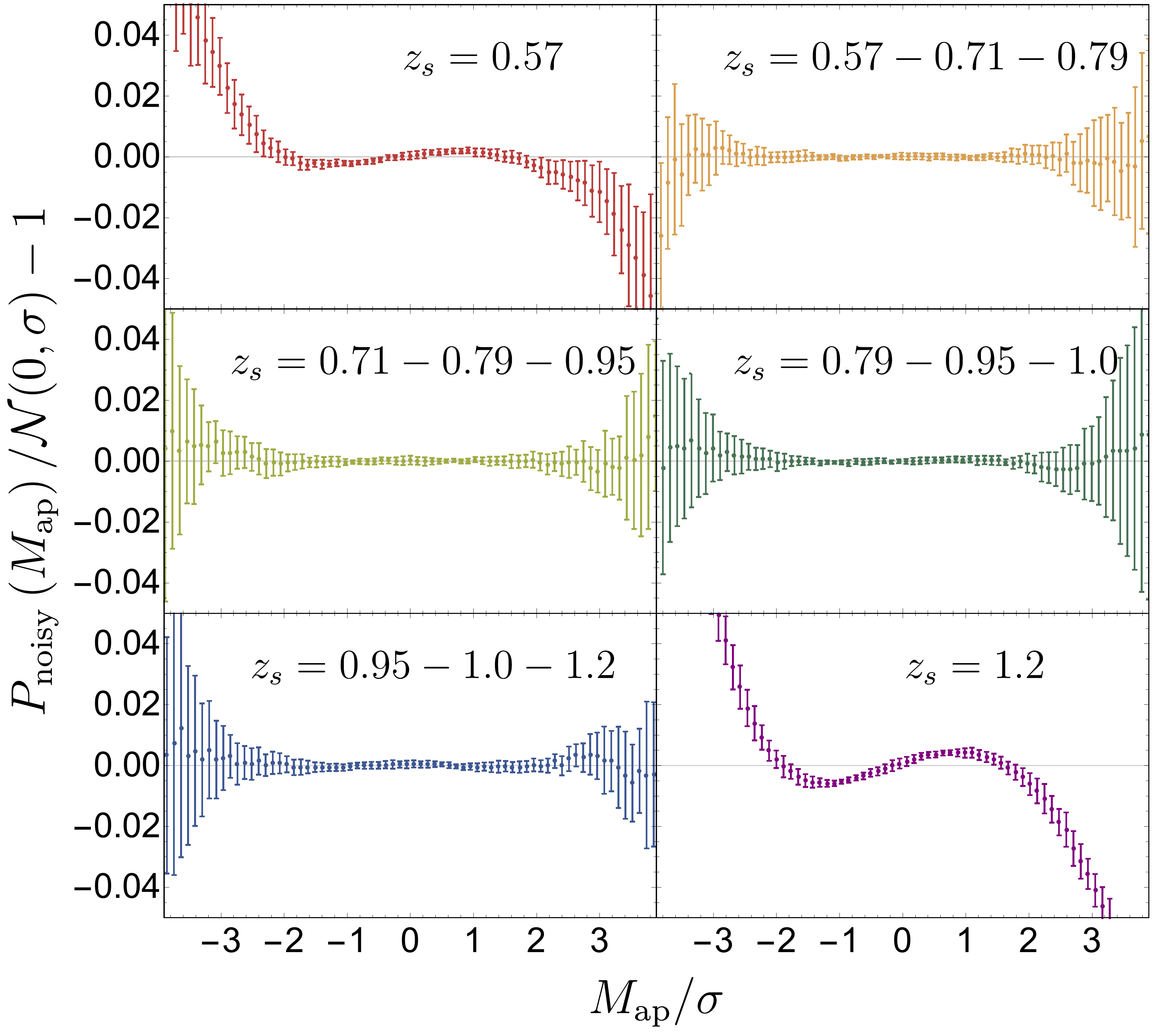}
    \caption{Residuals of the noisy $\map$ PDFs with respect to a Gaussian of the same variance. The opening angle is $\theta_1 = 15$ and $\theta_2 = 30$ arcmin. \red{The regular $\map$ fields are taken at source redshifts $z_s = 0.57$ and 1.2, and the BNT $\map$ fields have source redshifts located in between mimicking a Euclid binning}. Error bars are taken as the standard deviation between 10 realisations of the simulation with added noise to estimate the cosmic variance.}
    \label{fig::shapenoise}
\end{figure}

As an illustration of the effect of shape noise on the $\map$ PDF we show in Fig.~\ref{fig::shapenoise} the relative difference between the noisy PDFs $P_{SN}(\map)$ and a Gaussian of the same mean and total variance (noise + signal). This is done by adding Gaussian noise to each pixels of the simulated convergence maps of 10 realisations and at redshifts mimicking a Euclid binning, combining them so as to get nulled maps, measuring the resulting $\map$ PDFs and computing the standard deviation between the 10 realisations in each bin of the $\map$ PDF as an estimate of the cosmic variance. We fix the opening angle to $\theta_2 = 2 \theta_1 = 30$ arcmin and show both the residuals of the noisy $\map$ PDF with respect to a Gaussian of the same variance \red{for source redshifts $z_s = 0.57$ and 1.2 and nulled bins in between. Note that, as an additional tool, we also present in Appendix~\ref{appendix::error_bar} a fast and analytical method to estimate the detectability of non-Gaussian features in the Aperture mass PDF in a realistic setting and in the presence of shape noise.}

We observe that a part of the non-Gaussian signal still remains in the regular $\map$ fields while the signal of a single bin of the BNT $\map$ is so noisy that we only observe the zero-mean Gaussian of variance $\sigma^2_{SN, \map}$. This is explained by the fact that the amplitude of the BNT $\map$ signal is by construction way smaller than that of regular $\map$ since it boils down to reducing the number of lenses contributing to the effect by localising the lensing kernel at specific physical scales and also by the fact that the noise itself is increased. Though this could thus seem like the end of the line for the nulling strategy, this is fortunately not yet the case. Indeed, at the level of a tomographic analysis where multiple source redshifts are considered, the BNT strategy is a simple reorganisation of the signal that, similarly to a principle component analysis, drastically diminishes the redundancy of the information that is present in each map as a result of scale mixing. Thus no signal is lost and though the noise is seemingly increasing in each nulled bin, a joint analysis of all bins should enable to recover all the information while having increased our ability to theoretically probe this information. \red{There} remains to write down the formalism for the joint PDF of all $\map$ nulled bin which was hinted in \cite{Barthelemy20a} but is left for future work. On the other hand, looking at the regular $\map$ PDF at $z_s = 1.2$ and seeing that some non-Gaussianities remain can be misleading since i) a part of the signal comes from scales that are not well modelled which prevents us from extracting cosmological information out of it (as shown with the residual at $z_s = 0.57$) and ii) the signal in each redshift bin is very redundant and thus a false sense of accumulating information could come out of it while it is effectively not as much the case as one could imagine. The joint-analysis of BNT transformed redshift bins, where shape noise is properly taken into account and model, is thus the correct way to perform a tomographic analysis. We leave for a further work a precise analysis of its performance.

\section{Discussion and Conclusion}
\label{conclusion}
In this paper we have presented a formalism for the one-point statistics of the aperture mass seen as a difference of top-hat filters applied to the convergence field. The geometry and time-evolution within the light-cone is taken into account relying on the fact that the correlations of the underlying matter density field along the line of sight are negligible compared to transverse directions, which leads to redshift slices being treated as statistically independent. Within each redshift slice, the most likely non-linear dynamics of the matter density field filtered in concentric disks is on average and in the limit of small variance well approximated by the cylindrical collapse, a fact which makes it possible to treat the filtered field as a random variable satisfying a large deviation principle. 
The large deviation formalism then allows us to compute the joint CGF of the density filtered in disks of two different radii, thus the CGF of the density slope and finally the one-point statistics of the aperture mass $\map$ through projection effect. 
The formalism in itself only provides means to compute the $\map$ CGF and PDF if a prescription for the non-linear matter power spectrum is used as an input. Fortunately, this issue has received a lot of attention in the recent years with very reliable models such as emulators \citep{EuclidEmulator}. All subsequent non-Gaussian information is modelled through the cylindrical collapse dynamics in each redshift slice.

We also proposed to combine the present formalism with a nulling strategy -- the BNT transform -- that boils down to linearly combining maps in different redshift bins with weights chosen so as to effectively localise the redshift distribution of lenses contributing to the signal. This avoids the mixing of scales which is particularly important when one wants to leverage the influence of the small scales where theoretical models break down. \red{This strategy is also particularly relevant to mitigate the impact of baryonic physics since it was showed for example in \cite{2020arXiv201209614M} that trying so without any localisation of the signal along the line of sight requires to delay the small-scale information to a point where the inferred cosmological constraints are no longer competitive.}

The formalism we developed was tested against numerical simulations where it was found to, once combined with the nulling strategy, perform extremely well with no apparent deviation from the simulation given the estimated error bars, see Fig.~\ref{NulledPDF}, at least for sufficient large scales and redshifts. However more conclusive tests could not be performed for smaller scales due to some imprecision and inconsistencies that we found in the numerical suite like an inconsistent shift of the aperture mass skewness when increasing the map resolution. Since this issue is not observed at the level of the convergence field and since the aperture mass enhances subtle non-linear features of the convergence field -- see Fig.~\ref{measuredkXY} and equation~(\ref{decomposeS3}) -- the measured values are to be handled with caution and more numerical tests should be performed. Generally speaking, higher-order weak-lensing statistics are not tested in fine details in numerical simulations and there is a real need to provide theoretical prescriptions that enable those tests. We hope that the present formalism falls in this line.

Many recent works started the daunting task of evaluating the accuracy of weak-lensing simulations for non-Gaussian statistics \citep{2020MNRAS.493..305H}. \cite{2020AJ....159..284M} studied the impact of thickness of the lens planes used to build past light-cones and the mass resolution of the underlying N-Body simulation and thus proposed guiding lines for the design of future numerical suites. Another issue is the pseudo-independence of lensing maps \red{generated} from randomisation of the line of sight through replication of a small but better resolved N-body box. \cite{2016PhRvD..93f3524P} for example suggested that maps generated from 1 or 2 realisations of the same box could be considered independent but those results, to our knowledge, were not replicated nor extended to the case of the convergence PDF or moments and even less for the aperture mass. \red{Another important issue that must be taken into account, although not present in this paper, is the impact that small maps/patches constructed as planar projections of part of the sphere have on high-order statistics of the convergence/aperture mass fields. Such a study was recently performed in \cite{2018A&C....24...84V} and indeed showed a measurable impact on peak counts and Minkowski Functionals.} All of this suggests that indeed more work is needed if we are to extract subtle non-linear effects from weak-lensing simulated maps.

On a more theoretical side and for the specific case of the aperture mass, the fact that it enhances subtle non-linear features of the convergence field would call for evaluation of beyond cylindrical collapse contributions of the non-linear dynamics of the underlying density field, that is loop corrections in perturbation theory. This is not the main path chosen to improve agreement between theory and simulations since the BNT aperture mass is somewhat close to the density field filtered in long cylinders for which the cylindrical collapse dynamics were shown to perform extremely well \citep{cylindres}, but this is nonetheless an interesting work that should be carried out in the future.

Future theoretical work on the aperture mass PDF or moments could also include general relativistic corrections to the non-linear Newtonian dynamics of the underlying matter density field. Such effects were already taken into consideration in \cite{2020MNRAS.497.2078L} on the convergence PDF by means of ray-tracing through a relativistic N-body simulation but a detailed comparison with the Newtonian PDF was not performed. Nevertheless, should this be done and a significant effect be found, those type of corrections could be computed within the post-Friedmann formalism for which a re-definition of the usual weak-lensing fields was recently performed in \cite{2019JCAP...05..045G}. Another crucial correction to be taken into account comes from the fact that the observed field, the lensing-induced ellipticities of galaxies, correspond to the reduced shear $g = \gamma/(1+\kappa)$ rather than the shear itself. Fortunately, the leading correction to the observed $\map$ skewness that arises can be computed with perturbation theory as was done in \cite{Schneider1998} and in appendix~\ref{2ndOrder} and shown to be a few percent of the skewness value. This can also be performed at the level of the PDF of the density slope within a redshift slice by means of large deviation theory \citep{paolo}. Hence in principle, computing the observed $M_{\rm ap}^g$ PDF by fully taking into account the geometry of the light-cone is straightforward, but since it is numerically involved and that there are already challenges to compare the $\map$ PDF to simulations we leave it to future work.

Still based on \cite{paolo}, we also showed in appendix~\ref{sensitivity} how to account for compensated filters that are different from a difference of top-hat windows. Overall we show that even filters that deviate significantly from the top-hat lead to only a few percents difference on the skewness which states that our formalism could be readily implemented even with smoother filtering schemes, keeping the relative theoretical simplicity of top-hats and being able to model the systematic deviation coming from this approximation. This could be particularly useful knowing that some compensated filters do offer better cosmological constraints than others as is discussed for example in \cite{2016A&A...593A..88L}.

Finally, note that to be realistic, one would also need to account for the fact that the aperture mass is measured not from a single source redshift but from a given source galaxy distribution $n_s(z_s)$. This can be readily done in our formalism. Indeed, when aiming to predict the weak-lensing aperture mass measured from $n_s(z_s)$, one can simply replace the lensing kernel $\omega$ by
\begin{equation}
\label{eq:weightsource}
    \omega_{n_s}(\chi) = \frac{3\,\Omega_m\,H_0^2}{2\,c^2}\!\! \int_{\chi}^{\infty}\!\! d\chi_s \frac{\mathcal{D}(\chi)\,\mathcal{D}(\chi_s-\chi)}{\mathcal{D}(\chi_s) a(\chi)}\, n_s(z_s)\frac{dz_s}{d\chi_s},
\end{equation}
and only the implementation of the nulling strategy would be slightly different \citep{Nulling}.

\section*{Acknowledgements}
This work is partially supported by the SPHERES grant ANR-18-CE31-0009 of the French {\sl Agence Nationale de la Recherche} and by Fondation MERAC.
AB's work is supported by a fellowship from CNES.
This work has made use of the Horizon Cluster hosted by Institut d'Astrophysique de Paris. We thank St\'ephane Rouberol for running smoothly this cluster for us, Ken Osato for pointing us to the simulation we used in this paper as well as discussing its issues with us and Simon Prunet for fruitful discussions. 
AB thanks the Institut de Physique Th\'eorique for hosting several visits during the completion of this work.
We also thank Aoife Boyle, Oliver Friedrich, Cora Uhlemann, Martin Kilbinger and Rapha\"el Gavazzi
for fruitful discussions.

\bibliographystyle{mnras}
\bibliography{biblio} 

\begin{thebibliography}{}
\makeatletter
\relax
\def\mn@urlcharsother{\let\do\@makeother \do\$\do\&\do\#\do\^\do\_\do\%\do\~}
\def\mn@doi{\begingroup\mn@urlcharsother \@ifnextchar [ {\mn@doi@}
  {\mn@doi@[]}}
\def\mn@doi@[#1]#2{\def\@tempa{#1}\ifx\@tempa\@empty \href
  {http://dx.doi.org/#2} {doi:#2}\else \href {http://dx.doi.org/#2} {#1}\fi
  \endgroup}
\def\mn@eprint#1#2{\mn@eprint@#1:#2::\@nil}
\def\mn@eprint@arXiv#1{\href {http://arxiv.org/abs/#1} {{\tt arXiv:#1}}}
\def\mn@eprint@dblp#1{\href {http://dblp.uni-trier.de/rec/bibtex/#1.xml}
  {dblp:#1}}
\def\mn@eprint@#1:#2:#3:#4\@nil{\def\@tempa {#1}\def\@tempb {#2}\def\@tempc
  {#3}\ifx \@tempc \@empty \let \@tempc \@tempb \let \@tempb \@tempa \fi \ifx
  \@tempb \@empty \def\@tempb {arXiv}\fi \@ifundefined
  {mn@eprint@\@tempb}{\@tempb:\@tempc}{\expandafter \expandafter \csname
  mn@eprint@\@tempb\endcsname \expandafter{\@tempc}}}

\bibitem[\protect\citeauthoryear{{Bard}, {Kratochvil}  \& {Dawson}}{{Bard}
  et~al.}{2016}]{2016ApJ...819..158B}
{Bard} D.,  {Kratochvil} J.~M.,   {Dawson} W.,  2016, \mn@doi [\apj]
  {10.3847/0004-637X/819/2/158}, \href
  {https://ui.adsabs.harvard.edu/abs/2016ApJ...819..158B} {819, 158}

\bibitem[\protect\citeauthoryear{{Barthelemy}, {Codis}, {Uhlemann},
  {Bernardeau}  \& {Gavazzi}}{{Barthelemy} et~al.}{2020a}]{Barthelemy20a}
{Barthelemy} A.,  {Codis} S.,  {Uhlemann} C.,  {Bernardeau} F.,   {Gavazzi} R.,
   2020a, \mn@doi [\mnras] {10.1093/mnras/staa053}, \href
  {https://ui.adsabs.harvard.edu/abs/2020MNRAS.492.3420B} {492, 3420}

\bibitem[\protect\citeauthoryear{{Barthelemy}, {Codis}  \&
  {Bernardeau}}{{Barthelemy} et~al.}{2020b}]{Barthelemy20b}
{Barthelemy} A.,  {Codis} S.,   {Bernardeau} F.,  2020b, \mn@doi [\mnras]
  {10.1093/mnras/staa931}, \href
  {https://ui.adsabs.harvard.edu/abs/2020MNRAS.494.3368B} {494, 3368}

\bibitem[\protect\citeauthoryear{{Baugh}, {Gaztanaga}  \& {Efstathiou}}{{Baugh}
  et~al.}{1995}]{1995MNRAS.274.1049B}
{Baugh} C.~M.,  {Gaztanaga} E.,   {Efstathiou} G.,  1995, \mn@doi [\mnras]
  {10.1093/mnras/274.4.1049}, \href
  {https://ui.adsabs.harvard.edu/abs/1995MNRAS.274.1049B} {274, 1049}

\bibitem[\protect\citeauthoryear{{Bernardeau}}{{Bernardeau}}{1995}]{Bernardeau1995}
{Bernardeau} F.,  1995, \aap, \href
  {https://ui.adsabs.harvard.edu/abs/1995A&A...301..309B} {301, 309}

\bibitem[\protect\citeauthoryear{{Bernardeau} \& {Reimberg}}{{Bernardeau} \&
  {Reimberg}}{2016}]{seminalLDT}
{Bernardeau} F.,  {Reimberg} P.,  2016, \mn@doi [\prd]
  {10.1103/PhysRevD.94.063520}, \href
  {https://ui.adsabs.harvard.edu/abs/2016PhRvD..94f3520B} {94, 063520}

\bibitem[\protect\citeauthoryear{{Bernardeau} \& {Valageas}}{{Bernardeau} \&
  {Valageas}}{2000}]{BernardeauValageas}
{Bernardeau} F.,  {Valageas} P.,  2000, \aap, \href
  {https://ui.adsabs.harvard.edu/abs/2000A&A...364....1B} {364, 1}

\bibitem[\protect\citeauthoryear{{Bernardeau}, {van Waerbeke}  \&
  {Mellier}}{{Bernardeau} et~al.}{1997}]{Bernardeau1997}
{Bernardeau} F.,  {van Waerbeke} L.,   {Mellier} Y.,  1997, \aap, \href
  {https://ui.adsabs.harvard.edu/abs/1997A&A...322....1B} {322, 1}

\bibitem[\protect\citeauthoryear{{Bernardeau}, {Colombi}, {Gazta{\~n}aga}  \&
  {Scoccimarro}}{{Bernardeau} et~al.}{2002}]{BernardeauReview}
{Bernardeau} F.,  {Colombi} S.,  {Gazta{\~n}aga} E.,   {Scoccimarro} R.,  2002,
  \mn@doi [\physrep] {10.1016/S0370-1573(02)00135-7}, \href
  {https://ui.adsabs.harvard.edu/abs/2002PhR...367....1B} {367, 1}

\bibitem[\protect\citeauthoryear{{Bernardeau}, {Pichon}  \&
  {Codis}}{{Bernardeau} et~al.}{2014a}]{2014PhRvD..90j3519B}
{Bernardeau} F.,  {Pichon} C.,   {Codis} S.,  2014a, \mn@doi [\prd]
  {10.1103/PhysRevD.90.103519}, \href
  {https://ui.adsabs.harvard.edu/abs/2014PhRvD..90j3519B} {90, 103519}

\bibitem[\protect\citeauthoryear{{Bernardeau}, {Pichon}  \&
  {Codis}}{{Bernardeau} et~al.}{2014b}]{BCP13}
{Bernardeau} F.,  {Pichon} C.,   {Codis} S.,  2014b, \mn@doi [\prd]
  {10.1103/PhysRevD.90.103519}, \href
  {https://ui.adsabs.harvard.edu/abs/2014PhRvD..90j3519B} {90, 103519}

\bibitem[\protect\citeauthoryear{{Bernardeau}, {Nishimichi}  \&
  {Taruya}}{{Bernardeau} et~al.}{2014c}]{Nulling}
{Bernardeau} F.,  {Nishimichi} T.,   {Taruya} A.,  2014c, \mn@doi [\mnras]
  {10.1093/mnras/stu1861}, \href
  {https://ui.adsabs.harvard.edu/abs/2014MNRAS.445.1526B} {445, 1526}

\bibitem[\protect\citeauthoryear{{Boyle}, {Uhlemann}, {Barthelemy},
  {Friedrich}, {Codis}  \& {Bernardeau}}{{Boyle} et~al.}{2020}]{Boyle2020}
{Boyle} A.,  {Uhlemann} C.,  {Barthelemy} A.,  {Friedrich} O.,  {Codis} S.,
  {Bernardeau} F.,  2020, In prep

\bibitem[\protect\citeauthoryear{{Deshpande} et~al.,}{{Deshpande}
  et~al.}{2020}]{2020A&A...636A..95D}
{Deshpande} A.~C.,  et~al., 2020, \mn@doi [\aap] {10.1051/0004-6361/201937323},
  \href {https://ui.adsabs.harvard.edu/abs/2020A&A...636A..95D} {636, A95}

\bibitem[\protect\citeauthoryear{{Euclid Collaboration} et~al.,}{{Euclid
  Collaboration} et~al.}{2019}]{EuclidEmulator}
{Euclid Collaboration} et~al., 2019, \mn@doi [\mnras] {10.1093/mnras/stz197},
  \href {https://ui.adsabs.harvard.edu/abs/2019MNRAS.484.5509E} {484, 5509}

\bibitem[\protect\citeauthoryear{{Friedrich} et~al.,}{{Friedrich}
  et~al.}{2018}]{FriedrichDES17}
{Friedrich} O.,  et~al., 2018, \mn@doi [\prd] {10.1103/PhysRevD.98.023508},
  \href {https://ui.adsabs.harvard.edu/abs/2018PhRvD..98b3508F} {98, 023508}

\bibitem[\protect\citeauthoryear{Gavriliadis \& Athanassoulis}{Gavriliadis \&
  Athanassoulis}{2009}]{GAVRILIADIS20097}
Gavriliadis P.,  Athanassoulis G.,  2009, \mn@doi [Journal of Computational and
  Applied Mathematics] {https://doi.org/10.1016/j.cam.2008.10.011}, 229, 7

\bibitem[\protect\citeauthoryear{{Gressel}, {Bonvin}, {Bruni}  \&
  {Bacon}}{{Gressel} et~al.}{2019}]{2019JCAP...05..045G}
{Gressel} H.~A.,  {Bonvin} C.,  {Bruni} M.,   {Bacon} D.,  2019, \mn@doi
  [\jcap] {10.1088/1475-7516/2019/05/045}, \href
  {https://ui.adsabs.harvard.edu/abs/2019JCAP...05..045G} {2019, 045}

\bibitem[\protect\citeauthoryear{{Gruen} et~al.,}{{Gruen}
  et~al.}{2018}]{GruenDES17}
{Gruen} D.,  et~al., 2018, \mn@doi [\prd] {10.1103/PhysRevD.98.023507}, \href
  {https://ui.adsabs.harvard.edu/abs/2018PhRvD..98b3507G} {98, 023507}

\bibitem[\protect\citeauthoryear{{Hilbert} et~al.,}{{Hilbert}
  et~al.}{2020}]{2020MNRAS.493..305H}
{Hilbert} S.,  et~al., 2020, \mn@doi [\mnras] {10.1093/mnras/staa281}, \href
  {https://ui.adsabs.harvard.edu/abs/2020MNRAS.493..305H} {493, 305}

\bibitem[\protect\citeauthoryear{{Ivezi{\'c}} et~al.,}{{Ivezi{\'c}}
  et~al.}{2019}]{LSST}
{Ivezi{\'c}} {\v{Z}}.,  et~al., 2019, \mn@doi [\apj]
  {10.3847/1538-4357/ab042c}, \href
  {https://ui.adsabs.harvard.edu/abs/2019ApJ...873..111I} {873, 111}

\bibitem[\protect\citeauthoryear{{Kacprzak} et~al.,}{{Kacprzak}
  et~al.}{2016}]{2016MNRAS.463.3653K}
{Kacprzak} T.,  et~al., 2016, \mn@doi [\mnras] {10.1093/mnras/stw2070}, \href
  {https://ui.adsabs.harvard.edu/abs/2016MNRAS.463.3653K} {463, 3653}

\bibitem[\protect\citeauthoryear{{Kaiser}}{{Kaiser}}{1992}]{1992ApJ...388..272K}
{Kaiser} N.,  1992, \mn@doi [\apj] {10.1086/171151}, \href
  {https://ui.adsabs.harvard.edu/abs/1992ApJ...388..272K} {388, 272}

\bibitem[\protect\citeauthoryear{{Kaiser}}{{Kaiser}}{1995}]{kkaiser1994}
{Kaiser} N.,  1995, \mn@doi [\apjl] {10.1086/187730}, \href
  {https://ui.adsabs.harvard.edu/abs/1995ApJ...439L...1K} {439, L1}

\bibitem[\protect\citeauthoryear{{Kilbinger}}{{Kilbinger}}{2015}]{kilbinger15}
{Kilbinger} M.,  2015, \mn@doi [Reports on Progress in Physics]
  {10.1088/0034-4885/78/8/086901}, \href
  {https://ui.adsabs.harvard.edu/abs/2015RPPh...78h6901K} {78, 086901}

\bibitem[\protect\citeauthoryear{{Laureijs} et~al.,}{{Laureijs}
  et~al.}{2011}]{Euclid}
{Laureijs} R.,  et~al., 2011, arXiv e-prints, \href
  {https://ui.adsabs.harvard.edu/abs/2011arXiv1110.3193L} {p. arXiv:1110.3193}

\bibitem[\protect\citeauthoryear{{Lepori}, {Adamek}, {Durrer}, {Clarkson}  \&
  {Coates}}{{Lepori} et~al.}{2020}]{2020MNRAS.497.2078L}
{Lepori} F.,  {Adamek} J.,  {Durrer} R.,  {Clarkson} C.,   {Coates} L.,  2020,
  \mn@doi [\mnras] {10.1093/mnras/staa2024}, \href
  {https://ui.adsabs.harvard.edu/abs/2020MNRAS.497.2078L} {497, 2078}

\bibitem[\protect\citeauthoryear{Lewis \& Bridle}{Lewis \& Bridle}{2002}]{camb}
Lewis A.,  Bridle S.,  2002, \mn@doi [\prd] {10.1103/PhysRevD.66.103511}, 66,
  103511

\bibitem[\protect\citeauthoryear{{Lin}, {Kilbinger}  \& {Pires}}{{Lin}
  et~al.}{2016}]{2016A&A...593A..88L}
{Lin} C.-A.,  {Kilbinger} M.,   {Pires} S.,  2016, \mn@doi [\aap]
  {10.1051/0004-6361/201628565}, \href
  {https://ui.adsabs.harvard.edu/abs/2016A&A...593A..88L} {593, A88}

\bibitem[\protect\citeauthoryear{{LoVerde} \& {Afshordi}}{{LoVerde} \&
  {Afshordi}}{2008}]{ExtendedLimber}
{LoVerde} M.,  {Afshordi} N.,  2008, \mn@doi [\prd]
  {10.1103/PhysRevD.78.123506}, \href
  {https://ui.adsabs.harvard.edu/abs/2008PhRvD..78l3506L} {78, 123506}

\bibitem[\protect\citeauthoryear{{Martinet} et~al.,}{{Martinet}
  et~al.}{2018}]{2018MNRAS.474..712M}
{Martinet} N.,  et~al., 2018, \mn@doi [\mnras] {10.1093/mnras/stx2793}, \href
  {https://ui.adsabs.harvard.edu/abs/2018MNRAS.474..712M} {474, 712}

\bibitem[\protect\citeauthoryear{{Martinet}, {Harnois-D{\'e}raps}, {Jullo}  \&
  {Schneider}}{{Martinet} et~al.}{2020a}]{2020arXiv201007376M}
{Martinet} N.,  {Harnois-D{\'e}raps} J.,  {Jullo} E.,   {Schneider} P.,  2020a,
  arXiv e-prints, \href {https://ui.adsabs.harvard.edu/abs/2020arXiv201007376M}
  {p. arXiv:2010.07376}

\bibitem[\protect\citeauthoryear{{Martinet}, {Castro}, {Harnois-D{\'e}raps},
  {Jullo}, {Giocoli}  \& {Dolag}}{{Martinet}
  et~al.}{2020b}]{2020arXiv201209614M}
{Martinet} N.,  {Castro} T.,  {Harnois-D{\'e}raps} J.,  {Jullo} E.,  {Giocoli}
  C.,   {Dolag} K.,  2020b, arXiv e-prints, \href
  {https://ui.adsabs.harvard.edu/abs/2020arXiv201209614M} {p. arXiv:2012.09614}

\bibitem[\protect\citeauthoryear{{Matilla}, {Waterval}  \& {Haiman}}{{Matilla}
  et~al.}{2020}]{2020AJ....159..284M}
{Matilla} J. M.~Z.,  {Waterval} S.,   {Haiman} Z.,  2020, \mn@doi [\aj]
  {10.3847/1538-3881/ab8f8c}, \href
  {https://ui.adsabs.harvard.edu/abs/2020AJ....159..284M} {159, 284}

\bibitem[\protect\citeauthoryear{{Mead}, {Brieden}, {Tr{\"o}ster}  \&
  {Heymans}}{{Mead} et~al.}{2020}]{HMcode}
{Mead} A.,  {Brieden} S.,  {Tr{\"o}ster} T.,   {Heymans} C.,  2020, arXiv
  e-prints, \href {https://ui.adsabs.harvard.edu/abs/2020arXiv200901858M} {p.
  arXiv:2009.01858}

\bibitem[\protect\citeauthoryear{Mellier}{Mellier}{1999}]{kappadef}
Mellier Y.,  1999, \mn@doi [Annual Review of Astronomy and Astrophysics]
  {10.1146/annurev.astro.37.1.127}, 37, 127

\bibitem[\protect\citeauthoryear{{Munshi}, {Valageas}  \& {Barber}}{{Munshi}
  et~al.}{2004}]{2004MNRAS.350...77M}
{Munshi} D.,  {Valageas} P.,   {Barber} A.~J.,  2004, \mn@doi [\mnras]
  {10.1111/j.1365-2966.2004.07553.x}, \href
  {https://ui.adsabs.harvard.edu/abs/2004MNRAS.350...77M} {350, 77}

\bibitem[\protect\citeauthoryear{Patton, Blazek, Honscheid, Huff, Melchior,
  Ross  \& Suchyta}{Patton et~al.}{2017}]{Patton17}
Patton K.,  Blazek J.,  Honscheid K.,  Huff E.,  Melchior P.,  Ross A.~J.,
  Suchyta E.,  2017, \mn@doi [Monthly Notices of the Royal Astronomical
  Society] {10.1093/mnras/stx1626}, 472, 439

\bibitem[\protect\citeauthoryear{{Peebles}}{{Peebles}}{1980}]{Peebles}
{Peebles} P.~J.~E.,  1980, {The large-scale structure of the universe}

\bibitem[\protect\citeauthoryear{{Petri}, {Haiman}  \& {May}}{{Petri}
  et~al.}{2016}]{2016PhRvD..93f3524P}
{Petri} A.,  {Haiman} Z.,   {May} M.,  2016, \mn@doi [\prd]
  {10.1103/PhysRevD.93.063524}, \href
  {https://ui.adsabs.harvard.edu/abs/2016PhRvD..93f3524P} {93, 063524}

\bibitem[\protect\citeauthoryear{{Petri}, {Haiman}  \& {May}}{{Petri}
  et~al.}{2017}]{Petri17}
{Petri} A.,  {Haiman} Z.,   {May} M.,  2017, \mn@doi [\prd]
  {10.1103/PhysRevD.95.123503}, \href
  {https://ui.adsabs.harvard.edu/abs/2017PhRvD..95l3503P} {95, 123503}

\bibitem[\protect\citeauthoryear{{Planck Collaboration} et~al.,}{{Planck
  Collaboration} et~al.}{2020}]{planck}
{Planck Collaboration} et~al., 2020, \mn@doi [\aap]
  {10.1051/0004-6361/201833880}, \href
  {https://ui.adsabs.harvard.edu/abs/2020A&A...641A...1P} {641, A1}

\bibitem[\protect\citeauthoryear{{Porth}, {Smith}, {Simon}, {Marian}  \&
  {Hilbert}}{{Porth} et~al.}{2020}]{2020arXiv200608665P}
{Porth} L.,  {Smith} R.~E.,  {Simon} P.,  {Marian} L.,   {Hilbert} S.,  2020,
  \mn@doi [\mnras] {10.1093/mnras/staa2900}, \href
  {https://ui.adsabs.harvard.edu/abs/2020MNRAS.tmp.2730P} {}

\bibitem[\protect\citeauthoryear{{Reimberg} \& {Bernardeau}}{{Reimberg} \&
  {Bernardeau}}{2018}]{paolo}
{Reimberg} P.,  {Bernardeau} F.,  2018, \mn@doi [\prd]
  {10.1103/PhysRevD.97.023524}, \href
  {https://ui.adsabs.harvard.edu/abs/2018PhRvD..97b3524R} {97, 023524}

\bibitem[\protect\citeauthoryear{{Schneider}}{{Schneider}}{1996}]{schneider1996}
{Schneider} P.,  1996, \mn@doi [\mnras] {10.1093/mnras/283.3.837}, \href
  {https://ui.adsabs.harvard.edu/abs/1996MNRAS.283..837S} {283, 837}

\bibitem[\protect\citeauthoryear{{Schneider}, {van Waerbeke}, {Jain}  \&
  {Kruse}}{{Schneider} et~al.}{1998}]{Schneider1998}
{Schneider} P.,  {van Waerbeke} L.,  {Jain} B.,   {Kruse} G.,  1998, \mn@doi
  [\mnras] {10.1046/j.1365-8711.1998.01422.x}, \href
  {https://ui.adsabs.harvard.edu/abs/1998MNRAS.296..873S} {296, 873}

\bibitem[\protect\citeauthoryear{Schneider et~al.}{Schneider
  et~al.}{2019}]{Baryonification}
Schneider et~al., 2019, \mn@doi [JCAP] {10.1088/1475-7516/2019/03/020}, \href
  {https://ui.adsabs.harvard.edu/abs/2019JCAP...03..020S} {2019, 020}

\bibitem[\protect\citeauthoryear{{Takahashi}, {Oguri}, {Sato}  \&
  {Hamana}}{{Takahashi} et~al.}{2011}]{Takahashi11}
{Takahashi} R.,  {Oguri} M.,  {Sato} M.,   {Hamana} T.,  2011, \mn@doi [\apj]
  {10.1088/0004-637X/742/1/15}, \href
  {https://ui.adsabs.harvard.edu/abs/2011ApJ...742...15T} {742, 15}

\bibitem[\protect\citeauthoryear{{Takahashi}, {Sato}, {Nishimichi}, {Taruya}
  \& {Oguri}}{{Takahashi} et~al.}{2012}]{Halofit}
{Takahashi} R.,  {Sato} M.,  {Nishimichi} T.,  {Taruya} A.,   {Oguri} M.,
  2012, \mn@doi [\apj] {10.1088/0004-637X/761/2/152}, \href
  {https://ui.adsabs.harvard.edu/abs/2012ApJ...761..152T} {761, 152}

\bibitem[\protect\citeauthoryear{{Takahashi}, {Hamana}, {Shirasaki},
  {Namikawa}, {Nishimichi}, {Osato}  \& {Shiroyama}}{{Takahashi}
  et~al.}{2017}]{Simulation}
{Takahashi} R.,  {Hamana} T.,  {Shirasaki} M.,  {Namikawa} T.,  {Nishimichi}
  T.,  {Osato} K.,   {Shiroyama} K.,  2017, \mn@doi [\apj]
  {10.3847/1538-4357/aa943d}, 850, 24

\bibitem[\protect\citeauthoryear{{Taylor}, {Bernardeau}  \& {Huff}}{{Taylor}
  et~al.}{2020}]{x-cut}
{Taylor} P.~L.,  {Bernardeau} F.,   {Huff} E.,  2020, arXiv e-prints, \href
  {https://ui.adsabs.harvard.edu/abs/2020arXiv200700675T} {p. arXiv:2007.00675}

\bibitem[\protect\citeauthoryear{{Uhlemann}, {Pajer}, {Pichon}, {Nishimichi},
  {Codis}  \& {Bernardeau}}{{Uhlemann} et~al.}{2018a}]{NonGaussianities}
{Uhlemann} C.,  {Pajer} E.,  {Pichon} C.,  {Nishimichi} T.,  {Codis} S.,
  {Bernardeau} F.,  2018a, \mn@doi [\mnras] {10.1093/mnras/stx2623}, \href
  {https://ui.adsabs.harvard.edu/abs/2018MNRAS.474.2853U} {474, 2853}

\bibitem[\protect\citeauthoryear{{Uhlemann}, {Pichon}, {Codis}, {L'Huillier},
  {Kim}, {Bernardeau}, {Park}  \& {Prunet}}{{Uhlemann}
  et~al.}{2018b}]{cylindres}
{Uhlemann} C.,  {Pichon} C.,  {Codis} S.,  {L'Huillier} B.,  {Kim} J.,
  {Bernardeau} F.,  {Park} C.,   {Prunet} S.,  2018b, \mn@doi [\mnras]
  {10.1093/mnras/sty664}, \href
  {https://ui.adsabs.harvard.edu/abs/2018MNRAS.477.2772U} {477, 2772}

\bibitem[\protect\citeauthoryear{{Uhlemann}, {Friedrich},
  {Villaescusa-Navarro}, {Banerjee}  \& {Codis}}{{Uhlemann}
  et~al.}{2019}]{Uhlemann19}
{Uhlemann} C.,  {Friedrich} O.,  {Villaescusa-Navarro} F.,  {Banerjee} A.,
  {Codis} S.~r.,  2019, arXiv e-prints, \href
  {https://ui.adsabs.harvard.edu/abs/2019arXiv191111158U} {p. arXiv:1911.11158}

\bibitem[\protect\citeauthoryear{{Valageas}}{{Valageas}}{2002}]{Valageas}
{Valageas} P.,  2002, \mn@doi [\aap] {10.1051/0004-6361:20011663}, \href
  {https://ui.adsabs.harvard.edu/abs/2002A&A...382..412V} {382, 412}

\bibitem[\protect\citeauthoryear{{Vallis}, {Wallis}  \& {Kitching}}{{Vallis}
  et~al.}{2018}]{2018A&C....24...84V}
{Vallis} Z.~M.,  {Wallis} C.~G.~R.,   {Kitching} T.~D.,  2018, \mn@doi
  [Astronomy and Computing] {10.1016/j.ascom.2018.06.004}, \href
  {https://ui.adsabs.harvard.edu/abs/2018A&C....24...84V} {24, 84}

\bibitem[\protect\citeauthoryear{{Zorrilla Matilla}, {Sharma}, {Hsu}  \&
  {Haiman}}{{Zorrilla Matilla} et~al.}{2020}]{DeepLearning}
{Zorrilla Matilla} J.~M.,  {Sharma} M.,  {Hsu} D.,   {Haiman} Z.,  2020, arXiv
  e-prints, \href {https://ui.adsabs.harvard.edu/abs/2020arXiv200706529Z} {p.
  arXiv:2007.06529}

\bibitem[\protect\citeauthoryear{{Z{\"u}rcher}, {Fluri}, {Sgier}, {Kacprzak}
  \& {Refregier}}{{Z{\"u}rcher} et~al.}{2020}]{2020arXiv200612506Z}
{Z{\"u}rcher} D.,  {Fluri} J.,  {Sgier} R.,  {Kacprzak} T.,   {Refregier} A.,
  2020, arXiv e-prints, \href
  {https://ui.adsabs.harvard.edu/abs/2020arXiv200612506Z} {p. arXiv:2006.12506}

\makeatother
\end{thebibliography}

\appendix
\section{Non-linear covariance along the line of sight}
\label{nonlinearvariance}

As a consequence of extending the results of large deviation theory to finite, non-zero values of the variance, one needs a prescription to compute the (co)variances for example appearing in equation~(\ref{psicyl}). Past results (for instance \cite{cylindres}) on the 3D density field relied on modelling the non-linear variance by re-scaling the linear value
\begin{equation}
    \sigma_{nl}(R \rho^{1/2}) = \frac{\sigma_{nl}(R)}{\sigma_l(R)}\sigma_l(R \rho^{1/2}),
    \label{NonLinearParam}
\end{equation}
where $\sigma_{nl}^2(R)$ was acting as the driving parameter taken in its non-zero, finite value not predicted by the theory and thus left as a free parameter and measured in data or numerical simulations, or computed with some non-linear prescription for the power spectrum. 

This construction ensures that the reduced cumulants of the density field are exactly those obtained through standard tree-order PT and is also what one should typically use for projection of the density field, for example the convergence field as was shown in \cite{Barthelemy20a}. However, for the case of the joint statistics of the density field at different scales the choice of a unique driving parameter prevents us from imposing all the correct quadratic contributions in the CGF which is particularly problematic for the 1-point PDF of the aperture mass. Note indeed that sums of random variables satisfying a large deviation principle do not necessarily satisfy the same principle. As a consequence, we in this work choose to model the non-linear covariance, not by a re-scaling by the driving parameter which thus do not matter in the final expression of the CGF, but by a full non-linear prescription coming in our case from Halofit. This ensures that all quadratic contributions in the CGF are correctly modelled and modifies the standard tree-order PT results keeping their functional but using a non-linear power spectrum where the linear ones usually appears. One would then need to compare these results to high-order loop calculations to check the meaning of these corrections, this is left for future work.

\section{Technical comments on the effective mapping approach to the aperture mass (S)CGF}
\label{technical}

The procedure described in section~\ref{analyticalCGF} for the aperture mass CGF -- \textit{i.e} a projected quantity -- is from a mathematical point of view strictly identical to the construction of the density slope CGF in each redshift slice along the line of sight, solely changing the physics (cylindrical collapse VS effective collapse) and the initial conditions (smoothed VS un-smoothed Gaussian field). This in particular means that the effective procedure fails, at least on paper, to convey some of the information that the large-deviation approach to the aperture mass CGF does encode. Though this is not reflected in the successive cumulants of the field which are well reproduced by the effective approach, this is the case in the tails of the PDF or equivalently in the CGF near the critical points. Indeed it can be shown \citep{BCP13} by expanding the stationary condition, be it (\ref{stationnary}) or (\ref{EffectiveStationary}), near a critical point $\lambda_c$ that the density CGF behaves like $\phi_{\rho}(\lambda) \sim (\lambda-\lambda_c)^{3/2}$, which leads to the PDF exhibiting an exponential cut-off in $P_{\rho}(\rho) \sim \exp(-\lambda_c \rho)/(\rho-\rho_c)^{5/2}$. As for projected CGFs, projection effects do modify their asymptotic behaviour near critical points since they now behave like -- see for instance appendix A of \cite{BernardeauValageas} -- $\phi_{\rm proj}(\lambda) \sim (\lambda-\lambda_c)^{2}\log(\lambda-\lambda_c)$ which obviously changes the exponential cut-off of the PDF. Thus a more tedious but more appropriate effective approach to the analytical aperture mass CGF would rather be to fit an effective mapping at the level of each redshift slice along the line of sight. Fortunately, we did not in practice find any significant difference in the PDF tails between fitting an effective collapse at each slice or directly on the projected CGF, which is explained by the fact that the asymptotic result is reached rather far in the tails \citep{BCP13} and thus the main contributions to the shape of the PDF come from the few first cumulants.

A similar remark can be made regarding the degree of the effective mapping that directly influences the positions of the critical points along the real axis. Indeed we saw in the previous paragraph that the asymptotic behaviour of the PDF does change with their positions. However, we again find that this is not in practice a crucial issue and find that a polynomial mapping of degree 5 well describes the numerical generating functions.

Finally, we find it interesting to explicit that effectively, only the knowledge of the first few cumulants allows us to recover the entire PDF. This is as we saw enforced by the formal construction obtained through assumption that the $\map$ satisfies -- this is only true with the effective mapping -- a large deviation principle, and also by the fact that we impose our rate function to be convex through usage of the Legendre rather than Legendre-Fenchel transform, which effectively acts as if the rate function was linear beyond the critical points. The unchanging convexity of the rate function imposes that the PDF is uni-modal. This can be for example put into perspective with works in the field of mathematics, see for example \cite{GAVRILIADIS20097}, that present, in general, how much of the PDF can be recovered just from the information of a certain number of moments. It turns out that uni-modal PDFs can be very well approximated by a formal and general reconstruction coming from their first few cumulants. This partly explains why the effective mapping approach can work to such precision given that more constrains apply.

\section{Aperture mass cumulants at tree order}

We derive in this section the expression of the first cumulants of the aperture mass before integration along the line of sight. This derivation is greatly facilitated by use of the large deviation principle and is strictly equivalent to perturbation theory up to the first non-trivial term (i.e at tree-order). 

Let us first derive the expression for joint cumulants of the density field. The exact spherical collapse mapping can be written as 
\begin{equation}
    \rho(\tau) = 1 + \delta(\bar{\tau}) = \sum_{k \geqslant 0} \frac{\nu_k}{k!} \bar{\tau}^k
\end{equation}
where $\bar{\tau}$ is the mean linear density contrast given through the most probable mapping between the linear and late-time density field, $\delta$ its non-linear counterpart, and where $\nu_k$ are the spherically averaged perturbation theory kernels \citep{BernardeauReview}
\begin{equation}
    \nu_{k}= k! \int \mathrm{d} \Omega_{1} \ldots \mathrm{d} \Omega_{k} F_{i}\left(\mathbf{k}_{1}, \ldots, \mathbf{k}_{\mathbf{k}}\right).
    \label{vertices}
\end{equation}
Then the joint rate function of the density field filtered in two disks of radii $R_1$ and $R_2$ is
\begin{equation}
    \psi(\rho_1,\rho_2)=\frac{1}{2} \sum_{k,j}\Xi_{kj}(\rho_1^{1/2}R_1,\rho_2^{1/2}R_2)\bar{\tau}(\rho_k)\bar{\tau}(\rho_j),
\end{equation}
and its associated CGF is
\begin{equation}
    \begin{dcases}
    \phi(\lambda_1,\lambda_2) = \lambda_1\rho_1 + \lambda_2\rho_2 -\psi(\rho_1,\rho_2) \\ \lambda_1 = \frac{\partial}{\partial \rho_1} \psi(\rho_1,\rho_2) \\ \lambda_2 = \frac{\partial}{\partial \rho_2} \psi(\rho_1,\rho_2).
    \end{dcases}
\end{equation}
This allows to define the two operators
\begin{equation}
    \begin{dcases}
    D_1 = \frac{\partial}{\partial \lambda_1} =  \frac{\psi_{,22}}{\det\psi_{,ij}}\frac{\partial}{\partial \rho_1} - \frac{\psi_{,12}}{\det\psi_{,ij}}\frac{\partial}{\partial \rho_2} \\ D_2 = \frac{\partial}{\partial \lambda_2} =  \frac{\psi_{,11}}{\det\psi_{,ij}}\frac{\partial}{\partial \rho_2} - \frac{\psi_{,12}}{\det\psi_{,ij}}\frac{\partial}{\partial \rho_1}
    \end{dcases}
\end{equation}
which leads to 
\begin{equation}
    \left.\left\langle \rho_1^{1+p}\rho_2^{1+q} \right\rangle_c \!\!\!= \frac{\partial \phi(\lambda_1,\lambda_2)}{\partial \lambda_1^{1+p}\partial \lambda_2^{1+q}}\right|_{\lambda_{i}=0} \!\!\!\!\!\!\!\!\!= \left.D_{1}^{p} D_{2}^{q}\left(\frac{-\psi_{,12}}{\operatorname{det} \psi_{,i j}}\right)\right|_{\tau_{i}=0}\!\!\!\!\!\!\!.
    \label{jointkn}
\end{equation}

The same can be done in the simpler case of the 1-cell CGF and one gets
\begin{equation}
    \left\langle \rho^{2+p} \right\rangle_c = \left.\frac{\mathrm{d} \phi(\lambda)}{\mathrm{d} \lambda^{2+p}}\right|_{\lambda=0} \!\!\!\!\!\!= \left.\left[\frac{1}{\psi^{\prime \prime}(\rho)} \frac{\mathrm{d}}{\mathrm{d} \rho}\right]^{p}\left(\frac{1}{\psi^{\prime \prime}(\rho)}\right)\right|_{\tau=0}.
    \label{singlekn}
\end{equation}

Then combining equations~(\ref{jointkn}),(\ref{singlekn}) and (\ref{knXY}) one gets up to the integration along the line of sight
\begin{eqnarray}
    \left\langle \map^3 \right\rangle_c\!\!\!&\!\!\! =\!\!\!&\!\!\! \sigma^2_{\map}\left[\sigma^2(R_2)\Big(3\nu_2 + \frac{3}{2}\frac{\partial \log \sigma^2_{\map}}{\partial \log R_2}\Big) \right.\nonumber\\&-&\!\!\!\!\! \sigma^2(R_1)\Big(3\nu_2 + \frac{3}{2}\frac{\partial \log \sigma^2_{\map}}{\partial \log R_1}\Big)\nonumber \\&+&\!\!\!\!\!\left. \frac{3}{2}\sigma^2(R_1,R_2) \Big(\frac{\partial \log \sigma^2_{\map}}{\partial \log R_1}\! -\! \frac{\partial \log \sigma^2_{\map}}{\partial \log R_2}\Big)  \right]
    \label{mapskewness}
\end{eqnarray}
where 
\begin{equation}
    \sigma^2_{\map} = \sigma^2(R_1)+\sigma^2(R_2) - 2\sigma^2(R_1,R_2).
\end{equation}

Other cumulants can of course also be obtained in a similar fashion although their algebraic expression might become more and more complicated. The procedure can nonetheless be implemented in a symbolic calculus software and be used there.

\section{Effective mapping and aperture mass cumulants}
\label{EffectiveNu}

We here compute the relation between the $\mu$ coefficients of the effective mapping
\begin{equation}
    \zeta(\tau_{\rm eff}) = \sum_{k = 0}^n \frac{\mu_{k}}{k!} \tau_{\rm eff}^k,
\end{equation}
and the cumulants of the aperture mass whose SCGF is given by
\begin{equation}
    \varphi_{\map}(\lambda) = \lambda \zeta(\tau_{\rm eff}) - \frac{1}{2}\tau_{\rm eff}^2 ,
\end{equation}
with the stationary condition written as
\begin{equation}
    \lambda = \frac{\rm d}{\rm d \zeta} \frac{\tau_{\rm eff}^2}{2} = \tau_{\rm eff} \left(\frac{{\rm d}\zeta(\tau_{\rm eff})}{{\rm d}\tau_{\rm eff}}\right)^{-1}.
\end{equation}

Applying the same procedure leading to equation~(\ref{singlekn}) we arrive at 
\begin{align}
    S_{3, \map} &= 3\mu_{2} ,\\
    S_{4, \map} &= 12\mu_{2}^2 + 4\mu_{3}, \\
    S_{5, \map} &= 60\mu_{2}^3 + 60\mu_{2}\mu_{3} + 5\mu_{4} ,\\
    S_{6, \map} &= 360\mu_{2}^4 + 720\mu_{2}^2\mu_{3} + 90\mu_{3}^2 + 120\mu_{2}\mu_{4} + 6\mu_{5}.
\end{align}
These relations can be inverted and one finally arrives at
\begin{align}
    \mu_2 &= \frac{S_3}{3}, \\
    \mu_3 &= \frac{-4S_3^2+3S_4}{12}, \\
    \mu_4 &= \frac{40S_3^3 - 45 S_3 S_4 + 9 S_5}{45}, \\
    \mu_5 &= \frac{-560 S_3^4\!+\! 840 S_3^2 S_4 \!-\! 135 S_4^2 \!-\! 192 S_3 S_5 \!+\! 24 S_6}{144}.
\end{align}

\section{Sensitivity to the filtering scheme}
\label{sensitivity}

\begin{figure}
    \centering
    \includegraphics[width = \columnwidth]{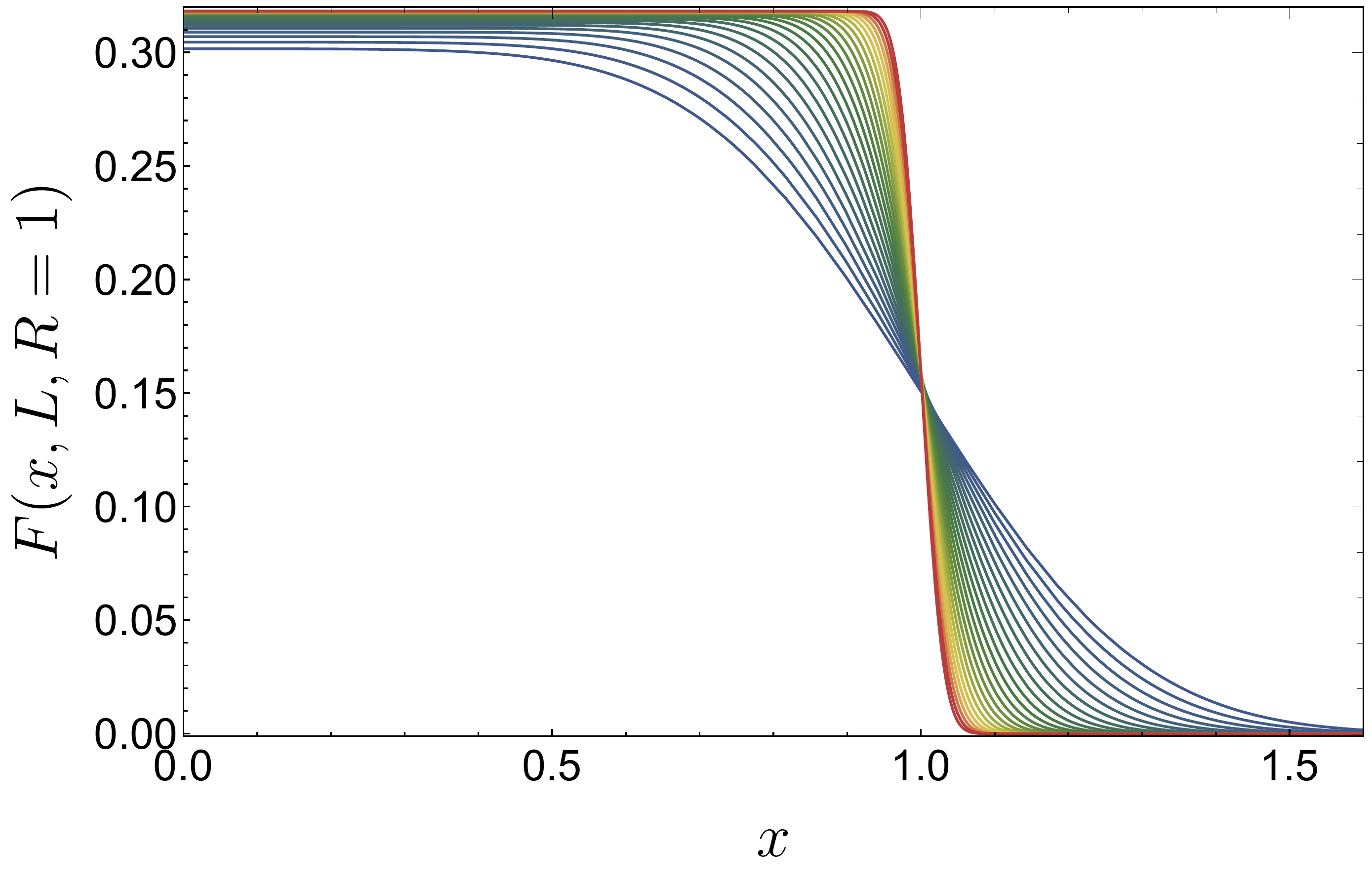}
    \caption{Smoothed top-hat window function for values of $L$ from 3 (blue) to 30 (red) and $R =1$.}
    \label{newfilter}
\end{figure}

One issue of top-hat filtering is its "sharpness" (non differentiability at the radius) which could make its precise implementation to real, pixelated data challenging. Hence the purpose of this section is to test the sensitivity of the predicted value of the aperture mass skewness to a slight modification of its filter by using a smoothed $C^{\infty}$ version of the top-hat defined by
\begin{equation}
    F(x,L,R) = \frac{1}{2} \left(1+ {\rm erf}\left[L \left(1 -\frac{x}{R}\right)\right]\right)/V(L,R),
    \label{eq::newfilter}
\end{equation}
where $R$ is the radius of the top-hat, $L$ is a parameter that influences the sharpness of the filter -- we recover a top-hat in the limit where $L$ tends to infinity -- and $V(L,R)$ is the normalisation of the filter defined by
\begin{equation}
    V(L,R) = \pi R^2 \frac{\left(2 e^{-L^2}\!/\!\sqrt{\pi} + (1+2L^2)(1+{\rm erf}(L))\right)}{4L^2}.
\end{equation}
We show in Fig.~\ref{newfilter} the shape of the filter depending on the value of its sharpness parameter $L$. Finally note that equation~(\ref{eq::newfilter}) is built so that the departure from an actual top-hat depends solely on $L$ and not on the radius so that the width needed for the filter to go from its maximum value to 0 is i) centred around the radius and ii) a fixed percentage of its value for a given $L$. Roughly, $L = 20$ corresponds to $\pm 10 \%$ around $R$, $L = 40$ is $\pm 5 \%$ and so on.

The principle behind the implementation of any filter using large deviation theory lies in its expression as a linear combination of top-hat kernels taken in the continuous limit \citep{seminalLDT,paolo}. First let us notice that since
\begin{equation}
    \kappa_{<\theta}\equiv \int_{0}^{\theta} \frac{\mathrm{d}^{2} \boldsymbol{\vartheta}}{\pi \theta^{2}} \kappa(\boldsymbol{\vartheta})
\end{equation}
then we obtain, by differentiating, 
\begin{equation}
    \kappa(\vartheta)=\kappa_{< \vartheta}+\frac{\vartheta}{2} \kappa_{< \vartheta}^{\prime}.
\end{equation}

Thus for any filter defining some quantity $\hat{\kappa}$ we get by integration by part
\begin{equation}
    \hat{\kappa}_=\int \mathrm{d}^{2} \boldsymbol{\vartheta} \, U(\vartheta) \kappa(\boldsymbol{\vartheta})=\int \mathrm{d} \vartheta \, \hat{U}(\vartheta) \kappa_{<\vartheta}
\end{equation}
with 
\begin{equation}
    \hat{U}(\vartheta) = -\pi \vartheta^2 U^{\prime}(\vartheta).
\end{equation}

\begin{figure}
    \centering
    \includegraphics[width = \columnwidth]{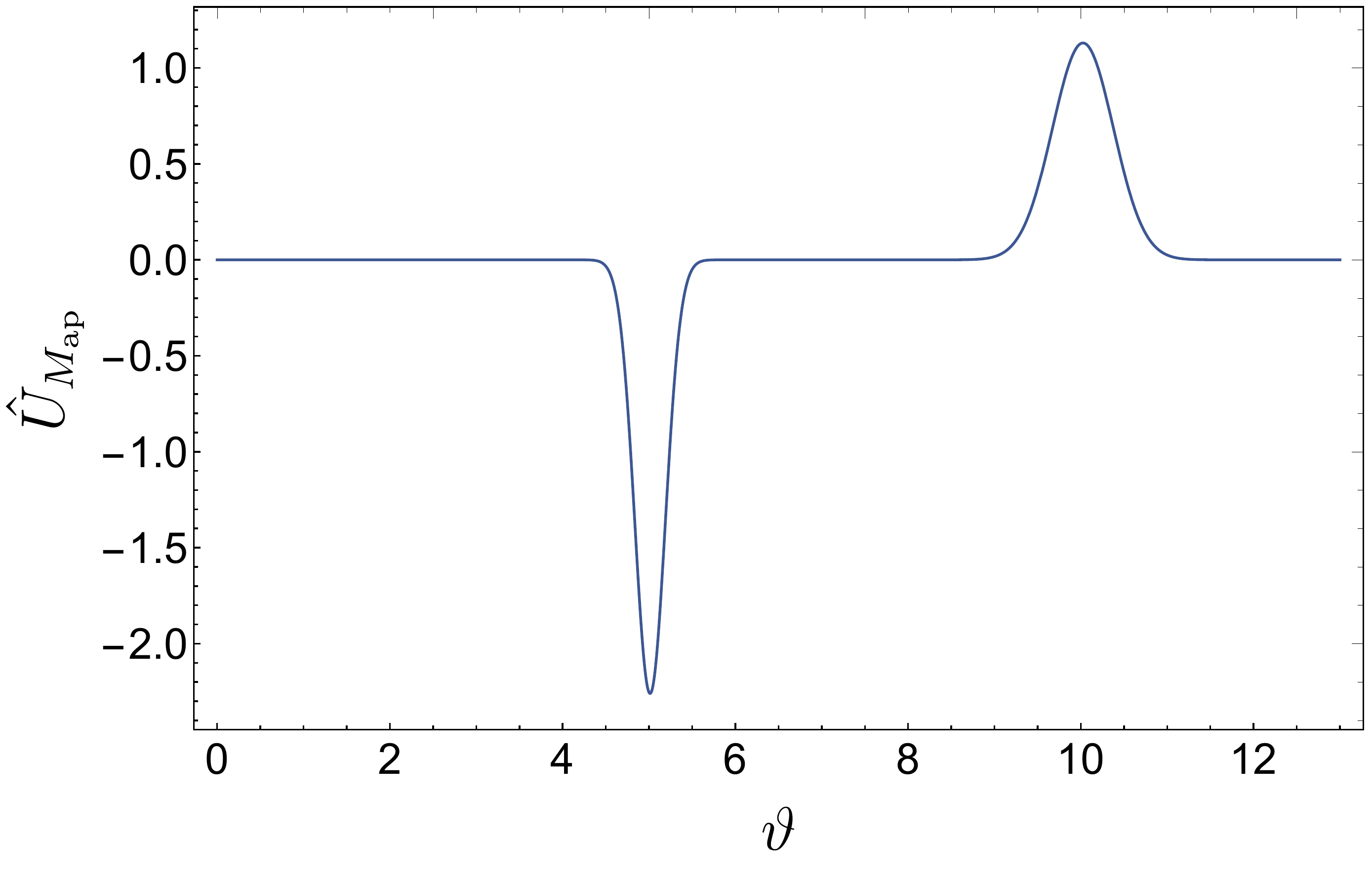}
    \caption{$\hat{U}_{\map}$ for $\theta_1 = 5$, $\theta_2 =10$ and $L = 20$. The departure from the exact top-hat for this $L$ is roughly $\pm 10 \%$ around the radius.}
    \label{hatU}
\end{figure}

For our definition of the aperture mass as the difference of the convergence smooth-top-hat filtered at two different scales,
\begin{multline}
    \hat{U}_{\map} (\vartheta) = \frac{4 \vartheta ^2 e^{L^2} L^3}{\sqrt{\pi } e^{L^2} \left(2 L^2+1\right) (\text{erf}(L)+1)+2 L} \\ \left(\frac{\exp(-L^2 (\theta_2-\vartheta )^2/\theta_2^2)}{\theta_2^3}-\frac{\exp(-L^2 (\theta_1-\vartheta )^2/\theta_1^2)}{\theta_1^3}\right)
\end{multline}
is shown in Fig.~\ref{hatU}. Note that $\hat{U}_{\map}$ would be a difference of two Dirac delta functions if we were using regular top-hats as is done in the main text. For simplicity we also use the same value of $L$ for the two different scales which leads to broader $\hat{U}_{\map}$ as $\theta_i$ increases though this might not be an important issue for radii relatively close one to another.

We now need to express the SCGF of the density in each slice along the line of sight. This accounts to write the continuous limit of equations~(\ref{varadhan}) and (\ref{psicyl}) re-written as
\begin{equation}
    \varphi(\lambda)=\sup_{\{\rho_i\}}\left[\!\lambda \sum_{i}^{N} \hat{U}_{i} \rho_i-\frac{\sigma^2\left(R_N\right)}{2} \sum_{k,j}\Xi_{kj}(\{\tau_{i}\})\tau_{k}\tau_{j}\!\right]\!.
\end{equation}
This is done by writing
\begin{multline}
\sum_{i=1}^{N} \hat{U}_{i}\left(R_{i}\right) \rho_{R_i} \rightarrow \int \mathrm{d}R \hat{U}(R) \zeta\left(\tau_r\right) \\
=\int \mathrm{d} r \frac{\mathrm{d} R}{\mathrm{d} r} \hat{U}\left[R(r)\right] \zeta\left(\tau_r\right)
\end{multline}
with $\rho_i = \zeta(\tau_i)$ the cylindrical (2D spherical) collapse, $R$ the smoothing radius of the non-linear density and $r$ its value for the linear field expressed through mass-conservation as
\begin{equation}
    r = R \rho^{1/2}.
\end{equation}
We also need a continuous limit to the cross-correlation matrix $\Xi$ which is obtained assuming the existence of an object $\xi\left(r^{\prime}, r^{\prime \prime}\right)$ defined by
\begin{equation}
    \int \mathrm{d} r^{\prime} \sigma^{2}\left(r, r^{\prime}\right) \xi\left(r^{\prime}, r^{\prime \prime}\right)=\delta_{\mathrm{D}}\left(r-r^{\prime \prime}\right).
\end{equation}
Therefore the continuous limit to the SCGF can be written as
\begin{multline}
    \varphi(\lambda)=\sup_{\tau}\Bigg[\lambda \int \mathrm{d} r \frac{\mathrm{d} R}{\mathrm{d} r} \hat{U}\left[R(r)\right] \zeta\left(\tau_r\right)\\-\frac{\sigma^2_{\rm slice}}{2} \int {\rm d}r {\rm d}r^{\prime} \tau_r \tau_{r^{\prime}} \xi(r,r^{\prime})\Bigg],
\end{multline}
and $\sigma^2_{\rm slice} = \int {\rm d}r {\rm d}r^{\prime} \sigma^2(r,r^{\prime}) \hat{U}(r) \hat{U}(r^{\prime})$. Note that the practical implementation of this extremization problem is not straightforward as shown in \cite{paolo}.

Thus expanding the values of $\varphi(\lambda)$ around 0 leads to the expression of the reduced cumulants. For the skewness in each slice one gets
\begin{equation}
    S_{3}^{\rm slice}\!\!=\!3 \nu_{2} \! \frac{\int \mathrm{d} x \hat{U}(x) \Sigma^{2}(x)}{\left[\int \mathrm{d} x \hat{U}(x) \Sigma(x)\right]^{2}}\!+\!3 \frac{\int \mathrm{d} x x \hat{U}(x) \Sigma(x) \Sigma^{\prime}(x)}{\left[\int \mathrm{d} x \hat{U}(x) \Sigma(x)\right]^{2}},
    \label{newS3}
\end{equation}
where the coefficient in front of the second term is actually 6 over the dimension of the collapse, $\nu_2$ is the usual spherical collapse coefficient and we have
\begin{equation}
    \Sigma(x)=\int \mathrm{d} y \sigma^{2}(x, y) \hat{U}(y).
\end{equation}

\begin{figure}
    \centering
    \includegraphics[width = \columnwidth]{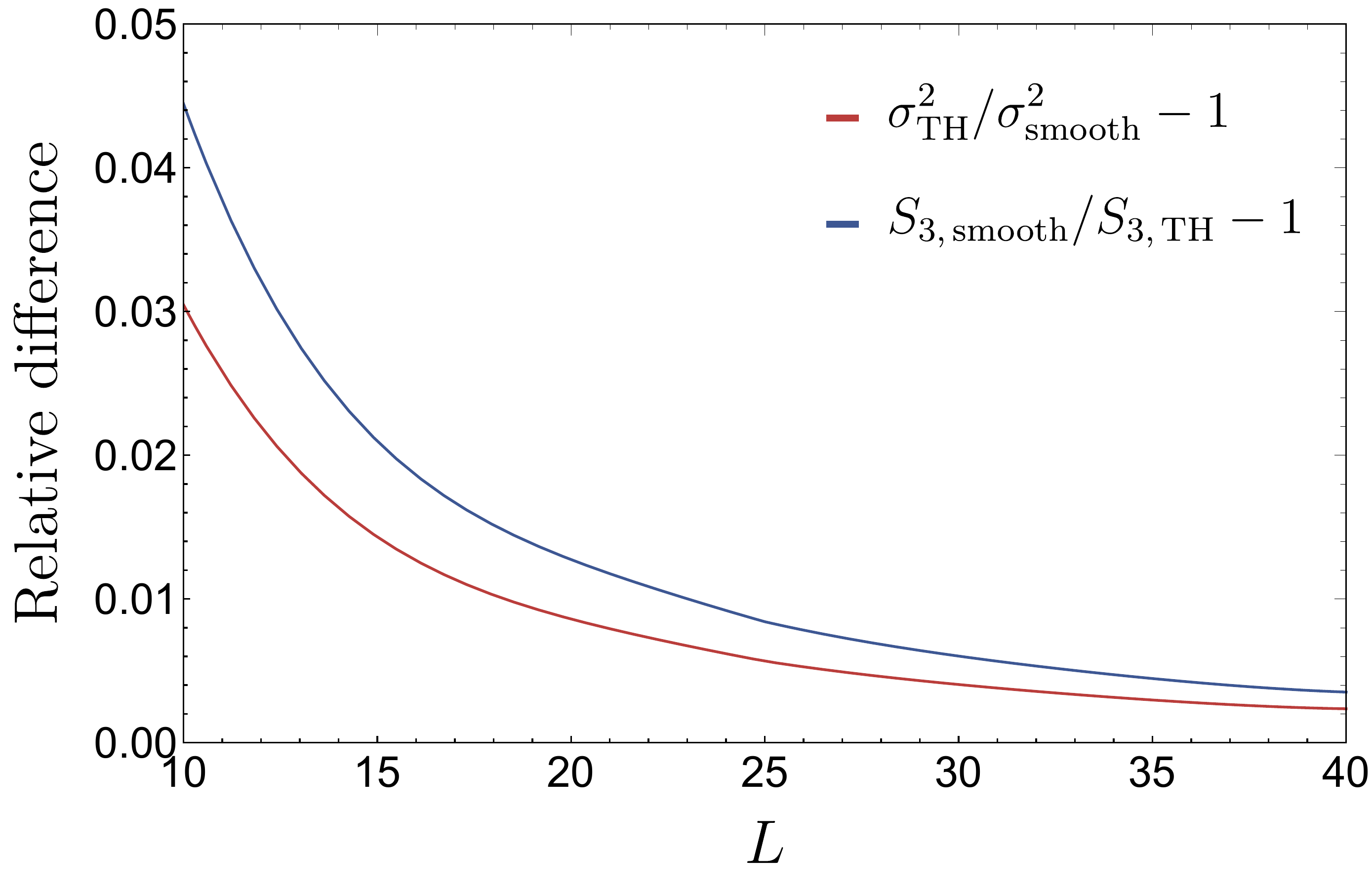}
    \caption{Red line: Relative difference between the $\map$ variance coming from exact top-hat windows and using its smoothed counterpart in equation~(\ref{eq::newfilter}). Blue line: Same thing but for the $\map$ reduced skewness. $\sigma^2_{\rm TH} =  2.17 \ 10^{-7}$ and $S_{3, {\rm TH}} = -596.08$.}
    \label{relativeS3}
\end{figure}

Now combining the result in equation~(\ref{newS3}) and the projection formula (\ref{projection}) one can get to the aperture mass skewness taking into account the geometry of the light-cone as well as the smoothed top-hat filter (\ref{eq::newfilter}). For a nulled aperture mass field with source redshifts located at $z_s = 1.2 - 1.4 - 1.65$ and smoothing angles $\theta_1 = 10$ and $\theta_2 = 20$ arcmin we both compute the variance and the reduced skewness using an exact top-hat window and its smoothed version for different values of $L$. We also use an Halofit power-spectrum. The relative difference between the two is displayed in Fig.~\ref{relativeS3} where on can see that the values of both the variance and skewness depend very weakly on the value of $L$. This is overall good news since approximate filtering schemes could thus be considered with still relatively good validity of our theoretical modelling.

\section{Second order corrections}
\label{2ndOrder}

\subsection{Post-born corrections}
Throughout this work we modelled the aperture mass as a weighted integral over the underlying non-linear density field, therefore assuming independent lenses and following light rays along un-perturbed lines of sight in a $\Lambda$CDM universe. 
Hence two ingredients were neglected: i) couplings between the lenses which state that the combination of lenses in geometrical optics is not linear, and ii) the fact that background lenses are themselves lensed by foreground lenses, thus changing the overall trajectory of light rays. These terms sometimes known as post-Born corrections will tend to Gaussianise the lensing fields since they characterise the introduction of random deflections along the light path which will in turn tend to diminish the impact of the non-linear clustering of matter. The mental image one could form is that of clustered chunks of matter blurred by these lensing terms. Those corrections were shown to be of little importance in the case of the PDF of the convergence field at scales relevant for perturbation theory ($\theta \sim 10$ arcmin) and relatively small redshifts relevant for upcoming cosmic shear surveys such as Euclid/LSST ($z \sim 1-2$) \citep{Barthelemy20b}.

Extending this result to the aperture mass PDF boils down to computing the post-Born corrections on the dominant non-Gaussian term that makes the PDF, namely its reduced skewness. Using equation~(41) of \cite{Barthelemy20b} the post-Born correction to the third moment of a filtered convergence field is given by
\begin{multline}
    \left\langle M_{\mathrm{ap}}^3\right\rangle_{\rm corr} \!\!\!=\!  \frac{-12}{(2 \pi)^4}\!\! \int_0^{z}\!\!\! \!\!\frac{{\rm d}z'}{H(z')} \, \omega_{(\rm null)}(z',z)^2\!\! \int_0^{z'}\!\!\!\!\! \frac{{\rm d}z''}{H(z'')} \, \omega_{(\rm null)}(z''\!,z)\\ \hskip 1. cm \omega(z''\!,z') \int \!\!\frac{{\rm d}^2{\bm \ell_1} {\rm d}^2{\bm \ell_2}}{\left[\mathcal{D}(z')\mathcal{D}(z'')\right]^2} \, P\!\left(\!\frac{\ell_1}{\mathcal{D}(z')},z'\!\right) P\!\left(\!\frac{\ell_2}{\mathcal{D}(z'')}, z''\!\right) \\  H({\bm \ell_1},{\bm \ell_2}) W(\ell_1)W(\ell_2)W(|{\bm \ell_1}+{\bm \ell_2}|),
    \label{k3}
\end{multline}
where
\begin{equation}
    H({\bm \ell_1},{\bm \ell_2}) = \frac{{\bm \ell_1}\cdot {\bm \ell_2}}{\ell_2^2} + \frac{({\bm \ell_1}\cdot {\bm \ell_2})^2}{\ell_1^2\ell_2^2},
    \label{H}
\end{equation}
and
\begin{equation}
    W(l) = 2\frac{J_1(l \theta_2)}{l\theta_2}-2\frac{J_1(l \theta_1)}{l\theta_1},
\end{equation}
is the $M_{\mathrm{ap}}$ window function in Fourier space and $J_1$ is the first order Bessel function of the first kind. The subscript (null) indicates where to input the nulled lensing kernel instead of the usual one if one implements nulling. Equation~(\ref{k3}) is general enough so that any window function applied on the convergence field can be used to get the post-Born corrective term to the skewness. Furthermore when involving top-hat windows, specific properties of Bessel functions can be used to further simplify the expression
\begin{multline}
    \int_{0}^{2 \pi} \mathrm{d} \varphi W_{TH}\left(\left|\boldsymbol{\ell}_{1}+\boldsymbol{\ell}_{2}\right|\right)\left[1-\cos ^{2}(\varphi)\right]\\=\pi W_{TH}\left(\ell_{1}\right) W_{TH}\left(\ell_{2}\right)
    \label{prop1}
\end{multline}
and
\begin{multline}
    \int_{0}^{2 \pi} \mathrm{d} \varphi W_{TH}\left(\left|\boldsymbol{\ell}_{1}+\boldsymbol{\ell}_{2}\right|\right) \left[1+\cos (\varphi) \frac{\ell_{1}}{\ell_{2}}\right] \\ =2 \pi W_{TH}\left(\ell_{2}\right)\left[W_{TH}\left(\ell_{1}\right)+\frac{\ell_{1}}{2} W_{TH}^{\prime}\left(\ell_{1}\right)\right],
    \label{prop2}
\end{multline}
with
\begin{equation}
    W_{TH}(l) = 2\frac{J_1(l)}{l}.
\end{equation}
Thus plugging equations~(\ref{prop1}) and (\ref{prop2}) yields
\begin{multline}
    \left\langle M_{\mathrm{ap}}^3\right\rangle_{\rm corr} = - \frac{12}{\pi^2 \theta_1^3 \theta_2^3} \int_0^{z} \frac{{\rm d}z'}{H(z')} \, \omega_{(\rm null)}(z',z)^2 \int_0^{z'} \frac{{\rm d}z''}{H(z'')} \\ \omega_{(\rm null)}(z'',z) \, \omega(z'',z') \int \frac{{\rm d}{\ell_1} {\rm d}{\ell_2}}{\left[\mathcal{D}(z')\mathcal{D}(z'')\right]^2} \, P\left(\frac{\ell_1}{\mathcal{D}(z')},z'\right) \\ P\left(\frac{\ell_2}{\mathcal{D}(z'')}, z''\right) \frac{1}{\ell_2}\bigg((\theta_1 J_1(\ell_1 \theta_2)-\theta_2 J_1(\ell_1 \theta_1)) (\theta_1 J_1(\ell_2 \theta_2) \\- \theta_2 J_1(\ell_2 \theta_1)) (\theta_2 (J_2(\ell_1 \theta_1)-J_0(\ell_1 \theta_1)) J_1(\ell_2 \theta_1)+\theta_1 (J_0(\ell_1 \theta_2)\\-J_2(\ell_1 \theta_2)) J_1(\ell_2 \theta_2))\bigg).
\label{map3}
\end{multline}

\begin{figure}
    \centering
    \includegraphics[width = \columnwidth]{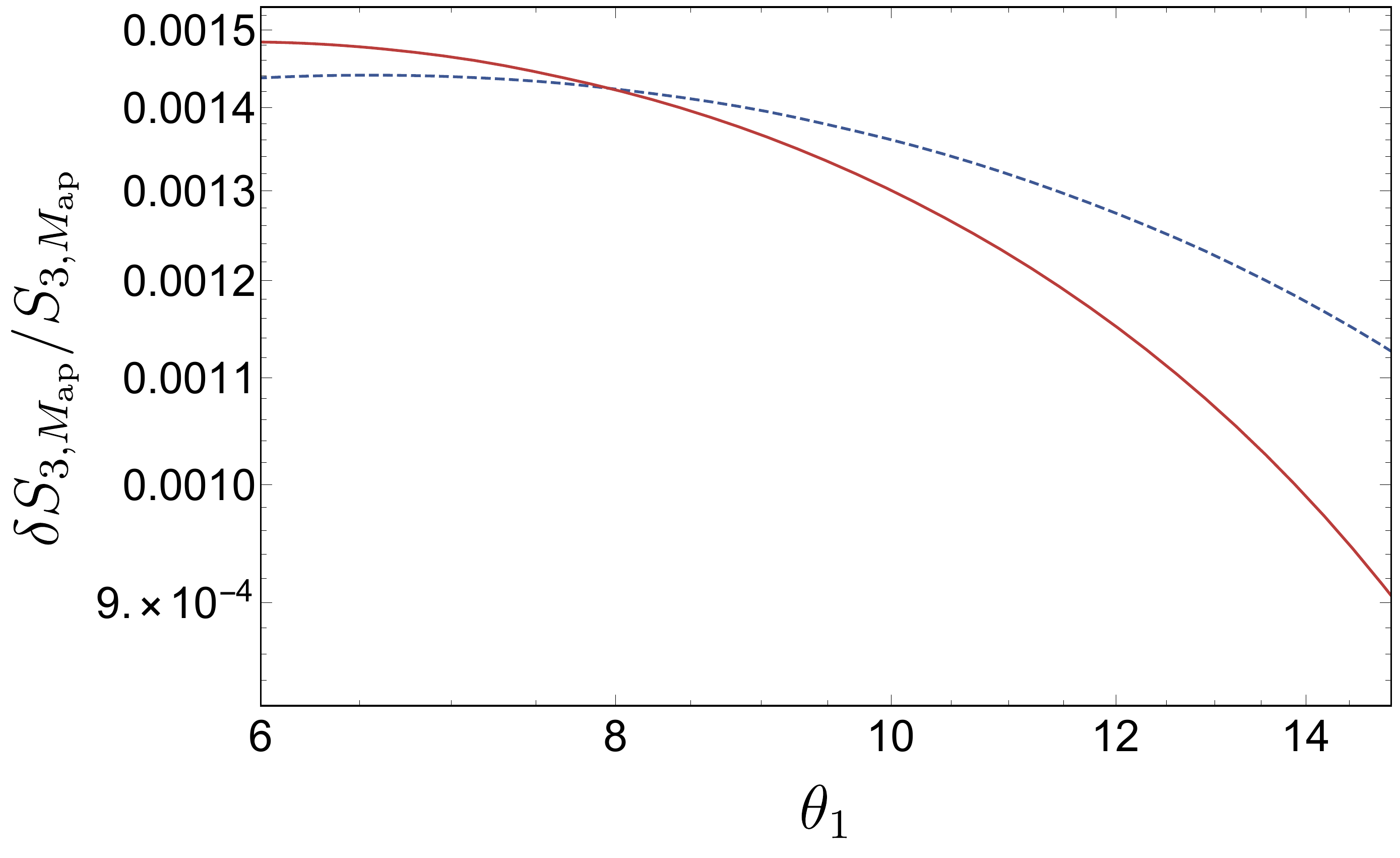}
    \caption{Impact of the leading order corrections to the BNT $M_{\rm ap}$ skewness induced by post-born (blue) and reduced shear (red) terms at source redshifts $z_s = 1.2, 1.4 \ \& \ 1.6$. The field is filtered by a difference of top-hats with opening angles $\theta_1$ and $\theta_2 = 2 \theta_1$ and we plot the evolution of the correction with respect to $\theta_1$ in arcmin. Dashed lines indicate negative values.}
    \label{PostBornMap}
\end{figure}

For a difference of top-hats with opening angles $\theta_1$ and $\theta_2 = 2 \theta_1$ we obtain the correction on the nulled $M_{\mathrm{ap}}$ skewness due to lens-lens coupling and geodesic deviation shown in Fig.~\ref{PostBornMap} for source redshifts located at $z_s = 1.2, 1.4 \ \& \ 1.6$ as a function of $\theta_1$. As expected, the correction is shown to i) Gaussianise the field and ii) be very small -- sub-percent -- which is not surprising since reducing the lensing kernel diminishes the importance of couplings between lenses.

\subsection{Reduced shear correction}

Assuming that the intrinsic ellipticity of galaxies has no preferred orientation, note that this is not the case in the presence of intrinsic galaxy alignments, the observed ellipticity is an unbiased estimator of the reduced shear $g = \gamma/(1-\kappa)$ rather than the shear itself. This implies that rather than equation~(\ref{Mapshear}), the observed physical aperture mass is given by
\begin{equation}
    \begin{split}
        M_{\mathrm{ap}}^g(\bm{\vartheta})&=\int \mathrm{d}^{2} \bm{\vartheta}^{\prime} Q_\theta\left(\vartheta^{\prime}\right) g_{t}\left(\bm{\vartheta}-\bm{\vartheta}^{\prime}\right) \\ &\approx M_{\mathrm{ap}}(\bm{\vartheta}) + \int \mathrm{d}^{2} \bm{\vartheta}^{\prime} Q_\theta\left(\vartheta^{\prime}\right) \gamma_{t}\left(\bm{\vartheta}-\bm{\vartheta}^{\prime}\right) \kappa\left(\bm{\vartheta}-\bm{\vartheta}^{\prime}\right) \\ &\approx M_{\mathrm{ap}}(\bm{\vartheta}) + \delta\map(\bm{\vartheta}).
         \label{Mapreducedshear}
    \end{split}
\end{equation}

We derive in this section the leading correction to the aperture mass skewness that accounts for the reduced shear. This is merely a perturbation theory calculation which still somewhat relies on the fact that the convergence $\kappa$ is small and described by linear perturbation theory. Though this is rather straightforward and very useful for an estimation of the amplitude of the effect, this is not at the same level that what can be performed using large deviation theory to account for the effect as is shown without any projection effects in \cite{paolo}.

First let us recall that one can deduce from their definition as derivatives of the projected gravitational potential that the convergence and the shear are equal in harmonic space up to a phase. Denoting  $\Tilde{f} $ quantities in harmonic space ($f$ being either the shear or the convergence field in our case) and introducing the wave vector ${\bm \ell} = (\ell \cos\varphi_s,\ell \sin\varphi_s)$ conjugate of ${\bm \vartheta} = (\vartheta \cos \varphi,\vartheta \sin \varphi)$ we have
\begin{equation}
    \Tilde{\gamma}({\bm \ell}) = \Tilde{\kappa}({\bm \ell}) e^{2 i \varphi_s}.
\end{equation}
Since the tangential shear is defined as
\begin{equation}
    \gamma_t({\bm \vartheta}) = - \Re\left(\gamma({\bm \vartheta})e^{-2 i \varphi}\right),
\end{equation}
then it can be expressed as a function of the convergence field through
\begin{equation}
    \gamma_t({\bm \vartheta}) = - \int \frac{{\rm d}^2{\bm l}}{(2 \pi)^2} \cos\left(2 (\varphi - \varphi_s)\right) e^{i {\bm \ell}\cdot {\bm \vartheta}} \Tilde{\kappa}({\bm \ell}).
\end{equation}

Up to leading order, the skewness of the physical aperture mass is then
\begin{equation}
    \left\langle (M_{\mathrm{ap}}^g)^3\right\rangle \approx  \left\langle \map^3\right\rangle + 3 \left\langle \map^2 \, \delta\map\right\rangle
\end{equation}
where the reduced shear correction is thus written as
\begin{multline}
    \left\langle \map^2 \, \delta\map\right\rangle = - \int {\rm d}^2 {\bm \vartheta_1} U_\theta(\vartheta_1) \int \frac{{\rm d}^2{\bm \ell_1}}{(2 \pi)^2} e^{i{\bm \ell_1}\cdot{\bm \vartheta_1}} \\ \int {\rm d}^2 {\bm \vartheta_2} U_\theta(\vartheta_2) \int \frac{{\rm d}^2{\bm \ell_2}}{(2 \pi)^2} e^{i{\bm \ell_2}\cdot{\bm \vartheta_2}} \int {\rm d}^2{\bm \vartheta} Q_\theta(\vartheta) \int \frac{{\rm d}^2{\bm \ell^{\prime}}}{(2 \pi)^2} e^{i{\bm \ell^{\prime}}\cdot{\bm \vartheta}} \\ \int \!\!\frac{{\rm d}^2{\bm \ell}}{(2 \pi)^2} e^{i{\bm \ell}\cdot{\bm \vartheta}} \cos\left(2(\varphi\!-\!\varphi_s)\right) \left\langle\Tilde{\kappa}({\bm \ell_1}) \Tilde{\kappa}({\bm \ell_2}) \Tilde{\kappa}({\bm \ell}) \Tilde{\kappa}({\bm \ell^{\prime}})\right\rangle.
    \label{corr1}
\end{multline}
Again stopping at leading order, we only consider the linear evolution of the matter density fluctuation which means that the $\kappa$ field is Gaussian. This allows to use Wick's theorem to compute the correlator in equation~(\ref{corr1}) which becomes
\begin{multline}
   \frac{ \left\langle\Tilde{\kappa}({\bm \ell_1}) \Tilde{\kappa}({\bm \ell_2}) \Tilde{\kappa}({\bm \ell}) \Tilde{\kappa}({\bm \ell^{\prime}})\right\rangle}{(2\pi)^4} =  C_{\ell}^{\kappa}(\ell_1)C_{\ell}^{\kappa}(\ell^{\prime})\delta_D(\ell_1+\ell_2)\delta_D(\ell^{\prime}+\ell) \\\hskip 1.9 cm +  C_{\ell}^{\kappa}(\ell_1)C_{\ell}^{\kappa}(\ell_2)\delta_D(\ell_1+\ell)\delta_D(\ell^{\prime}+\ell_2) \\\hskip 1 cm +  C_{\ell}^{\kappa}(\ell_1)C_{\ell}^{\kappa}(\ell_2)\delta_D(\ell_1+\ell^{\prime})\delta_D(\ell+\ell_2).
\end{multline}
The first term will yield zero because of the integration of the cosine and the 2 other terms yield the same contribution. Carrying out the integration other ${\bm \ell_1}$ and ${\bm \ell_2}$ is straightforward thanks to the presence of the Dirac delta functions and we arrive at
\begin{multline}
    \left\langle \map^2 \, \delta\map\right\rangle = -2 \!\int {\rm d}^2 {\bm \vartheta_1} U_\theta(\vartheta_1)\! \int {\rm d}^2 {\bm \vartheta_2} U_\theta(\vartheta_2)\! \int\! {\rm d}^2{\bm \vartheta} Q_\theta(\vartheta) \\\hskip 3 cm \int \!\!\frac{{\rm d}^2{\bm \ell}}{(2 \pi)^2} e^{i{\bm \ell}\cdot({\bm \vartheta}-{\bm \vartheta_1})} C_{\ell}^{\kappa}(\ell) \cos\left(2(\varphi\!-\!\varphi_s)\right)\\
    \int \!\!\frac{{\rm d}^2{\bm \ell^{\prime}}}{(2 \pi)^2} e^{i{\bm \ell^{\prime}}\cdot({\bm \vartheta}-{\bm \vartheta_2})} C_{\ell}^{\kappa}(\ell^{\prime}).
\end{multline}
The integration over ${\bm \vartheta_1}$ ${\bm \vartheta_2}$ leads to the Fourier expression of the aperture mass (convergence-) filter and all that remains is to perform the angular integrations over ${\bm \vartheta}$, ${\bm \ell}$ and ${\bm \ell^{\prime}}$. These yield expressions that correspond to the integral definition of the Bessel functions of the first kind
\begin{equation}
    J_n(|x|) = \frac{1}{2 \pi (i)^n} \!\! \int_{-\pi}^{\pi} \!\! {\rm d}\varphi \, e^{i (x\cos(\varphi) -n \varphi)}, \ n \in \mathbb{N}, \ x \in \mathbb{R}.
\end{equation}
The expression for the reduced shear correction to the skewness of the aperture mass is thus
\begin{multline}
     \left\langle (M_{\mathrm{ap}}^g)^3\right\rangle_{\rm corr} = \frac{3}{\pi} \int {\rm d}\ell \ell C_\ell^\kappa(\ell) W(\ell) \int {\rm d}\ell^{\prime} \ell^{\prime} C_\ell^\kappa(\ell^{\prime}) W(\ell^{\prime}) \\ \int {\rm d}\vartheta \vartheta Q_\theta(\vartheta) J_0(\vartheta \ell^{\prime})J_2(\vartheta \ell).
     \label{RedShearCorr}
\end{multline}

We now evaluate this correction in the same configuration as the post-born correction, that is nulling with evolving opening angles and plot the resulting correction in Fig.~\ref{PostBornMap}. Finally, note that in the case the nulling, the procedure is only exact for the convergence and shear fields and thus another term accounting for this issue should also be present when introducing the reduced shear correction. Taking into account all corrections that arise because of inaccuracies of the procedure in realistic settings is left for future work.

\subsection{Magnification bias correction}

Individual galaxies can be (de)magnified and thus their flux is (de)increased. At the flux limit of a survey, this can cause fainter sources to be included in the observed sample while they would, in the absence of lensing, be excluded. At the same time, the density of galaxies in the small region around this source appears reduced (increased) since it is also (de)magnified. As such, the net effect depends on the slope of the intrinsic, unlensed, galaxy luminosity function at the survey's flux limit. This is known as the magnification bias. We follow the prescription given in \cite{2020A&A...636A..95D} for the resulting "observed" shear (and not the reduced shear since we are looking at the leading order correction) which reads
\begin{equation}
    \gamma_{\rm obs} = \gamma + \gamma \delta^{{\rm g}} + (5s - 2)\gamma \kappa,
\end{equation}
where $\delta^{\rm g}$ is the intrinsic, unlensed, galaxy overdensity at the source (or in the redshift bin),
\begin{equation}
    s = \left.\frac{\partial \log_{10}(n(z_s,m))}{\partial m}\right|_{m_{\rm lim}},
\end{equation}
$n(z_s,m)$ is the true distribution of galaxies, evaluated at the source (central redshift of the bin) and at a given magnitude (luminous flux) $m$, and $m_{\rm lim}$ is the survey's limiting magnitude. The leading order correction to the skewness coming from this effect is exactly the same as in the case of the reduced shear in equation~(\ref{RedShearCorr}) up to a factor $(5s - 2)$. Indeed the $\gamma\delta^{\rm g}$ term will yield a zero contribution in any correlator if the lenses and sources do not overlap. This is the case when considering source planes or very narrow redshift bins but more importantly this is also the case when applying the BNT transform to any tomographic bins \citep{Nulling}. 

\section{Theoretical error bar on the skewness}
\label{appendix::error_bar}

\red{The purpose of this section is to present a fast and accurate estimate of the error bar one could expect in a realistic measurement of third cumulant of the Aperture mass field. This can be used both to estimate the precision one needs in one's modelling or more pragmatically to quantify to which extent non-gaussianities in the Aperture mass field can be detected in a given survey.}

\red{In practice the $3^{\rm rd}$ cumulant of the field is measured by an estimator that we choose here to be given by the so-called k-statistics. Thus defining the sums of the $r^{\rm th}$ powers of the $n$ independent data points (seen as $n$ effective independent realisations of the $\map$ values denoted $X_i$) as 
\begin{equation}
    s_r \equiv \sum_{i = 1}^n X^r_i,
\end{equation}
an unbiased estimator of $\langle\map^3\rangle_c$ that we call $k_3$ is given by
\begin{equation}
   k_3 = \frac{n^2 s_3-3 n s_2 s_1+2 s_1^3}{(n-2) (n-1) n}.
\end{equation}
We can thus compute the variance of the estimator, $\sigma^2_{k_3}$,
\begin{multline}
    \left\langle\left(k_3 - \langle\map^3\rangle_c\right)^2\right\rangle = \frac{6 \langle\map^2\rangle_c^3 n}{(n-2) (n-1)}+ \frac{9 \langle\map^4\rangle_c \langle\map^2\rangle_c}{n-1} \\ +\frac{9 \langle\map^3\rangle_c^2}{n-1}+\frac{\langle\map^6\rangle_c}{n},
    \label{vark3}
\end{multline}
which can then be used to estimate the error bar on the measured cumulant.}

\red{In the presence of Gaussian shape noise considered statistically independent from the $\map$ field, only $\langle\map^2\rangle_c$ is affected and one would only need to replace $\langle\map^2\rangle_c^3 n$ in equation~(\ref{vark3}) by $$ \langle\map^2\rangle_c \rightarrow \langle\map^2\rangle_c + \sigma_{SN, \map}^2 $$ given in equation~(\ref{sig2SN}).}

\red{The form given in equation~(\ref{vark3}) thus renders apparent how the number of data points, the shape noise and the amplitude of the signal which depends on the scales and redshifts probed will make the measurement of $\langle\map^3\rangle_c$ compatible with zero, that is whether or not non-Gaussian features will be detected in the PDF. There only remains to estimate the number of data points that are independent: Though not exact, since the correlation between disks will rapidly decay as their distance grows, we can put an upper limit on the number of data points by computing the number of non-overlapping disks one can draw on the surface area. The error bar computed in this fashion can be shown to be very close to the ones estimated from the numerical simulation.}

\red{Unfortunately for the scope of this paper and as seen in Fig.~\ref{fig::shapenoise}, the shape noise contribution on a single nulled bin of the BNT $\map$ is dominant and the complete study of how one could mitigate this fact in the context of a full tomographic analysis is beyond the scope of this paper and left for future work.}

\label{lastpage}
\end{document}